\documentclass[twocolumn]{aastex6}
\usepackage{graphicx, natbib, hyperref}
\hypersetup{colorlinks=true, urlcolor=black}
\usepackage[flushleft]{threeparttable}

\shorttitle{ASPECS: Continuum imaging in the UDF}
\shortauthors{Aravena et al.}

\def\lsim{\mathrel{\rlap{\lower 3pt \hbox{$\sim$}} \raise 2.0pt \hbox{$<$}}}
\def\gsim{\mathrel{\rlap{\lower 3pt \hbox{$\sim$}} \raise 2.0pt \hbox{$>$}}}

\begin{document}

\title{
The ALMA spectroscopic survey in the Hubble Ultra Deep Field: Continuum number counts, resolved 1.2-mm extragalactic background, and properties of the faintest dusty star forming galaxies}

\author{
M.~Aravena\altaffilmark{1}
R.~Decarli\altaffilmark{2}, 
F.~Walter\altaffilmark{2,3,4},
E.~Da~Cunha\altaffilmark{6},
F.~E.~Bauer\altaffilmark{7,8,9}
C.~L.~Carilli\altaffilmark{4,5},
E.~Daddi\altaffilmark{10},
D.~Elbaz\altaffilmark{10},
R.\,J.~Ivison\altaffilmark{11,12},
D.~A.~Riechers\altaffilmark{13},
I. Smail\altaffilmark{14,15},
A. M. Swinbank\altaffilmark{14,15},
A. Weiss \altaffilmark{16},
T. Anguita\altaffilmark{17, 8},
R. J. Assef\altaffilmark{1},
E. Bell\altaffilmark{18},
F. Bertoldi\altaffilmark{19},
R. Bacon\altaffilmark{20},
R. Bouwens\altaffilmark{21,22},
P. Cortes\altaffilmark{23,4},
P. Cox\altaffilmark{23},
J. G\'onzalez-L\'opez\altaffilmark{7},
J. Hodge\altaffilmark{21},
E. Ibar\altaffilmark{24},
H. Inami\altaffilmark{20},
L. Infante\altaffilmark{7},
A. Karim\altaffilmark{19},
O. Le F{\`e}vre\altaffilmark{25},
B. Magnelli\altaffilmark{19},
K. Ota\altaffilmark{26},
G. Popping\altaffilmark{11},
K. Sheth\altaffilmark{27},
P. van der Werf\altaffilmark{21},
J. Wagg\altaffilmark{28}
}
\altaffiltext{1}{N\'{u}cleo de Astronom\'{\i}a, Facultad de Ingenier\'{\i}a, Universidad Diego Portales, Av. Ej\'{e}rcito 441, Santiago, Chile. E-mail: {\sf manuel.aravenaa@mail.udp.cl}}
\altaffiltext{2}{Max-Planck Institut f\"{u}r Astronomie, K\"{o}nigstuhl 17, D-69117, Heidelberg, Germany.}
\altaffiltext{3}{Astronomy Department, California Institute of Technology, MC105-24, Pasadena, California 91125, USA}
\altaffiltext{4}{NRAO, Pete V.\,Domenici Array Science Center, P.O.\, Box O, Socorro, NM, 87801, USA}
\altaffiltext{5}{Astrophysics Group, Cavendish Laboratory, J. J. Thomson Avenue, Cambridge CB3 0HE, UK}
\altaffiltext{6}{Centre for Astrophysics and Supercomputing, Swinburne University of Technology, Hawthorn, Victoria 3122, Australia}
\altaffiltext{7}{Instituto de Astrof\'{\i}sica, Facultad de F\'{\i}sica, Pontificia Universidad Cat\'olica de Chile Av. Vicu\~na Mackenna 4860, 782-0436 Macul, Santiago, Chile}
\altaffiltext{8}{Millennium Institute of Astrophysics, Chile}
\altaffiltext{9}{Space Science Institute, 4750 Walnut Street, Suite 205, Boulder, CO 80301, USA}
\altaffiltext{10}{Laboratoire AIM, CEA/DSM-CNRS-Universite Paris Diderot, Irfu/Service d'Astrophysique, CEA Saclay, Orme des Merisiers, 91191 Gif-sur-Yvette cedex, France}
\altaffiltext{11}{European Southern Observatory, Karl-Schwarzschild Strasse 2, D-85748 Garching bei M\"unchen, Germany}
\altaffiltext{12}{Institute for Astronomy, University of Edinburgh, Blackford Hill, Edinburgh EH9 3HJ, UK}
\altaffiltext{13}{Cornell University, 220 Space Sciences Building, Ithaca, NY 14853, USA}
\altaffiltext{14}{Centre for Extragalactic Astronomy, Department of Physics, Durham University, South Road, Durham DH1 3LE, UK}
\altaffiltext{15}{Institute for Computational Cosmology, Durham University, South Road, Durham DH1 3LE, UK}
\altaffiltext{16}{Max-Planck-Institut f\"ur Radioastronomie, Auf dem H\"ugel 69, 53121 Bonn, Germany}
\altaffiltext{17}{Departamento de Ciencias F\'{\i}sicas, Universidad Andres Bello, Fernandez Concha 700, Las Condes, Santiago, Chile}
\altaffiltext{18}{Department of Astronomy, University of Michigan, 500 Church St, Ann Arbor, MI 48109, USA}
\altaffiltext{19}{Argelander Institute for Astronomy, University of Bonn, Auf dem H\"{u}gel 71, 53121 Bonn, Germany}
\altaffiltext{20}{Universit\'{e} Lyon 1, 9 Avenue Charles Andr\'{e}, 69561 Saint Genis Laval, France}
\altaffiltext{21}{Leiden Observatory, Leiden University, NL-2300 RA Leiden, The Netherlands}
\altaffiltext{22}{UCO/Lick Observatory, University of Califronia, Santa Cruz, CA 95064, USA}
\altaffiltext{23}{Joint ALMA Observatory - ESO, Av. Alonso de C\'ordova, 3104, Santiago, Chile}
\altaffiltext{24}{Instituto de F\'{\i}sica y Astronom\'{\i}a, Universidad de Valparaiso, Avda. Gran Breta\~na 1111, Valparaiso, Chile}
\altaffiltext{25}{Aix Marseille Universit\'{e}, CNRS, LAM (Laboratoire d'Astrophysique de Marseille) UMR 7326, 13388, Marseille, France}
\altaffiltext{26}{Kavli Institute for Cosmology, University of Cambridge, Madingley Road, Cambridge CB3 0HA, UK ; Cavendish Laboratory, University of Cambridge, 19 J.J. Thomson Avenue, Cambridge CB3 0HE, UK }
\altaffiltext{27}{Science Mission Directorate, NASA Headquarters, Washington, DC 20546-0001, USA}
\altaffiltext{28}{SKA Organization, Lower Withington Macclesfield, Cheshire SK11 9DL, UK}

\begin{abstract} 

We present an analysis of a deep (1$\sigma$=13\,$\mu$Jy) cosmological 1.2-mm continuum map based on ASPECS, the ALMA Spectroscopic Survey in the Hubble Ultra Deep Field. In the 1\,arcmin$^2$  covered by ASPECS we detect nine sources at $>3.5\sigma$ significance at 1.2-mm.  Our ALMA--selected sample has a median redshift of $z=1.6\pm0.4$, with only one galaxy detected at z$>$2 within the survey area. This value is significantly lower than that found in millimeter samples selected at a higher flux density cut-off and similar frequencies. Most galaxies have specific star formation rates similar to that of main sequence galaxies at the same epoch, and we find median values of stellar mass and star formation rates of $4.0\times10^{10}\ M_\sun$ and $\sim40~M_\sun$ yr$^{-1}$, respectively. Using the dust emission as a tracer for the ISM mass, we derive depletion times that are typically longer than 300\,Myr, and we find molecular gas fractions ranging from $\sim$0.1 to 1.0. As noted by previous studies, these values are lower than using CO--based ISM estimates by a factor $\sim$2. The 1\,mm number counts (corrected for fidelity and completeness) are in agreement with previous studies that were typically restricted to brighter sources. With our individual detections only, we recover $55\pm4\%$ of the extragalactic background light (EBL) at 1.2\,mm measured by the {\it Planck} satellite, and we recover $80\pm7\%$ of this EBL if we include the bright end of the number counts and additional detections from stacking. The stacked contribution is dominated by galaxies at $z\sim1-2$, with stellar masses of (1--3)$\times$10$^{10}$\,M$_\odot$. For the first time, we are able to characterize the population of galaxies that dominate the EBL at 1.2\,mm. 
\end{abstract}

\section{Introduction}

One of the most fundamental discoveries with regard to the cosmic evolution of galaxies has been the determination that a substantial fraction of the integrated Extragalactic Background Light (EBL) arises at infrared-to-millimeter wavelengths: the Cosmic Infrared Background (CIB). Quantitative observations of the CIB began with the Cosmic Background Explorer (COBE). At a low angular resolution ($0.7\deg$), COBE provided the first large-scale measurement of the spectral energy distribution (SED) of the EBL from the far-infrared to the (sub)millimeter \citep{puget96,fixsen98}. The CIB consists of the combined flux of all extragalactic sources, and contains much information about the history and  formation of galaxies, and of the large scale structure of the Universe. 

The observation that the cosmic density of star-formation was an order of magnitude higher at cosmological redshifts, $z\sim2-4$ \citep[e.g.,][]{madau96, lilly96}, opened the possibility that most of the CIB arose from dust re-processed UV-light from distant galaxies. These studies used the Lyman dropout technique to identify normal galaxies at high-redshift, being mostly insensitive to dust obscured star formation. Later, sensitive maps obtained with submillimeter/millimeter bolometer arrays were thus able to directly detect and identify luminous dusty star forming galaxies (DSFGs), which were soon found to contribute a fraction to the EBL at these wavelengths \citep[e.g.,][]{smail97}.


Since then, a number of groups have conducted (sub)millimeter surveys of the sky, currently yielding up to hundreds of sources in contiguous areas of the sky \citep[e.g.,][]{hughes98, barger98, eales00, bertoldi00, scott02, cowie02, voss06, bertoldi07,scott08,greve08, weiss09,austermann10,vieira10,aretxaga11,hatsukade11,scott12,mocanu13}. These blank field bolometer (sub)millimeter surveys discovered a population of luminous DSFGs at high redshift that were not accounted for in optical studies. These galaxies -- also called ``submillimeter galaxies'' (SMGs) due to the region of the electromagnetic spectrum in which they were first discovered -- have been characterised as massive starburst galaxies with typical stellar and molecular gas masses of $\sim10^{11}\ M_\sun$, typically located at $z=1-3$ \citep[e.g.,][]{chapman05} with a tail out to $z\sim6$ \citep{weiss13, riechers13}, and most likely driven by relatively bright mergers \citep{engel10}. As such, these galaxies are found to be gas/dust rich, with gas fractions typically exceeding 0.2 \citep[e.g.][]{daddi10a, tacconi10, magdis12, tacconi13,bothwell13}. Despite their large SFRs implied by the large IR luminosities ($>10^{12.0-12.5}\ L_\sun$) and significant abundance at high-redshift, these galaxies (e.g. $S_{\rm 1.2mm}>2-3$ mJy) were found to contribute only a minor fraction of the EBL at submillimeter wavelengths \citep{barger99,eales99,smail02,coppin06,knudsen08,weiss09,scott12, chen13}. Hence, questions about the properties of the population of galaxies that dominate this EBL remain. 
\begin{figure*}[ht]
\centering
\includegraphics[scale=0.55]{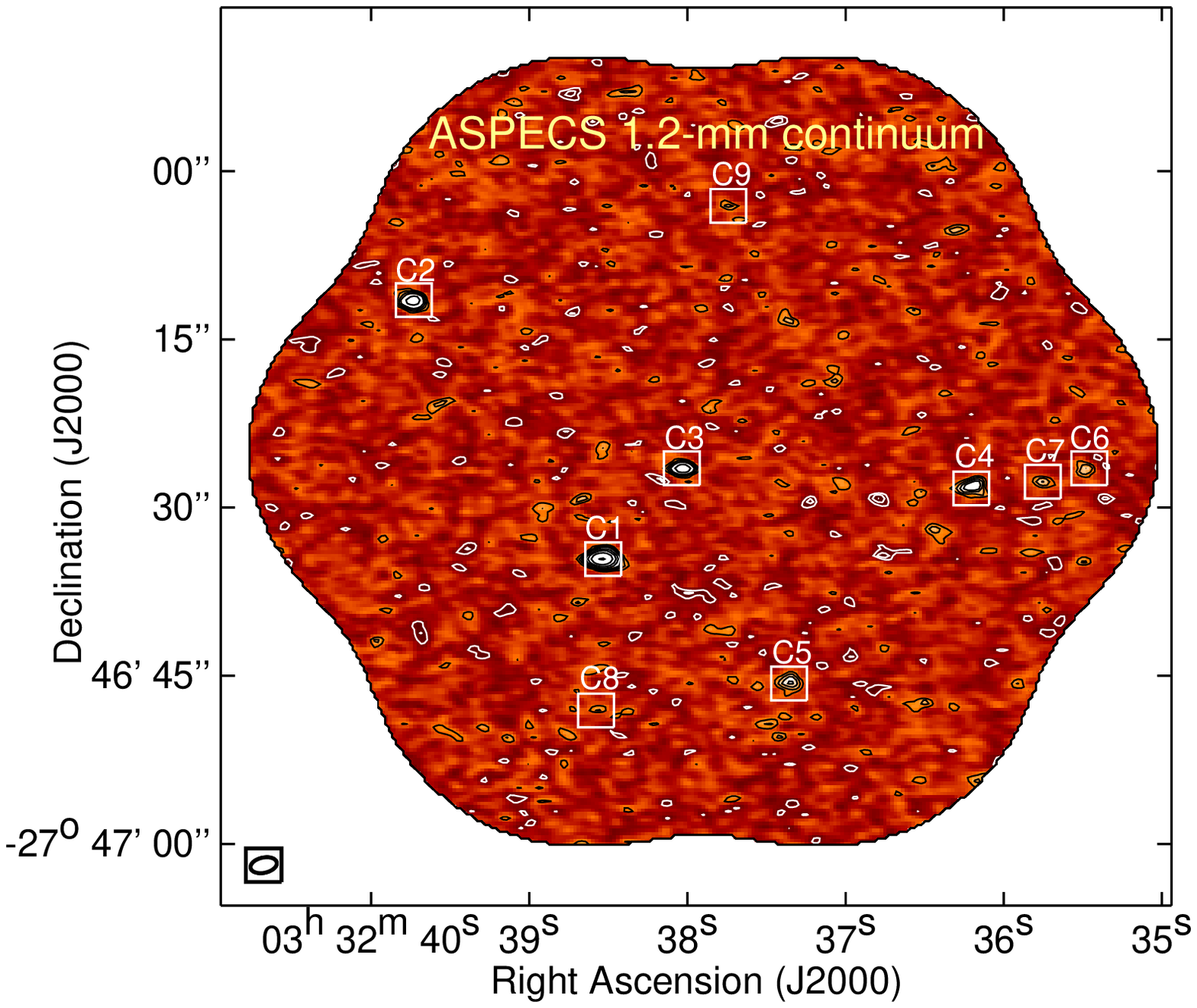}\hspace{4mm}
\includegraphics[scale=0.55]{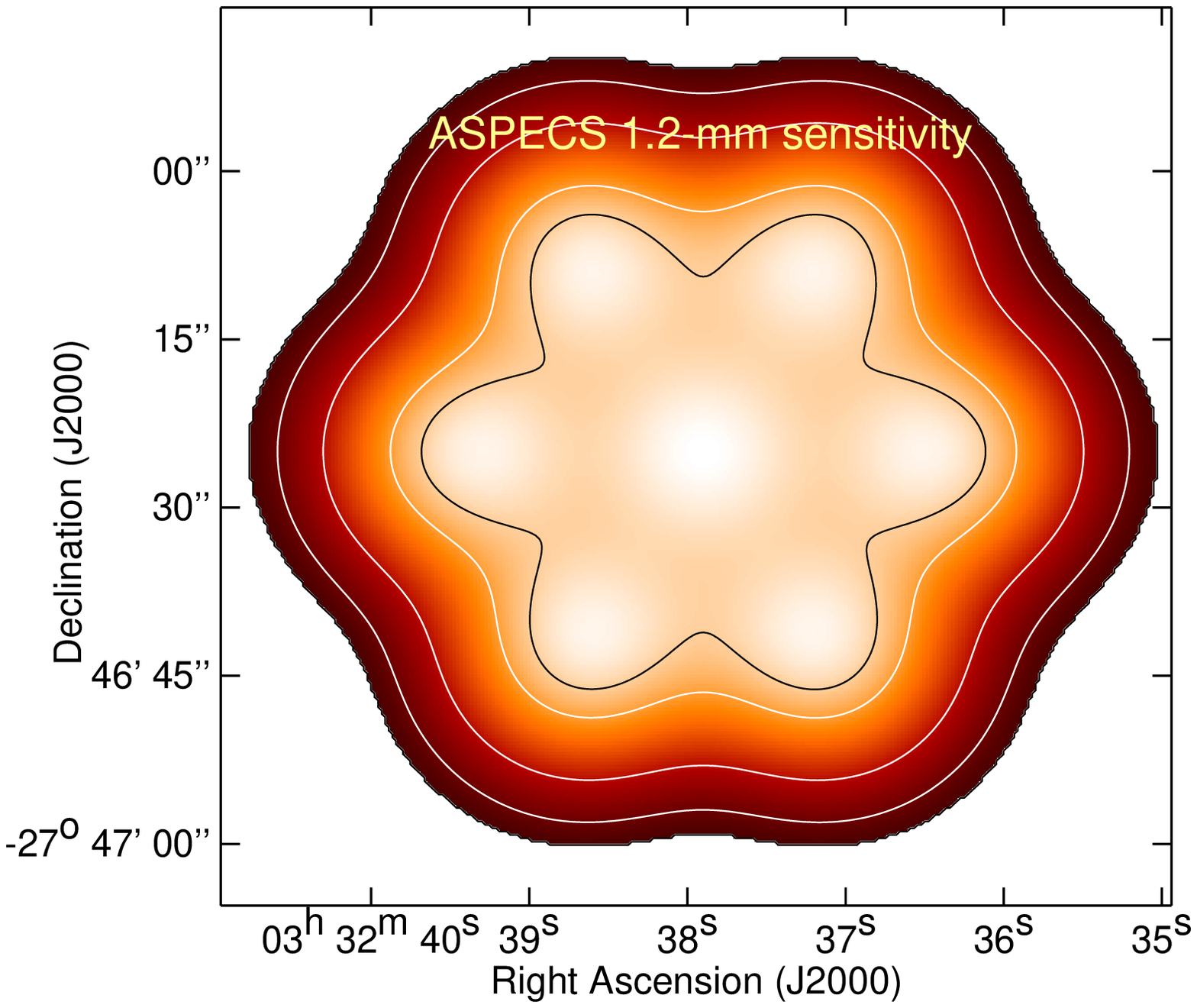}
\caption{({\it Left:}) ALMA 1.2-mm signal-to-noise continuum mosaic map obtained in the HUDF. Black and white contours show positive and negative emission, respectively. Contours are shown at $\pm2,3,4,5,8,12,20$ and $40\sigma$, with $\sigma=12.7\mu$Jy beam$^{-1}$ at the field center. The boxes show the position of the sources detected with our extraction procedure at $S/N>3.5$. The synthesized beam ($1''\times2''$) is shown in the lower left. ({\it Right:}) ALMA 1.2-mm observations primary beam (PB) pattern to represent the sensitivity obtained across the covered HUDF region. PB levels are shown by the black/white contours at levels 0.3, 0.5, 0.7 and 0.9 of the maximum. Both the signal-to-noise and PB maps are shown down to PB$=0.2$.\label{fig_map1mm}}
\end{figure*}

\begin{figure*}[ht]
\centering
\includegraphics[scale=0.55]{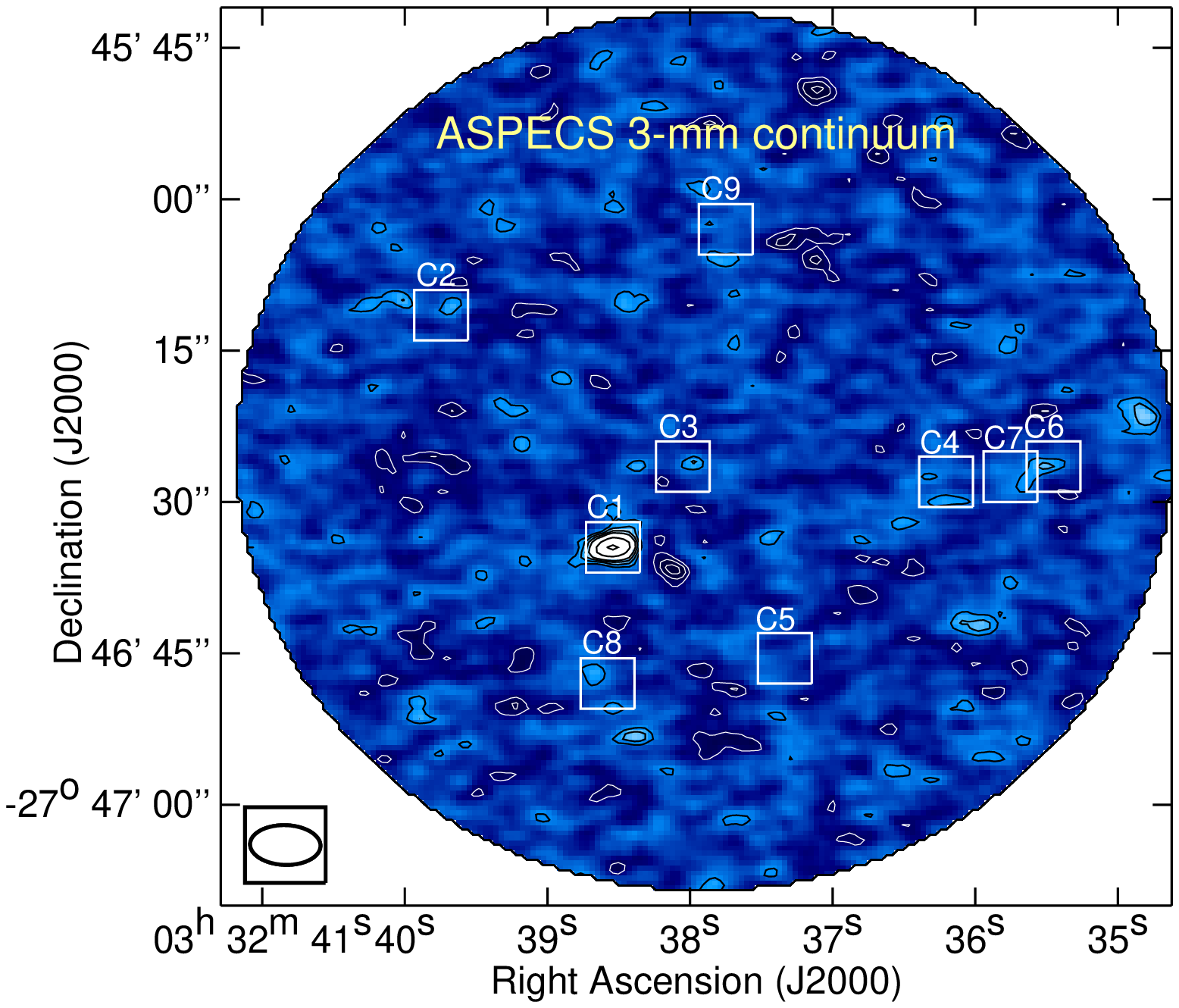}\hspace{4mm}
\includegraphics[scale=0.55]{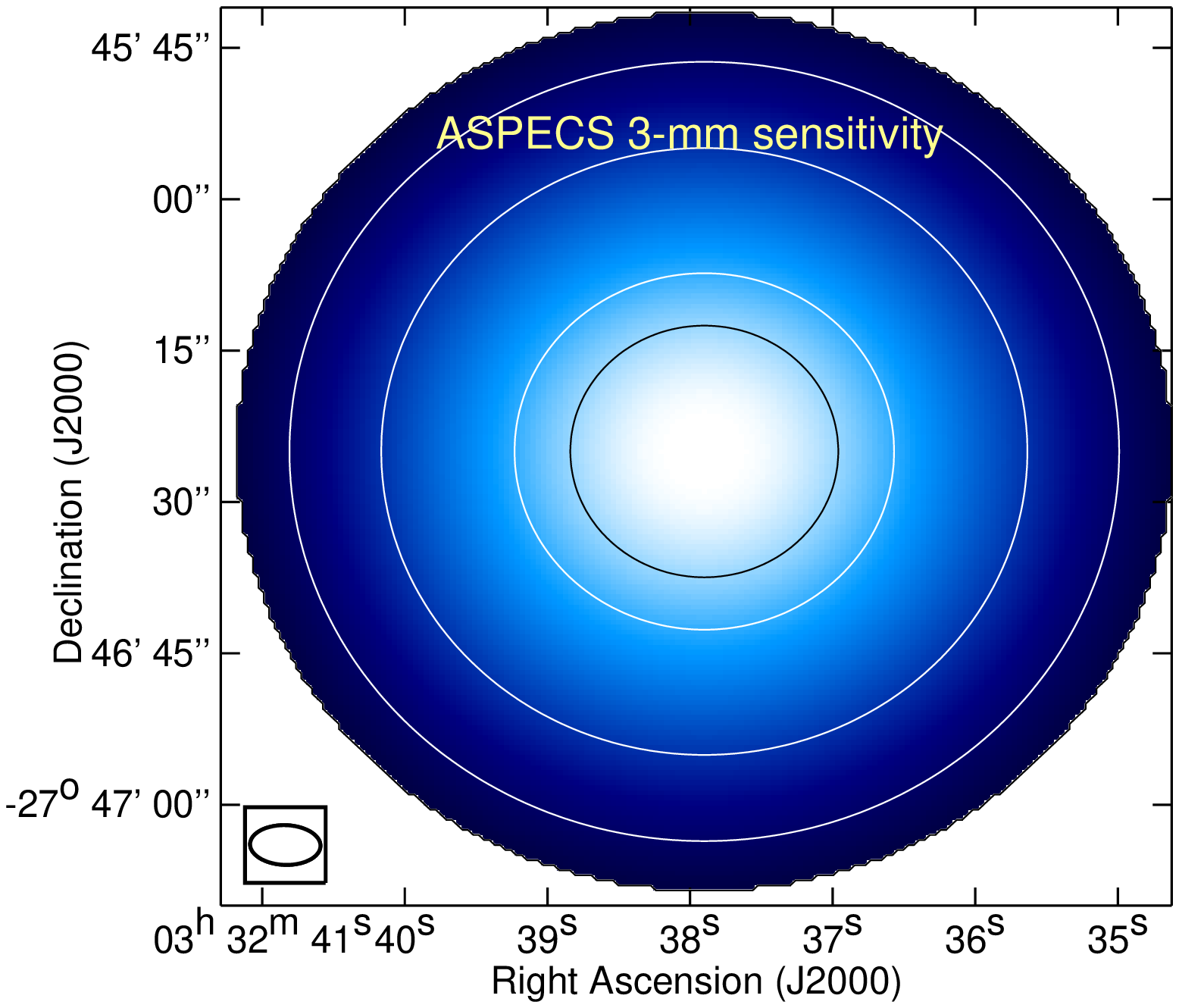}
\caption{({\it Left:}) ALMA 3-mm signal-to-noise continuum mosaic map obtained in the HUDF. Black and white contours show the positive and negative signal, respectively. Contours are shown at $\pm2,3,4,5,8,12,20$ and $40\sigma$, with $\sigma=3.8\mu$Jy beam$^{-1}$ at the field center. The boxes show the position of the sources detected in the 1.2-mm map, with our extraction procedure at $S/N>3.5$. The synthesized beam ($2''\times3''$) is shown in the lower left. ({\it Right:}) ALMA 3-mm observations primary beam (PB) pattern. PB levels are shown by the black/white contours at levels 0.3, 0.5, 0.7 and 0.9. Both the signal-to-noise and PB maps are shown down to PB$=0.2$.\label{fig_map3mm}}
\end{figure*}

To locate and characterise the population of faint DSFGs that make up most of the EBL at (sub)millimeter wavelengths, we must overcome several observational limitations. First, the poor resolution of (sub)millimeter bolometer maps taken with single-dish telescopes, typically with beam sizes between $10-30''$, makes the identification of an optical counterpart difficult and thus limits the characterisation of submillimeter sources. In addition, this affects the number counts, since the brightest sources are seen to split into multiple components in high-resolution (sub)millimeter images \citep{younger07, wang11, smolcic12, hodge13, karim13, miettinen15}. Secondly, the sensitivity of single dish bolometer maps, typically down to $0.5-1.0$ mJy, along with confusion at the faint levels limits our view to the most luminous sources. An important approach to reach fainter galaxies has been the use of gravitational lensing enabled by massive galaxy clusters \citep[e.g.,][]{smail97, smail02, sheth04,knudsen08,noble12,johansson12, chen13}. However, these surveys suffer severely from cosmic variance, due to the small areas covered in the source plane, source confusion, and the need for accurate lens models and magnification maps. A parallel approach has been to perform stacking of the submillimeter emission using pre-selected samples of optical/infrared galaxies. This approach has successfully resolved significant amounts of the EBL at (sub)millimeter wavelengths, reaching down to sources with $S_{\rm 1.2mm}>0.1$ mJy \citep{webb04,knudsen05,greve10,decarli14}. The major limitation of this approach is that it yields average properties over a population of galaxies that must be assumed to have similar (sub)millimeter properties. 

The advent of the Atacama Millimeter/submillimeter Array (ALMA) is opening up a new window for the study of the faint DSFG population. Its significantly higher angular resolution compared to single-dish telescopes ($<3''$), and the unparalleled sensitivity allow us to reach flux density levels in (sub)millimeter continuum maps even deeper than those achieved by studies of galaxy cluster fields or based on stacking analysis. Several recent studies have individually pinpointed (sub)millimeter sources down to 0.1 mJy in the 1-mm band \citep{hatsukade13,ono14,carniani15,oteo15,hatsukade16, dunlop16}. Some of these surveys have used clever approaches by taking advantage of archival data \citep{ono14,carniani15,fujimoto16}, including ALMA calibration fields \citep{oteo15}. Recently, \citet{fujimoto16} were able to reach down to a flux limit of 15$\mu$Jy at 1.2-mm, providing the deepest measurements of the number counts to date, and allowing them to resolve most of the CIB into individual sources. Despite the substantial progress, the current studies are still affected significantly by cosmic variance and are not ``blank-field'' in nature (as some of them target overdense fields). Most importantly, the lack of sufficiently deep complementary data have only permitted the characterisation of a handful of sources \citep{hatsukade15,fujimoto16, yamaguchi16}. 

Using ALMA in Cycle 2, we have conducted a deep ALMA Spectroscopic Survey (ASPECS) of a region of the {\it Hubble} Ultra Deep Field (UDF), covering the full 3-mm and 1-mm bands. In this paper, we present the {\it deepest} millimeter continuum images obtained to date in a contiguous 1 arcmin$^2$ area. This is the Paper~II in the ASPECS series. A full description of the survey and spectral line search is presented in Paper~I \citep{walter16}. Measurements of the CO luminosity function and cosmic density of molecular gas are shown in Paper~III \citep{decarli16a}. A detailed analysis of the CO brightest objects is presented in Paper~IV \citep{decarli16b}. A search for {\sc [CII]} line emission is shown in Paper~V \citep{aravena16b}.  This paper is organised as follows: in \S \ref{sec_obs}, we summarise the ALMA observations and multi-wavelength ancillary data available. Here, we also present the obtained ALMA continuum maps at 1.2-mm and 3-mm. In \S \ref{sec_results}, we present the detected sources and compute the fidelity and completeness of our extraction procedures in the 1.2-mm map. In \S \ref{sec_counts}, we derive the number counts at 1.2-mm. In \S \ref{sec_prop}, we characterise the multi-wavelength properties of the individually detected sources, including their typical stellar masses, SFRs and redshifts, and discuss whether our sources are starbursts or more quiescent star forming galaxies. In \S \ref{sec_stack}, we conduct a stacking analysis to determine the average properties of the faintest population of galaxies, not detected individually by our survey. In \S \ref{sec_ism}, we investigate the ISM properties of the individually detected sources based on measurements of the ISM masses from the 1.2-mm fluxes. We estimate their gas masses, depletion timescales and fractions. In \S \ref{sec_cib}, we determine the contribution of both our individually-detected and stacked sample to measure the fraction of the EBL at 1.2-mm resolved by our observations. We discuss the properties of the galaxies that dominate the CIB. Finally, in \S \ref{sec_concl}, we summarise the main results of this paper. Throughout the paper, we assume a standard $\Lambda$CDM cosmology with $H_0=70$ km s$^{-1}$ Mpc$^{-1}$, $\Omega_\Lambda=0.7$ and $\Omega_{\rm M}=0.3$.

\section{Observations}
\label{sec_obs}

\begin{table*}[t]
\begin{flushleft}
\caption{Sources detected in the ASPECS 1.2-mm continuum map. Columns: (1), (2) Source full and short names; (3), (4) Position of the 1.2-mm continuum detection in the ALMA 1.2-mm map; (5) Signal to noise ratio (SNR) of the 1.2-mm detection; (6) Flux density at 1.2-mm, corrected for PB; (7) Primary beam correction at the location of the detection in the 1.2-mm mosaic; (8) Flux density at 3.0-mm of the ALMA 1.2-mm continuum detection. Upper limits are given at the $3\sigma$ level; (9) Primary beam correction at the location of the 1.2-mm detection in the 3.0-mm map; (10) Is there an optical counterpart identification for this source? Yes or no; }\label{tab_sources}
\end{flushleft}
\begin{tabular}{lccccccccc}
\hline
IAU name & Short name  & RA$_{\rm 1.2mm}$ & Dec$_{\rm 1.2mm}$ & SNR & $S_{\rm 1.2mm}$ & PB$_{\rm 1.2mm}$ & $S_{\rm 3mm}$ & PB$_{\rm 3mm}$ & OID?\\
ALMA\ldots & ASPECS\ldots & (J2000) & (J2000) &    &    ($\mu$Jy) &   & ($\mu$Jy) &   \\
 (1) & (2) & (3) &  (4) & (5) &  (6) & (7) &  (8) & (9) &  (10) \\
\hline\hline
\multicolumn{10}{c}{Main sample at $>3.5\sigma$ significance}\\
\hline
MMJ033238.54-274634.6$^\dagger$ & C1  & 03:32:38.54 & $-27$:46:34.6 & 39.9 & $553\pm 14$ & 0.92 &  $31.1\pm5.0$ & $0.89$ & Y \\
MMJ033239.73-274611.6$^\dagger$& C2  & 03:32:39.73 & $-27$:46:11.6 & 10.3 & $223\pm 22$ & 0.59  & $<21$  &  0.56 & Y\\
MMJ033238.03-274626.5& C3 & 03:32:38.03 & $-27$:46:26.5&  9.6 & $145\pm 12$ & 0.95  &  $<12$ &  1.00 & Y\\
MMJ033236.20-274628.2& C4 & 03:32:36.20 & $-27$:46:28.2 &  6.1 & $ 87\pm 14$ & 0.89 &  $<17$ &  0.68  & Y\\
MMJ033237.35-274645.7& C5 & 03:32:37.35 & $-27$:46:45.7 &  5.2 & $ 71\pm 14$ & 0.92 & $<16$  &  0.70  & Y\\
MMJ033235.47-274626.6$^\dagger$& C6  & 03:32:35.47 & $-27$:46:26.6 &  3.9 & $ 97\pm 25$ & 0.51 &  $<25$ &  0.45  & Y\\
MMJ033235.75-274627.7& C7 & 03:32:35.75 & $-27$:46:27.7 &  3.7 & $ 70\pm 19$ & 0.67 & $<21$  &  0.55  & Y\\
MMJ033238.57-274648.0& C8 & 03:32:38.57 & $-27$:46:48.0 &  3.6 & $ 46\pm 13$ & 0.99 &  $<18$ &  0.62  & N\\
MMJ033237.74-274603.0 & C9 & 03:32:37.74 & $-27$:46:03.0 &  3.5 & $ 55\pm 16$ & 0.80 &  $<16$ &  0.70  & N\\
\hline
\multicolumn{10}{c}{Supplemetary sample at $3.0-3.5\sigma$ significance}\\
\hline
MMJ033237.36-274613.2& C10 & 03:32:37.36 & $-27$:46:13.2 &  3.3 & $ 45\pm 14$ & 0.93 &  $<13$ &  0.88  & N\\
MMJ033238.77-274650.1& C11 & 03:32:38.77 & $-27$:46:50.1 &  3.2 & $ 47\pm 14$ & 0.88 & $<21$  &  0.55  & N\\
MMJ033237.42-274650.4& C12 & 03:32:37.42 & $-27$:46:50.4 &  3.2 & $ 59\pm 18$ & 0.69 &  $<19$ &  0.60  & Y\\
MMJ033236.50-274647.4& C13 & 03:32:36.50 & $-27$:46:47.4 &  3.2 & $ 67\pm 21$ & 0.60 &  $<22$ &  0.52  & Y\\
MMJ033236.43-274632.1& C14 & 03:32:36.43 & $-27$:46:32.1 &  3.1 & $ 46\pm 15$ & 0.85 & $<16$  &  0.73  & Y\\
MMJ033237.49-274649.3& C15 & 03:32:37.49 & $-27$:46:49.3 &  3.1 & $ 52\pm 17$ & 0.76 & $<18$  &  0.63  & N\\
MMJ033237.75-274609.6& C16 & 03:32:37.75 & $-27$:46:09.6 &  3.0 & $ 41\pm 14$ & 0.93 &  $<14$ &  0.85  & N\\
\hline\hline
\end{tabular}
\flushleft
\noindent $^\dagger$ Sources ASPECS C1, C2 and C6 in this paper correspond to sources 3mm.1, 3mm.2 and 3mm.5 in Decarli et al. (Paper~IV).
\end{table*}

\subsection{ALMA observations and data reduction}

The ASPECS survey setup and data reduction steps are described in detail in Paper~I (Walter et al. 2016). Here we repeat the most relevant information for the study presented here.

ALMA band 3 and band 6 observations were obtained during Cycle-2 as part of projects 2013.1.00146.S (PI: F. Walter) and 2013.1.00718.S (PI: M. Aravena). Observations in band-3 were conducted between July 01, 2014 to January 05, 2015, and observations in band 6 were conducted between December 12, 2014 to April 21, 2015 under good weather conditions.

Observations in band 3 were performed in a single pointing in spectral scan mode, using 5 frequency tunings to cover $84.2-114.9$ GHz. Over this frequency range the ALMA half power beam width (HPBW), which corresponds to a primary beam (PB) response of 0.5, varies between $61''$ and $45''$. Observations in band 6 were performed in a 7-point mosaic, using a hexagonal pattern (Fig. \ref{fig_map1mm}): the central pointing overlaps the other 6 pointings by about half the ALMA PB, i.e., close to Nyquist sampling. We scanned band 6 using eight frequency tunings, covering $212.0-272.0$ GHz. The ALMA PB in individual pointings ranges between $30''$ and $23''$.

Observations in bands 3 and 6 were taken with ALMA's compact array configurations, C34-2 and C34-1, respectively. The observations used between 30 and 35 antennas in each band, resulting in synthesized beam sizes of $3.6''\times2.1''$ and $1.7''\times0.9''$ from the low to high frequency ends of bands 3 and 6, respectively.

Flux calibration was performed on planets or Jupiter's moons, with passband and phase calibration determined from nearby quasars, and should be accurate within $\pm10\%$. Calibration and imaging was done using the Common Astronomy Software Application package ({\sc CASA}). The calibrated visibilities were inverted using the {\sc CASA} task \verb CLEAN \, using natural weighting. To obtain continuum maps, we collapsed along the frequency axis in the uv-plane and inverted the visibilities using the {\sc CASA} task \verb CLEAN \, using natural weighting and mosaic mode. We use the Multi-frequency Imaging Synthesis (MFS) algorithm with \verb nterms=1 \, as the joint implementation of \verb nterms>1 \, and mosaic mode are not yet available in {\sc CASA}. This implies assuming a first order polymial fit for point sources along the frequency axis, which is the best assumption for low signal to noise data (most sources with S/N$<10$) as in this paper \citep[see {\sc CASA} cookbook and ][]{rau11}. We also tested the effect of using different frequency weightings in the visibility plane, however no significant changes were seen in the final collapsed images.

In this process, we produced `clean' maps masking with tight boxes all the continuum sources previously detected in the `dirty' maps with significances above $5\sigma$, and cleaning down to a $2.5\sigma$ threshold. Given the large bandwidth covered by our observations, the contamination by line emission in the continuum map becomes negligible.

The final maps are shown in Figs. \ref{fig_map1mm} and \ref{fig_map3mm}. The sensitivity in each map declines with respect to the distance from the phase pointing center, and, given the smaller PB, declines particularly sharply for the 1.2-mm observations at the outskirts of the mosaicked region. We reach an rms sensitivity of $12.7\mu$Jy and $3.8\mu$Jy in the centres of the 1.2-mm and 3-mm maps, respectively. The final map average frequencies over the frequency ranges covered are 242 and 95 GHz, respectively.

Finally, we note that while source confusion for individual detections is negligible in these deep ALMA maps, it is at the level where it  becomes important for stacking analyses. With an ALMA beam size at 1.2-mm of $1.7''\times0.9''$, there are $8.47\times10^6$ beams per deg$^2$. At the bottom flux bin of our number count measurements (see \S \ref{sec_counts}), we find $1.32\times10^5$ sources per deg$^2$. This translates into one source per $\sim 64$ beams, and implies that confusion is not an issue. The same logic applies for the stacking analysis presented below (see \S \ref{sec_stack}). The deepest stacks considered reach a $3\sigma$ level of 8 $\mu$Jy at 1.2-mm. Extrapolating the number counts to this flux level, we find about $6.0\times10^5$ sources per deg$^2$. This results in one source per $14$ beams. According to \citet{helou90}, bright source confusion becomes important at one source per 22 beams, suggesting that stacking experiments in these ALMA deep maps will be affected. However, this confusion limit depends on the slope of the number counts, and since this slope appears to flatten at these faint flux levels, it is possible that confusion would have a lesser impact at these depths, and in particular on stacking analyses.

\subsection{Multi-wavelength data}

Our ALMA observations cover a $\sim1$ arcmin$^2$ region within the deepest 4.7 arcmin$^2$ of the {\it Hubble} UDF: the eXtremely Deep Field (XDF). Available data includes {\it HST} Advanced Camera for Surveys (ACS) and Wide Field Camera 3 IR data from the HUDF09, HUDF12 and Cosmic Assembly Near-infrared Deep Extragalactic Legacy Survey (CANDELS) programs as well as public photometric and spectroscopic catalogs \citep{coe06,xu07,rhoads09,schenker13,mclure13, skelton14,bouwens14,morris15,momcheva15}. In this study, we make use of this optical and infrared coverage of the XDF, including the photometric and spectroscopic redshift information available from \citet{skelton14}. In addition to the {\it HST} coverage, a wealth of optical and infrared coverage from ground based telescopes is available in this field \citet[see ][]{skelton14}. The HUDF was also covered by the {\it Spitzer} Infrared Array Camera (IRAC) and Multiband Imaging Photometer (MIPS), as well as by the {\it Herschel} Photodetector Array Camera and Spectrometer (PACS) and the Spectral and Photometric Imaging Receiver (SPIRE) \citep{elbaz11}.  

\section{Results}
\label{sec_results}

\subsection{Source detection and flux measurements}

Source detection was performed using SExtractor \citep{bertin96} in the ALMA 1.2-mm and 3-mm maps prior to PB correction. We use a minimum area of 5 pixels ($1.5''$) for detection, extracting sources down to 2.5$\sigma$, where $\sigma$ is evaluated locally for each source. Source extraction in the 1.2-mm map was performed beyond the HPBW of our mosaic, out to PB $=0.3$, however most sources are detected within PB $=0.5$, in the central region of the mosaic. Although we extract all sources down to $2.5\sigma$, we consider as individual detections only sources above $>3.5\sigma$ significance. This significance level cut corresponds to roughly $50-60\%$ fidelity of the sample (see \S~\ref{sec_purity}). These sources are highlighted with boxes in Figs. 1 and 2, and are listed in Table~1.

Nine sources are detected in the 1.2-mm map at a significance above 3.5$\sigma$. For reference, Table~1 also lists another 7 sources with significances between $3.0-3.5\sigma$ (our supplementary sample). Given the lower significance of these sources, we choose not to use them to study the multi-wavelength properties of this population. Nevertheless we can use them to constrain the number counts of faint sources, after correcting for fidelity and completeness. Only one source is detected in the 3-mm map at the $>3.5\sigma$ significance level, corresponding to the brightest detection at 1.2-mm. For this reason, we only show the 1.2-mm detected sources in Figs.~1--2. 

We compute fluxes based on 2-dimensional Gaussian fit centered at the location of the SExtractor detection. In all but one case (discussed below) the sources are unresolved at the resolution and depth of the 1.2-mm observations. We therefore list the flux as the peak flux density value at the source position delivered by the fitting routine. These fitted values are in agreement with the actual pixel values at the position of the sources. We cannot discard the possibility that sources with low significances are indeed being resolved given the relatively small beam size. It is thus unclear what fraction of the flux is being unaccounted for in individual sources.

Only the brightest source in the map is marginally spatially resolved with a measured angular size of $(0.52\pm0.14)''\times(0.43\pm0.26)''$ (PA$=49\deg$), and we record the integrated flux in Table~1. More details on this source's properties are given in Paper~IV \citep{decarli16b}. Since only one source is detected in the 3-mm map, in what follows we concentrate on characterising the properties of the 1.2-mm sources.

\begin{figure*}[ht]
\centering
\includegraphics[scale=0.4]{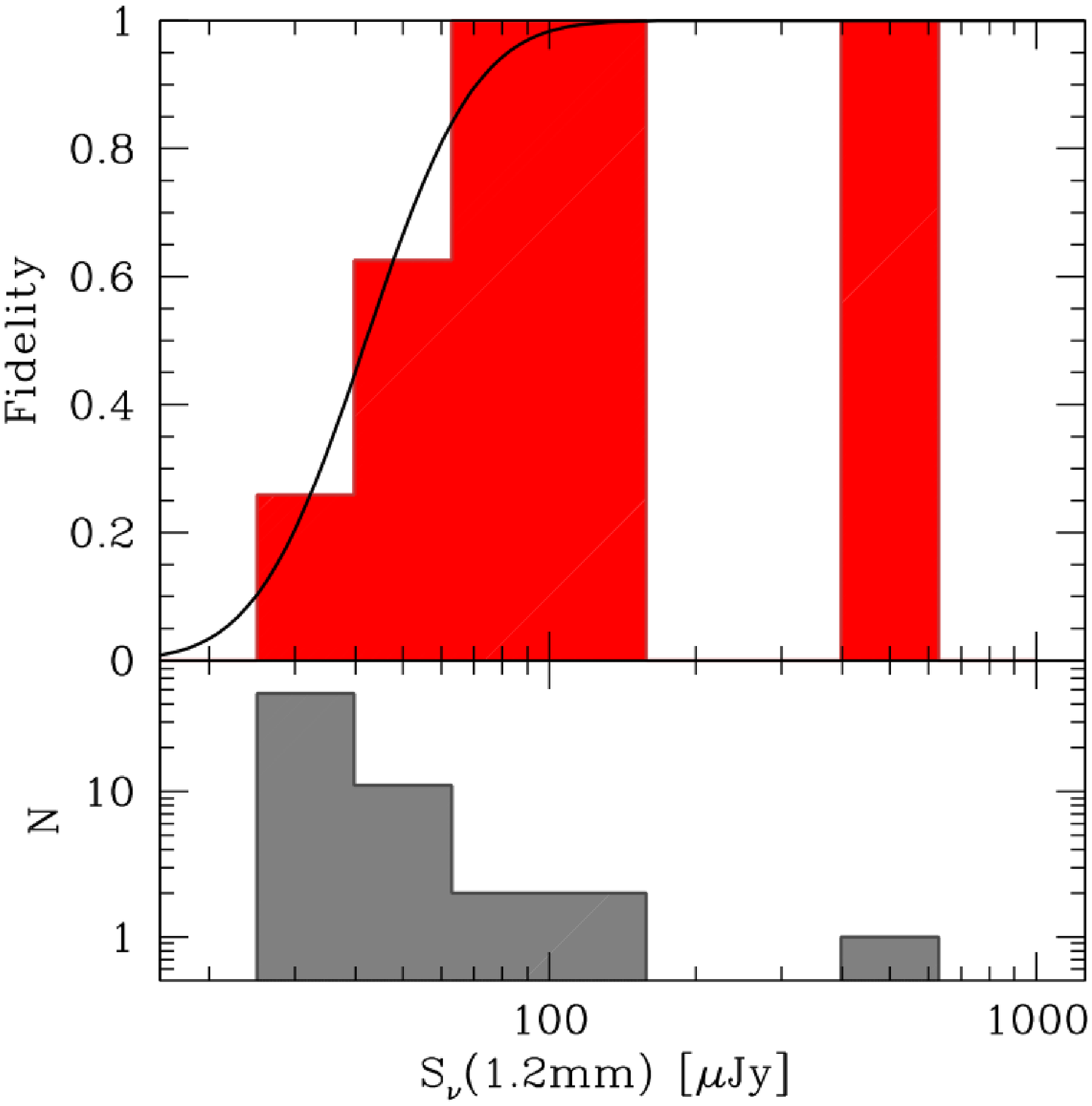}
\includegraphics[scale=0.4]{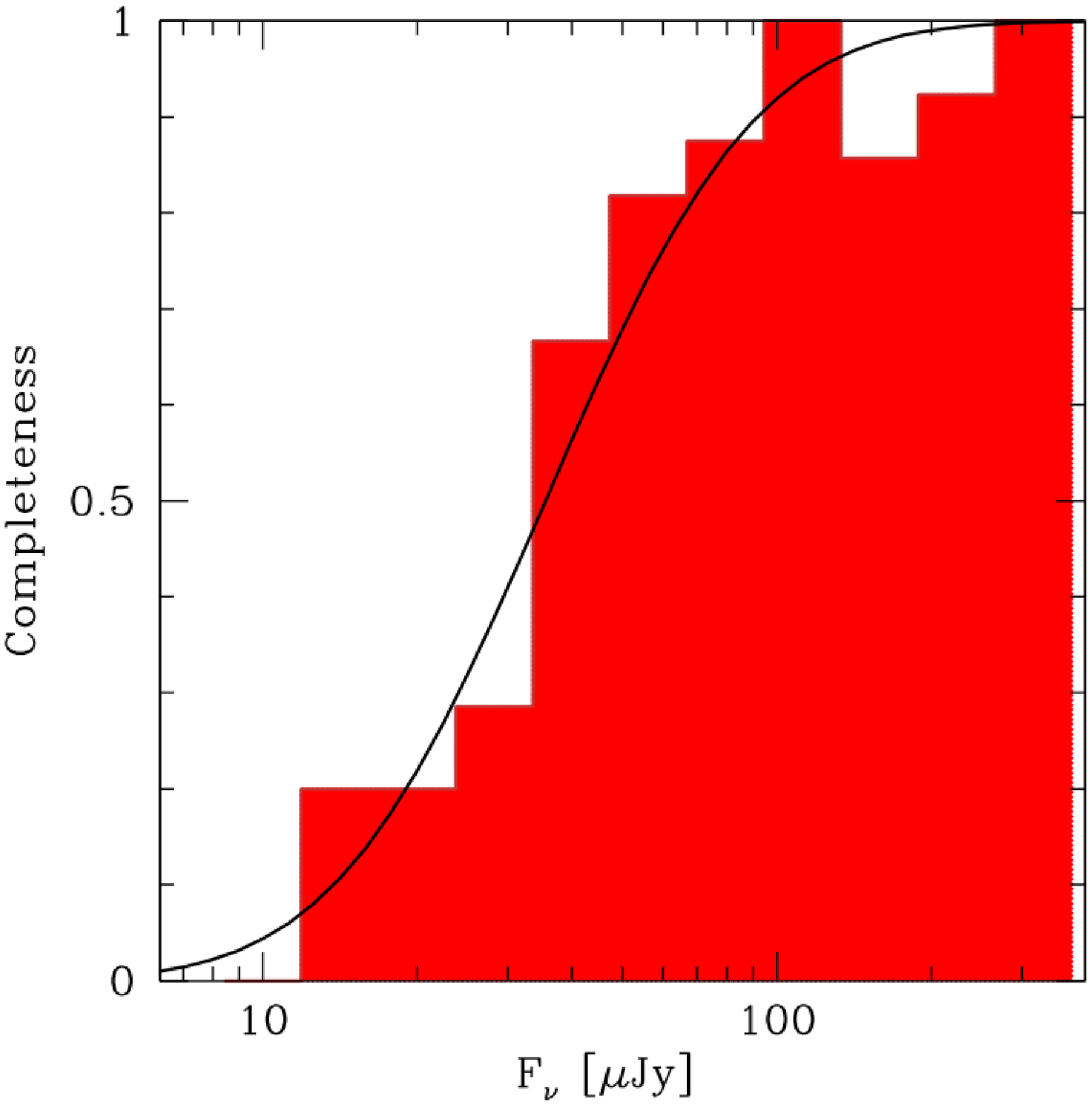}
\caption{({\it Left:}) Fidelity ({\em top panel}) and number of detections ({\em bottom panel})  as a function of 1.2-mm flux density of the ASPECS sample (non-cumulative). The solid curve is a model for the fidelity. Our sample shows 100\% fidelity at $S_{\rm 1.2mm}\sim100\mu$Jy and 50\% fidelity at $\sim40\mu$Jy (3.0$\sigma$). ({\it Right:}) Completeness of our 1.2-mm continuum sample detection as a function of 1.2-mm flux density. The solid curve shows a model for the completeness behaviour as a function of 1.2-mm flux density. Our sample shows 100\% completeness at $S_{\rm 1.2mm}\sim300\mu$Jy and 50\% completeness at $\sim40\mu$Jy (3.0$\sigma$).}\label{fig_purity}\label{fig_completeness}
\end{figure*}

\subsection{Fidelity and completeness}\label{sec_purity}

We quantify the occurrence of spurious sources in our 1.2-mm sample by applying the detection routine explained in the previous section to the inverted `negative' map. We thus compute the fidelity $P$ of our sample as:

\begin{equation}
P({\rm S_{\rm 1.2mm}})=1-\frac{N_{\rm neg}({\rm S_{\rm 1.2mm}})}{N_{\rm pos}({\rm S_{\rm 1.2mm}})}, 
\end{equation}

where $N_{\rm neg}$ and $N_{\rm pos}$ are the number of negative and positive sources, respectively, detected in the map as a function of 1.2-mm flux density. 
\begin{figure*}[ht]
\centering
\includegraphics[scale=0.6]{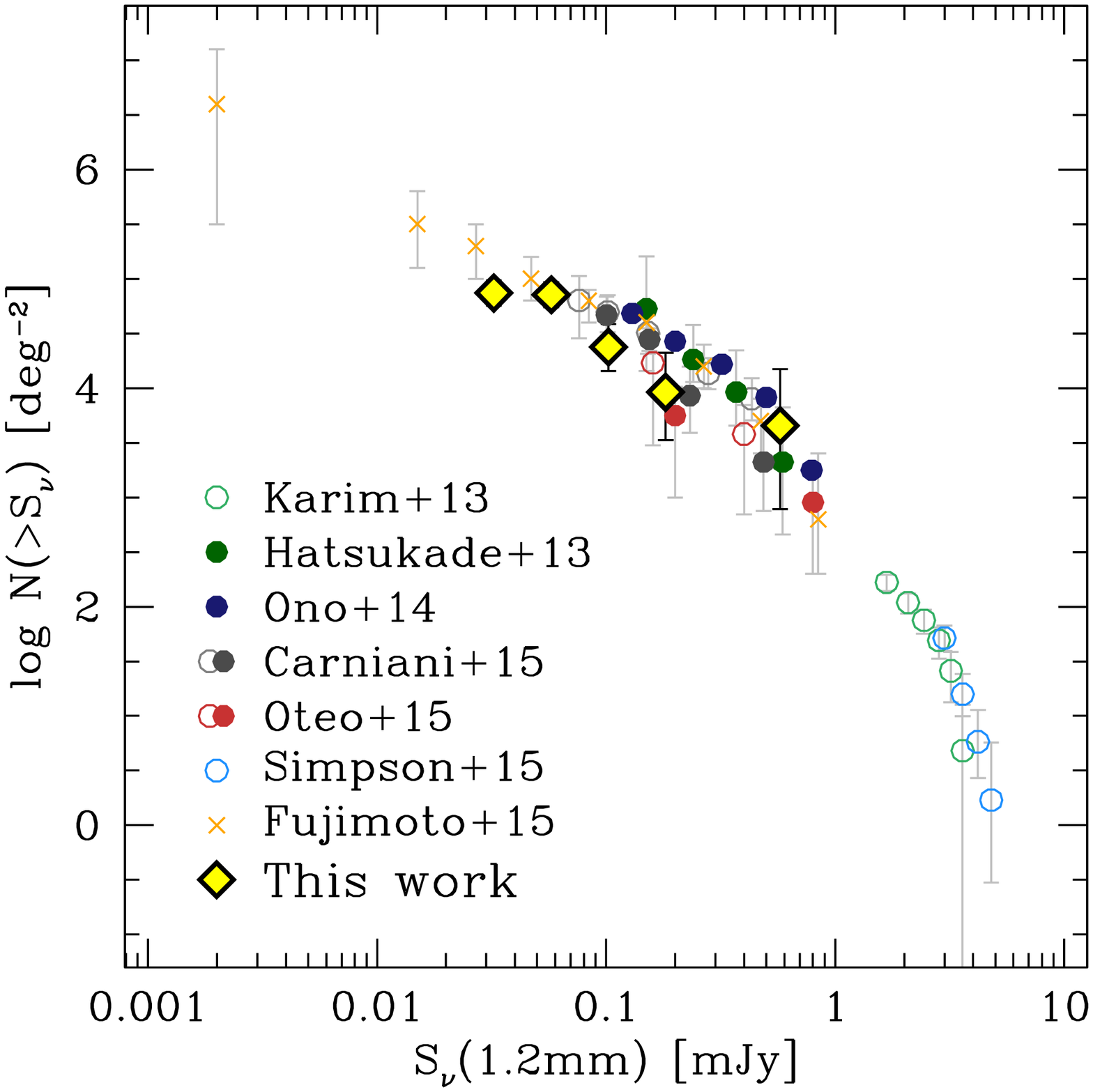}
\caption{Number counts of ALMA 1.2mm continuum sources in the UDF compared with values from the literature. Our data have been corrected to account for completeness and fidelity in the source identification, as discussed in the text. Uncertainties in each number count measurements correspond to Poisson errors. Our measurements span almost two orders of magnitude in flux density. Filled circles represent literature measurements obtained at 1.2-mm. Open circles represent measurements from different wavelengths than 1.2-mm and converted to this wavelength. Most of the measurements from the literature at the faint levels are not blank field and are thus biased, since their observations target bright sources in the field (they measure counts around other sources). The \citet{fujimoto16} data pointing towards lower flux densities are based on lensed galaxy clusters.}\label{fig_ncounts}
\vspace{10mm}	
\end{figure*}

\begin{table}[h]
\begin{flushleft}
\caption{ALMA UDF 1.2-mm number counts. Columns: (1) Flux density bin center (in units of mJy); (2) Number of entries per bin (before fidelity and completeness correction); (3) Number of sources per square degrees; (4), (5) Lower and upper uncertainties (error bars) on $N(>S_\nu)$.}\label{tab_ncounts}
\end{flushleft}
\begin{tabular}{ccccc}
\hline
log($S_\nu$) & $dN/dlog(S_\nu)$ & $N(>S_\nu)$ & $\delta N_{-}$ & $\delta N_{+}$ \\
 (mJy)     &   (mJy$^{-1}$) & (deg$^{-2}$) & (deg$^{-2}$) & (deg$^{-2}$)\\
 (1) & (2) & (3) & (4) & (5) \\
 \hline\hline
  $-1.49$ & 23 &   132000 & 3700 & 43000\\
  $-1.24$ & 10  &   71500 & 16600 & 21500\\
  $-0.99$ & 3  &   23700  & 9400 & 14700\\
  $-0.74$ & 1  &    9200 &  5800 & 11900\\
  $-0.24$ & 1  &    4500  &  3800 & 10400\\
\hline\hline	
\end{tabular}
\end{table}

Figure \ref{fig_purity} shows the fidelity and number of positive detections in our map as a function of 1.2-mm flux density.  Not surprisingly, we find that the fidelity of our sample is a strong function of the 1.2-mm flux density. We reach 100\% fidelity at $100\mu$Jy (7.8$\sigma$) and 50\% fidelity at $40\mu$Jy ($\sim3.0\sigma$). This means that at the $3\sigma$ level, half of our sources are expected to be spurious, which motivates our choice of 3.5$\sigma$ cut for the main sample.

We parametrise the fidelity with 1.2-mm flux density as:
\begin{equation}
P(S_{\rm 1.2mm})=\frac{1}{2} {\rm erf} \{ \frac{{\rm log}_{10}(S_{\rm 1.2mm})-A}{B} \} +1.0
\end{equation}
where $A={\rm log}_{10}(42)$ and $B=1/4$, and $S_{\rm 1.2mm}$ is in units of $\mu$Jy. We use this parametrisation to compute the fidelity level or reliability of our individual detections.

We compute the completeness of our observations by running Monte Carlo simulations on our continuum map. We ingest 10 artificial point-like sources with randomly generated flux levels (between $10-300 \mu$Jy) in the ALMA map. We then run our source detection procedure to identify and compute the fraction of recovered sources (versus the input sources). Recovered artificial sources are matched with the input positions within a radius of $1''$, roughly the size of our synthesized beam. Similar to the findings of \citet{fujimoto16}, the input and recovered flux densities agree well within individual source uncertainties. We repeat this process 10 times, for a total of 100 artificial sources. Note that we do not inject all 100 sources in a single step since this would result in significant source blending in the ALMA image.

Figure \ref{fig_completeness} shows the resulting completeness as a function of extracted 1.2-mm flux density. We find that our sample is 100\% complete at $S_{\rm 1.2mm}\sim300\mu$Jy (23$\sigma$) and 50\% complete at $\sim40\mu$Jy (3.0$\sigma$). This indicates that at the $3\sigma$ level, we recover only half of real input sources.

We parametrize the completeness with 1.2-mm flux density as:
\begin{equation}
C(S_{\rm 1.2mm})=\frac{1}{2} {\rm erf} \{ \frac{{\rm log}_{10}(S_{\rm 1.2mm})-A'}{B'} \} +1.0
\end{equation}
where $A'={\rm log}_{10}(35)$ and $B'=0.45$, and $S_{\rm 1.2mm}$ is in units of $\mu$Jy. We use this parametrisation to compute the completeness level of our individual detections.

\section{Number counts}

\label{sec_counts}

\begin{figure*}[ht]
\centering
\includegraphics[scale=0.9]{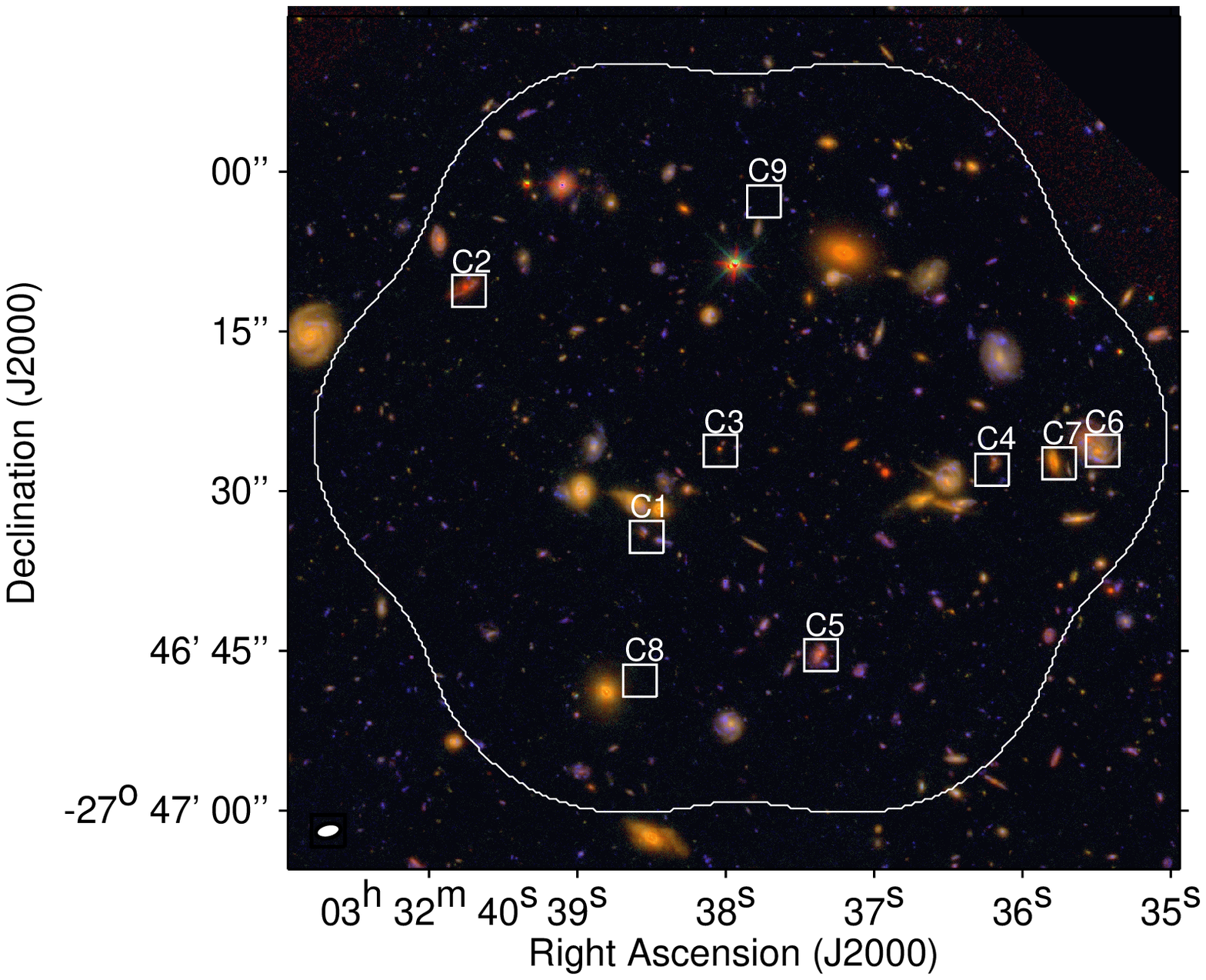}
\caption{HUDF multi--color image (F435W, F850LP, F105W) of the region covered by our 1-mm ALMA observations. The boxes show the position of the 1.2-mm sources detected with our extraction procedure at $S/N>3.5$. The white contour shows the coverage of our ALMA observations down to PB$=0.2$. }\label{fig_color}
\end{figure*}

\begin{figure*}
\centering
\includegraphics[scale=1.5]{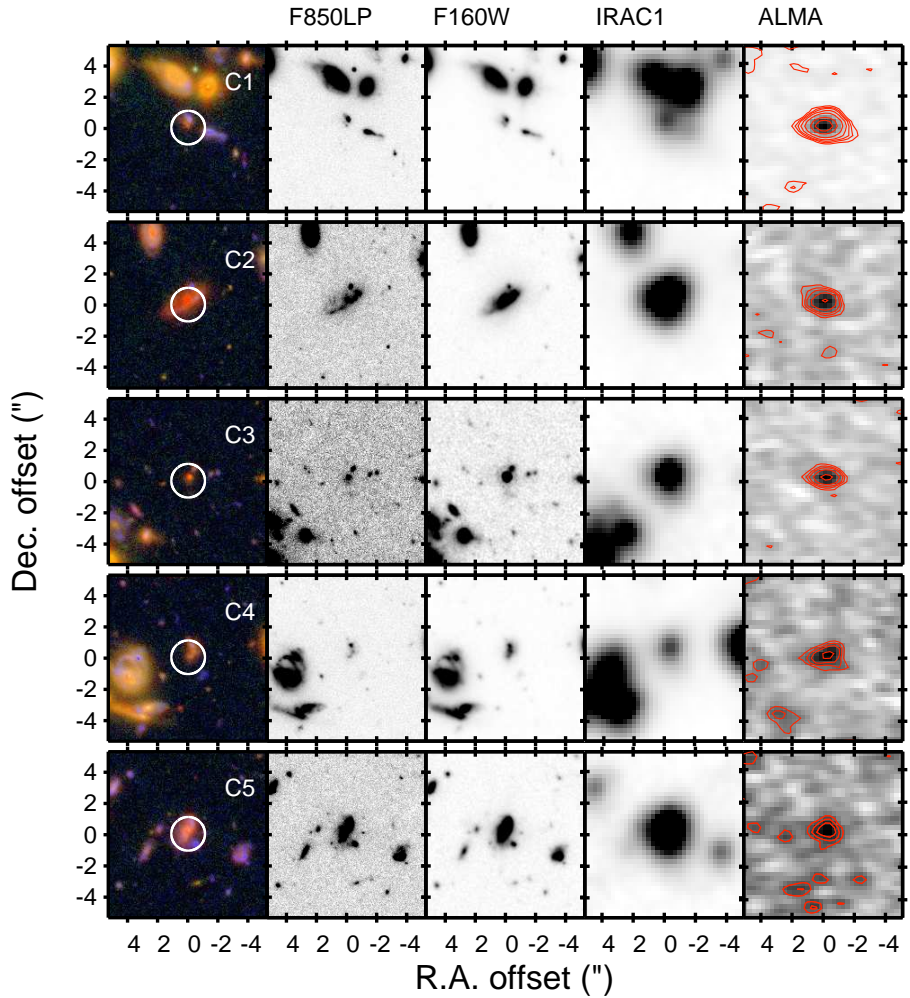}
\caption{Multi-wavelength image thumbnails toward the ALMA 1.2-mm continuum detections ($>3.5\sigma$). From left to right, we show an optical-near infrared false color composite (F435W/F850LP/F105W), and individual images in the F850LP, F160W, IRAC channel 1 and $10"\times10"$ in size. }\label{fig_thumb}
\end{figure*}

\begin{figure*}
\centering
\figurenum{6}
\includegraphics[scale=1.5]{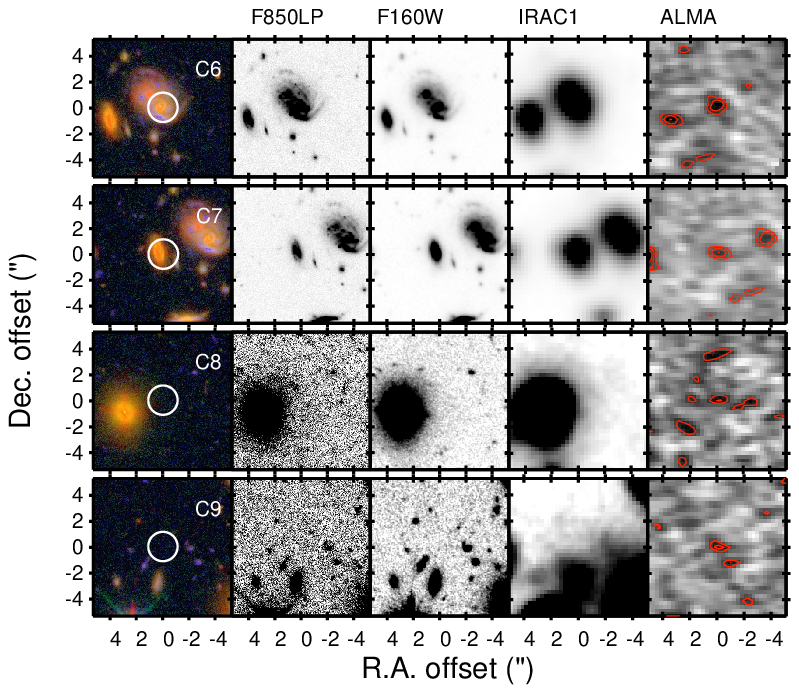}
\caption{continued}
\end{figure*}

We use the sources detected in our ALMA UDF map to compute the number counts at 1.2mm. We compute the number counts ($N(S_i)$) in each flux density bin $S_i$ as:

\begin{equation}
N(S_i) = \frac{1}{A} \sum_{j=1}^{X_i} \frac{P_j}{C_j},
\end{equation}

where $A$ is the effective area of our ALMA mosaic and $X_i$ is the number of sources in each particular bin $i$. The parameters $P_j$ and $C_j$ correspond to the fidelity and completeness at the flux bin $i$. Since we are limited by the modest number of detections, we compute the cumulative number counts rather than computing differential counts by summing up each $N(S_i)$ over all measurements $>S_i$. In addition, we extend our number count measurements down to significances of $3\sigma$. While at this level there is substantial contamination and low detection rate, we can statistically correct the values for fidelity and completeness. As pointed out in the previous section, at the $3\sigma$ level we reach 50\% fidelity as well as 50\% completeness in our sample detection. This implies that these effects cancel out when we compute the number counts. Thus, while we obtain correct number counts at the $3\sigma$ level, the identification of real sources is correct only in half of the cases.

The uncertainties in the number counts are computed by including the Poissonian errors as well as flux uncertainties in each individual measurement. The uncertainties in each bin are dominated by the Poissonian errors on $X_i$, however at the lowest significance levels the flux uncertainties start to have a significant contribution. 

The cumulative number counts ($N(>S_\nu)$) are shown in Fig. \ref{fig_ncounts}. The actual measurements are listed in Table \ref{tab_ncounts}. For comparison, we show number count measurements from the literature \citep{karim13, hatsukade13,ono14,carniani15,oteo15,simpson15,fujimoto16}. We scale the flux densities of the different studies as $S_{\rm 1.2mm}=0.4~S_{\rm 870}$, $S_{\rm 1.2mm}=0.8~S_{\rm 1.1mm}$ and $S_{\rm 1.2mm}=1.3~S_{\rm 1.3mm}$ \citep[for consistency with ][]{fujimoto16}.

Our ALMA UDF observations appear to be in general agreement with these earlier measurements, in particular with the counts obtained by \citet{carniani15} and \citet{oteo15}. However, our counts are lower by about a factor of 2 in the flux range $S_{\rm 1.2mm}=0.06-0.4$ mJy compared to other studies in the literature \citep[][]{hatsukade13,ono14,fujimoto16}. These difference could be explained by the fact that these studies might be biased as they used pointed observations toward brighter sources in the field to derive the number counts (i.e., these studies are not unbiased blank field surveys). 

Another possibility is that cosmic variance does play an important role among the different analyses; e.g. the ECDFS, where the UDF resides, is believed to be underdense of submillimeter sources above $\sim3$ mJy (at 345 GHz) by a factor of $\sim2$ \citep{weiss09}. As indicated by several studies, the ECDFS appears to be underdense in other galaxy populations as well, including $BzK$ galaxies, X-ray  and radio sources  \citep[e.g.,][]{lehmer05,blanc08}. However, as already noted by \citet{weiss09}, the underdensity appears to be seen only in the brightest sources, given the steep slope at fainter fluxes \citep[see also ][]{karim13}. Another possibility is that the differences in number counts between studies come from scatter induced by different analysis techniques and methods. This effect was seen to be a dominant compared to statistical fluctuations in radio surveys \citep{condon07}.

\section{Multi-wavelength properties of the ALMA 1.2-mm sources}

\label{sec_prop}

\subsection{Astrometric offset}

Using the identified mm/optical counterpart positions (see below), we measure a systematic astrometric offset of the {\it HST} positions of $\approx0.3"$ to the north of the ALMA positions. To check the ALMA registration we inspected the millimeter calibrators used, finding good astrometric solutions, accurate within $0.01"$ with respect to the catalogued radio-based values. Based on the GOODS 2008 data release documentation\footnote{\url{https://archive.stsci.edu/pub/hlsp/goods/v2/}\\ \url{h_goods_v2.0_rdm.html}}, it is clear that a consistent offset ($0.32''$) was applied to the GOODS-North astrometric solution but not to the GOODS-South data. Hence, we correct the {\it HST} positions by $0.3"$ to match the ALMA millimeter registration throughout. This is consistent with results from a shallower ALMA millimeter continuum survey of the full HUDF \citep{dunlop16, rujopakarn16}.

\subsection{Identification and SED fitting}

Figure \ref{fig_color} shows the location of the 1.2-mm continuum sources with respect to the optical galaxies in the field. Our blank-field observations encompass a significant number of optical galaxies, however this contrast the galaxies detected in the millimeter regime. Our sources do not appear to be clustered. 

For each individual 1.2-mm continuum detection, we identify optical counterparts within a radius of $1''$ from the millimeter position. We choose this search radius since it is well matched to the ALMA 1.2-mm synthesized beam ($1.7''\times0.9''$). Figure \ref{fig_thumb} presents multi-wavelength cutouts for individual detections. Seven of the continuum sources with significances $>3.5\sigma$ have an obvious counterpart in the {\it HST} images, and five of these have an available spectroscopic redshift \citep[see Table \ref{tab_prop};][]{skelton14}. The other two millimeter detections, with lower significances in our sample ($\sim3.5-3.6\sigma$), do not show an obvious counterpart. Four out of seven sources with significances between $3.0-3.5\sigma$ do not have an optical counterpart (Table \ref{tab_sources}), consistent with the fidelity level at this significance, and indicating that some or all of these are likely spurious millimeter detections. Another possibility would be that these are faint dusty galaxies at higher redshifts \citep[as in HDF850.1; see ][]{walter12}.

We fit the spectral energy distribution (SED) of the continuum--detected galaxies using the high-redshift extension of \texttt{MAGPHYS} \citep{dacunha08, dacunha15}. We use the available 26 broad and medium band filters in the optical and infrared regimes, from the $U$ band to {\it Spitzer} IRAC 8$\mu$m. We here also include the ALMA 1.2-mm data flux densities, however we note that the optical/infrared data has a much stronger weight given the tighter constrains in this part of the spectra. We do not include {\it Herschel} photometry in the fits since its angular resolution is very poor, being almost the size of our target field for some of the IR bands. The {\it Herschel} photometry is thus heavily blended. 

\begin{figure}
\centering
\includegraphics[scale=0.3]{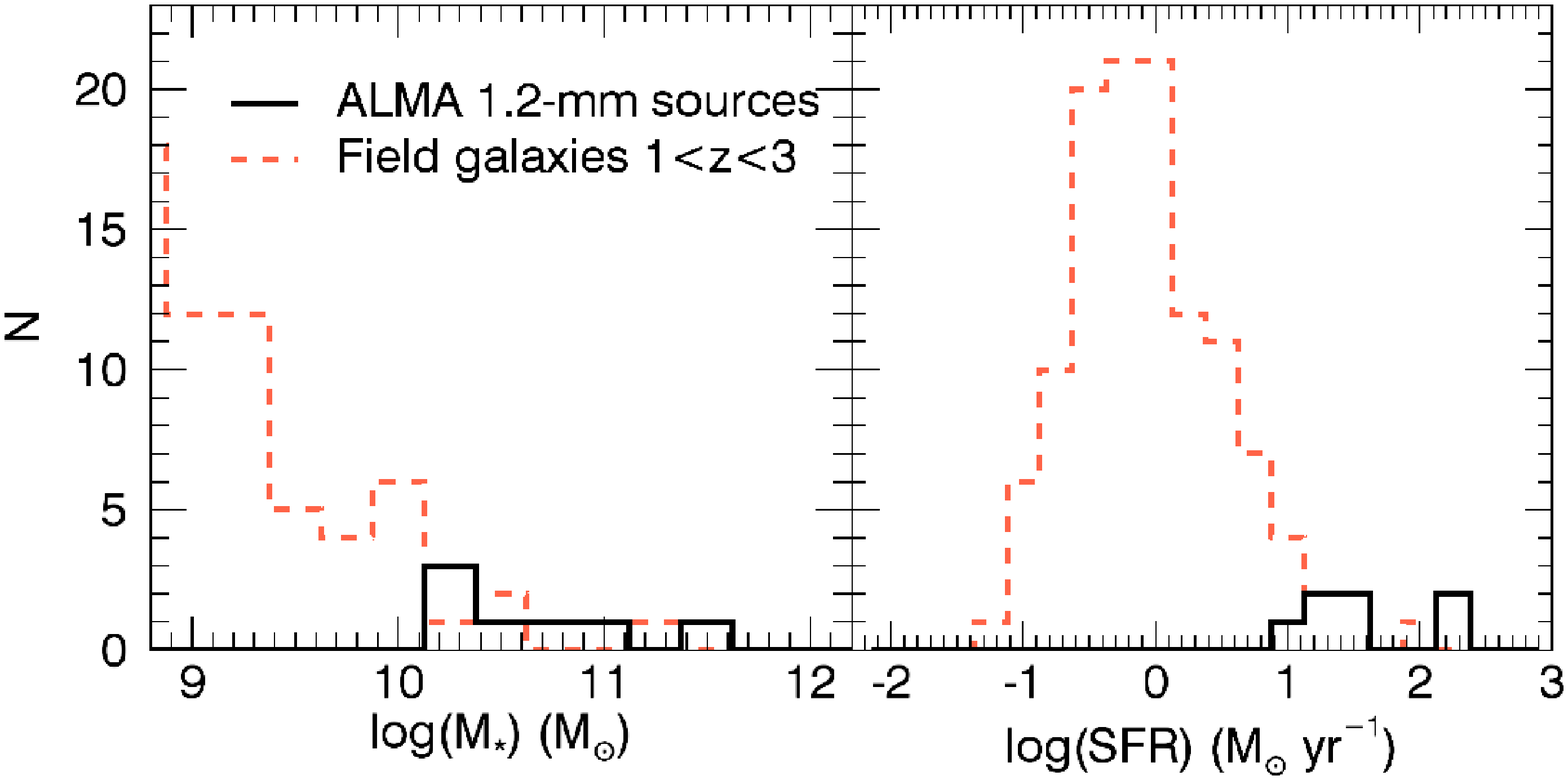}
\caption{Distribution of stellar masses and SFRs (obtained from SED fitting) for the galaxies detected in our ALMA 1.2-mm continuum map. For comparison, the distribution of field galaxies in the relevant redshift range is shown.}\label{fig_distr}
\end{figure}

For each individual galaxy, we perform SED fits to the photometry fixed at the best available redshift. \texttt{MAGPHYS} delivers estimates for the stellar masses, star formation rate (SFR), dust mass and IR luminosity. Even though for most galaxies we do not have photometric constraints on the observed IR SED, \texttt{MAGPHYS} employs a physically-motivated prescription to balance the energy output at different wavelengths. Thus, estimates on the IR luminosity, and/or dust mass, come from constraints on the dust re-processed UV light, which is well sampled by the UV-to-infrared photometry.  For some galaxies with faint optical/near-infrared fluxes or with weak constraints in the photometry, \texttt{MAGPHYS} is able to output only some of the parameters with enough accuracy (e.g., stellar masses). However all the optical counterparts of our millimeter detected sample are sufficiently bright to yield good parameters derived by \texttt{MAGPHYS}. The properties derived for individual sources detected in our ALMA 1.2-mm continuum are shown in Table \ref{tab_prop}.

Figure \ref{fig_distr} shows the distribution of stellar masses and SFRs of our ALMA 1.2-mm continuum sources. For comparison, we show the stellar masses and SFRs derived in the same way for field galaxies located within the field of view of our ALMA map (within PB=0.4), and selected to be in a redshift range that matches the redshifts of our ALMA continuum sources. We limit the comparison sample to sources with $m_{\rm F850LP}$ and $m_{F160W} < 27.5$ mag AB, in order to ensure good SED fits and derived properties. We find that the faint DSFG population, as revealed by our ALMA 1.2-mm sources, have higher stellar masses and SFRs than the field galaxy population at similar redshifts, yet much lower values than those found in brighter DSFGs (i.e. SMGs). Our sources show a median stellar mass of $4.0\times10^{10}\ M_\odot$ and a median SFR of $40\ M_\sun$ yr$^{-1}$, which are significantly lower than the typical values for SMGs, with stellar masses in the range $(0.8-3.0)\times10^{11}\ M_\sun$ \citep[e.g.,][]{michalowski10,hainline11,michalowski12,simpson14,dacunha15, koprowski16}, and SFRs well above $100\ M_\sun$ yr$^{-1}$ \citep[e. g.,][]{casey14}. 

\begin{table*}
\begin{flushleft}
\caption{Derived properties for the ALMA UDF 1.2-mm sources. Columns:  (1) Source name; (2) Best available redshift estimate. If spectroscopic, we quote three decimal places. If photometric, we quote only 2 decimal digits. References: CO based redshifts, confirmed with optical spectroscopy for C1, C2 and C6 (Walter et al. 2016, Paper~I; Decarli et al. 2016b, Paper~IV; Skelton et al. 2014). Optical redshifts for C3, C4, C5, and C7 \citep{skelton14}. Photometric redshifts for C3 and C4 from \citet{coe06} and \citet{skelton14}. (3), (4) AB magnitudes in the F850LP and F160W {\it HST} bands. Uncertainties in quoted values range between 0.01-0.05 mag; (5) Stellar mass derived through SED fitting; (6) SFR derived through SED fitting; (7) Specific SFR (SFR/$M_*$); (8) IR luminosity output from \texttt{MAGPHYS}; (9) ISM mass derived from the dust mass delivered by \texttt{MAGPHYS} and a gas-to-dust ratio $\delta_{\rm GDR}=200$; (10) ISM mass obtained from the 1.2-mm flux and the calibrations from \citet{scoville14}.}
\end{flushleft}
\begin{tabular}{cccccccccc}
\hline
ID & z$_{\rm best}$ & $m_{\rm F850LP}$ & $m_{\rm F160W}$ & log$_{10}(M_{*})$ & log$_{10}$(SFR) & log$_{10}$(sSFR) & log$_{10}(L_{\rm IR})$ & log$_{10}(M_{\rm ISM,d})$ & log$_{10}(M_{\rm ISM,1mm})$ \\
ASPECS &  & (AB mag) & (AB mag) & $(M_{\odot})$ & $(M_{\odot}$ yr$^{-1})$   &  (Gyr$^{-1})$ & ($L_\sun$)  & $(M_{\odot})$ & $(M_{\odot})$  \\
(1) & (2) & (3) & (4) & (5) & (6) & (7) & (8) & (9) & (10) \\
\hline\hline
C1 & 2.543 & 24.0 & 23.2 & $10.36_{-0.00}^{0.12}$ & $ 2.26_{ -0.00}^{ 0.09}$ & $ 1.03_{ -0.10}^{ 0.00}$ & $12.74_{ -0.01}^{ 0.14}$ & $10.59_{ -0.08}^{ 0.06}$ & $10.69\pm 0.01$ \\
C2 & 1.552 & 24.4 & 21.7 & $11.53_{-0.12}^{0.02}$ & $ 1.65_{ -0.00}^{ 0.08}$ & $-0.88_{ -0.00}^{ 0.20}$ & $11.90_{ -0.01}^{ 0.04}$ & $10.45_{ -0.15}^{ 0.12}$ & $10.32\pm 0.04$ \\
C3 & 1.65 & 25.8 & 23.6 & $11.02_{-0.09}^{0.12}$ & $ 2.36_{ -0.34}^{ 0.08}$ & $ 0.38_{ -0.50}^{ 0.10}$ & $12.54_{ -0.32}^{ 0.04}$ & $ 10.05_{ -0.09}^{ 0.06}$ & $10.04\pm 0.05$ \\
C4 & 1.89 & 24.5 & 23.1 & $10.36_{-0.06}^{0.01}$ & $ 1.64_{ -0.35}^{ 0.00}$ & $ 0.27_{ -0.40}^{ 0.00}$ & $11.65_{ -0.33}^{ 0.00}$ & $ 10.00_{  -0.25}^{ 0.25}$ & $ 9.91\pm 0.07$ \\
C5 & 1.846 & 23.4 & 22.0 & $10.61_{-0.06}^{0.06}$ & $ 1.43_{ -0.18}^{ 0.26}$ & $-0.18_{ -0.20}^{ 0.25}$ & $11.46_{ -0.20}^{ 0.36}$ & $ 9.92_{  -0.29}^{ 0.28}$ & $ 9.83\pm 0.08$ \\
C6 & 1.088 & 22.1 & 21.1 & $10.48_{-0.10}^{0.10}$ & $ 1.41_{ -0.14}^{ 0.28}$ & $-0.07_{ -0.20}^{ 0.35}$ & $11.53_{ -0.15}^{ 0.26}$ & $ 10.10_{  -0.22}^{ 0.25}$ & $ 9.95\pm 0.11$ \\
C7 & 1.094 & 22.8 & 21.4 & $10.88_{-0.10}^{0.11}$ & $ 1.21_{ -0.37}^{ 0.38}$ & $-0.68_{ -0.34}^{ 0.35}$ & $11.49_{ -0.35}^{ 0.33}$ & $ 9.88_{  -0.40}^{ 0.30}$ & $ 9.81\pm 0.12$ \\
C8 & \ldots & $>30.6$ & $>30.6$ & \ldots & \ldots & \ldots & \ldots & \ldots & \ldots \\
C9 & \ldots & $>30.6$ & $>30.6$ & \ldots & \ldots & \ldots & \ldots & \ldots & \ldots \\
\hline \label{tab_prop}
\end{tabular}
\end{table*}

\begin{figure}[h]
\centering
\includegraphics[scale=0.4]{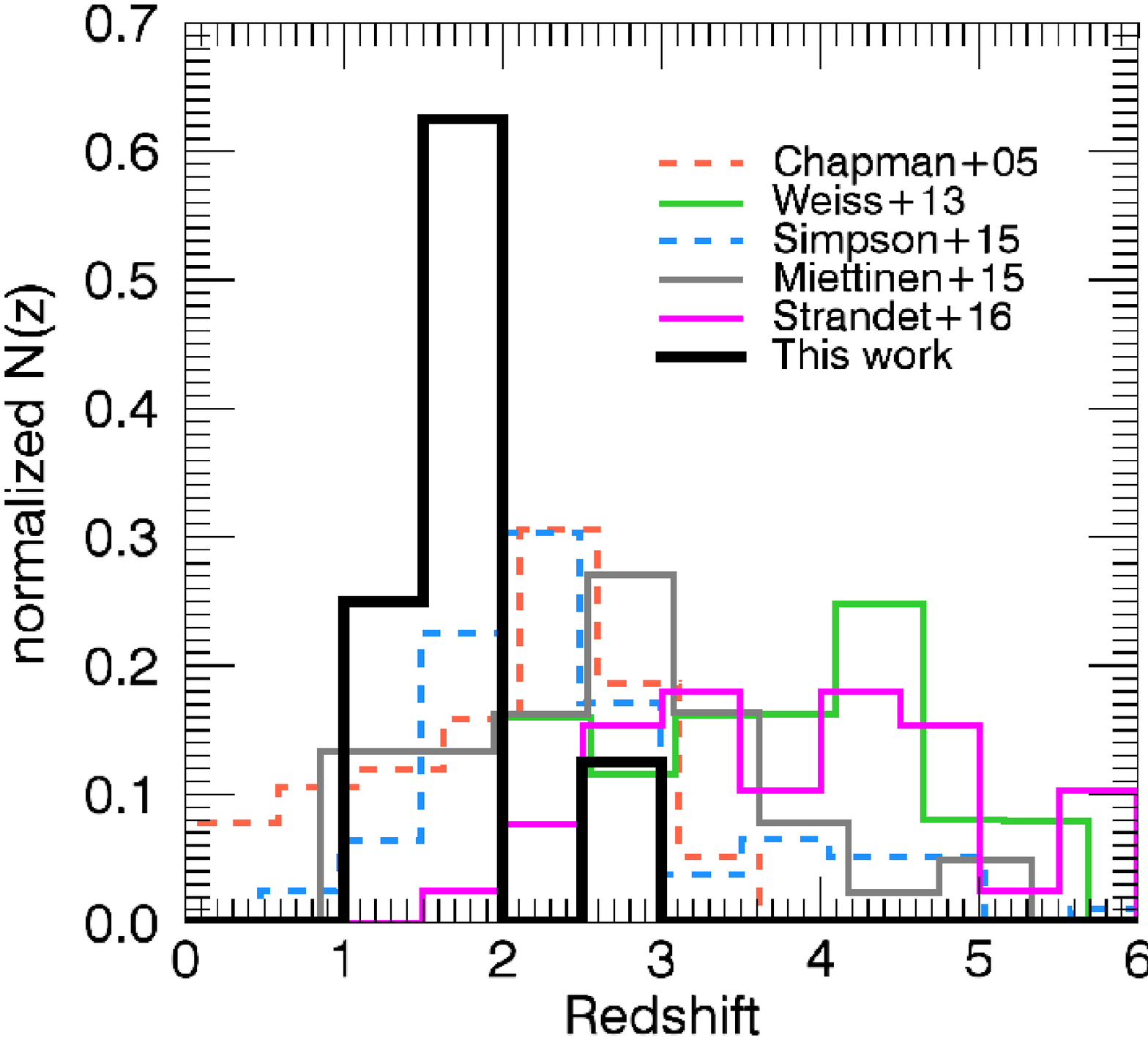}
\caption{Redshift distribution for (sub)millimeter selected galaxies. The y-axis shows the number of galaxies in each bin, normalized to the total number of galaxies in each sample. The black solid line shows the redshift distribution of our ALMA UDF 1.2-mm detections ($>3\sigma$). The gray and green solid lines show the redshift distribution for the 1.2/1.4-mm selected samples of SMGs in the COSMOS \citep{miettinen15} and SPT surveys \citep{weiss13}, respectively. The dashed orange and blue lines show the 850/870-$\mu$m selected SMGs from \citet{chapman05} and from the ECDFS \citep{simpson14}.} \label{fig_zdistr}
\end{figure}

\subsection{Redshift distribution}

Since most of the galaxies detected at $>3.5\sigma$ in our sample have available spectroscopic redshifts from the various surveys of the UDF, we investigate the redshift distribution of our sample. 

Figure \ref{fig_distr} shows the redshift distribution for our ALMA continuum sources that have an optical counterpart compared with various millimeter selected samples of bright DSFGs from the literature. 

We find that all the 1.2-mm continuum sources detected above $3.5\sigma$ in our sample are located in the redshift range $z=1-3$, and none are associated convincingly with a galaxy at $z>3$. This excludes the source candidates without counterparts. While this may only reflect the low number statistics due to the small area of the sky covered, it also supports the idea that the population of galaxies discovered in our deep ALMA 1.2-mm continuum map significantly differs from the population of DSFGs found in shallower but wider (sub)millimeter surveys. The DSFGs samples from the literature are found to have a median redshifts ranging from $z\sim2.1$ and $z\sim3.1$, respectively, with a possible tail extending out to $z\sim6$ \citep{chapman05,yun12, smolcic12, weiss13, riechers13, simpson14, miettinen15, strandet16, dunlop16}. We find that our faint ALMA millimeter-selected galaxies, however, have a median redshift $z=1.7\pm0.4$. The uncertainty here corresponds to the scatter in the redshifts. This median redshift is significantly lower than the typical redshift of bright DSFGs, irrespective of the nature the DSFG samples (lensed or unlensed) or the selection wavelength (870-$\mu$m or 1.2-mm). Statistically, this would not be significantly affected if the two sources without counterparts were located at $z>2$ given the small scatter in the redshift distribution.

\begin{figure*}[ht]
\centering
\includegraphics[scale=0.45]{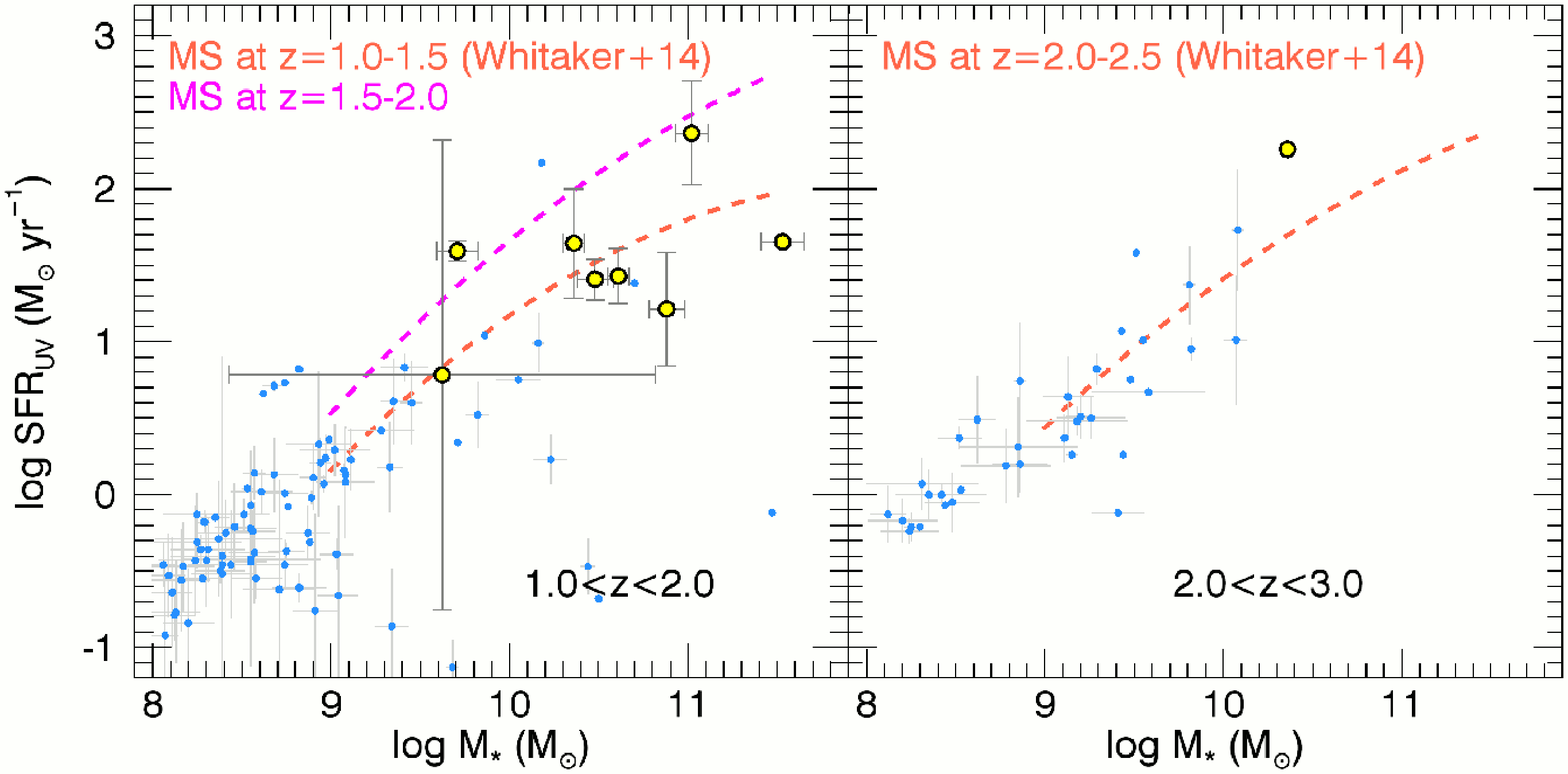}
\caption{Stellar mass versus SFR for the galaxies covered in our ALMA UDF 1.2-mm map in the two relevant redshift bins. The large yellow circles show the ALMA 1.2-mm continuum sources ($>3.5\sigma$). The small blue circles show field galaxies in either the $1.0<z<2.0$ or $2.0<z<3.0$ redshift bins. Field galaxies are restricted to be brighter than 27.5 AB mag in the F850LP and F160W bands. For comparison, the orange and magenta curves represent the best second order polynomial fits of the star formation sequence at $1.0<z<1.5$, $1.5<z<2.0$ and $2.0<z<2.5$ for the left and right panels, respectively \citep{whitaker14}.}\label{fig_msz}
\end{figure*}

While the SMG and fainter-mm source populations are obviously different as reflected by the significantly lower 1.2-mm fluxes, this is the first time that we are able to evaluate the redshift distribution of the faintest 1.2-mm emitters in a contiguous blank field (below $S_{\rm 1.2mm}=0.5$ mJy). Other studies reaching down to the faint mm flux regime, are mostly based on archival data of different individual fields where the faint mm emitters are not the main targets \citep[e.g.,][]{oteo15, carniani15, fujimoto16} or do not have the excellent deep multi wavelength coverage of the HUDF in order to address this issue. 

The decline in the median redshift with decreasing flux density for millimeter selected sources was recently predicted by phenomenological models of galaxy evolution \citep{bethermin15b}. Even though the prediction does not assess the redshifts for populations with 1.2-mm flux densities below 0.2 mJy, already at this flux level they find a median redshift of $\sim2$ compared to the much higher $z\sim3$ predicted for brighter SMGs selected at 1.2-mm. By extrapolating their prediction down to a flux density cut of $\sim35\mu$Jy (our 3$\sigma$ cut), we find an expected median redshift of $\sim1.5$. This value is in good agreement with our measurements, and supports the fact that the redshift distribution of millimeter-selected galaxies is affected by the flux density cut.

\subsection{Starburst versus Main sequence}

An important result from multi-wavelength surveys in the last decade has been the determination that typical star-forming galaxies form a tight linear relationship in the SFR-$M_{\rm stars}$ plane out to $z\sim3$ \citep[e.g., ][]{brinchmann04, elbaz07, noeske07,daddi07,pannella09,karim11,rodighiero11,whitaker12}. Sources that lie close to this star formation relationship have been termed main sequence galaxies. Galaxies lying above this sequence are called starbursts, as they have excess star formation activity with respect to most galaxies in the main-sequence for the same stellar mass, or higher specific star formation rates (sSFRs). This sequence has been observed to evolve with redshift, with higher SFRs for a given stellar mass at increasing redshifts \citep{whitaker12}, and it has also been claimed to flatten at the high stellar mass end \citep{whitaker12, whitaker14,pannella15,lee15}.

Figure \ref{fig_msz} shows the stellar mass versus SFR derived using \texttt{MAGPHYS} for all HST-detected galaxies at $1<z<3$ contained within our ALMA UDF survey area (within PB $=0.4$ of our 1.2-mm map), and restricted to be brighter than 27.5 AB mag in the F850LP and F160W bands. We show the sources detected in our 1.2-mm observations ($>3.5\sigma$), and compare with the main-sequence fit derived by Whitaker et al. (2014). We find that all the millimeter detected galaxies at $z<2$ are located within the scatter of the main sequence at $z\sim1-2$ and taking into account the uncertainties in the derived properties. Similarly, the only millimeter detection at $z>2$ (ASPECS C1) is also well within the scatter of the main sequence at $z=2.0-2.5$. We thus conclude that our faint ALMA 1.2-mm continuum sources are main-sequence galaxies at $z\sim1-3$.

\begin{figure*}[ht]
\centering
\includegraphics[scale=0.3]{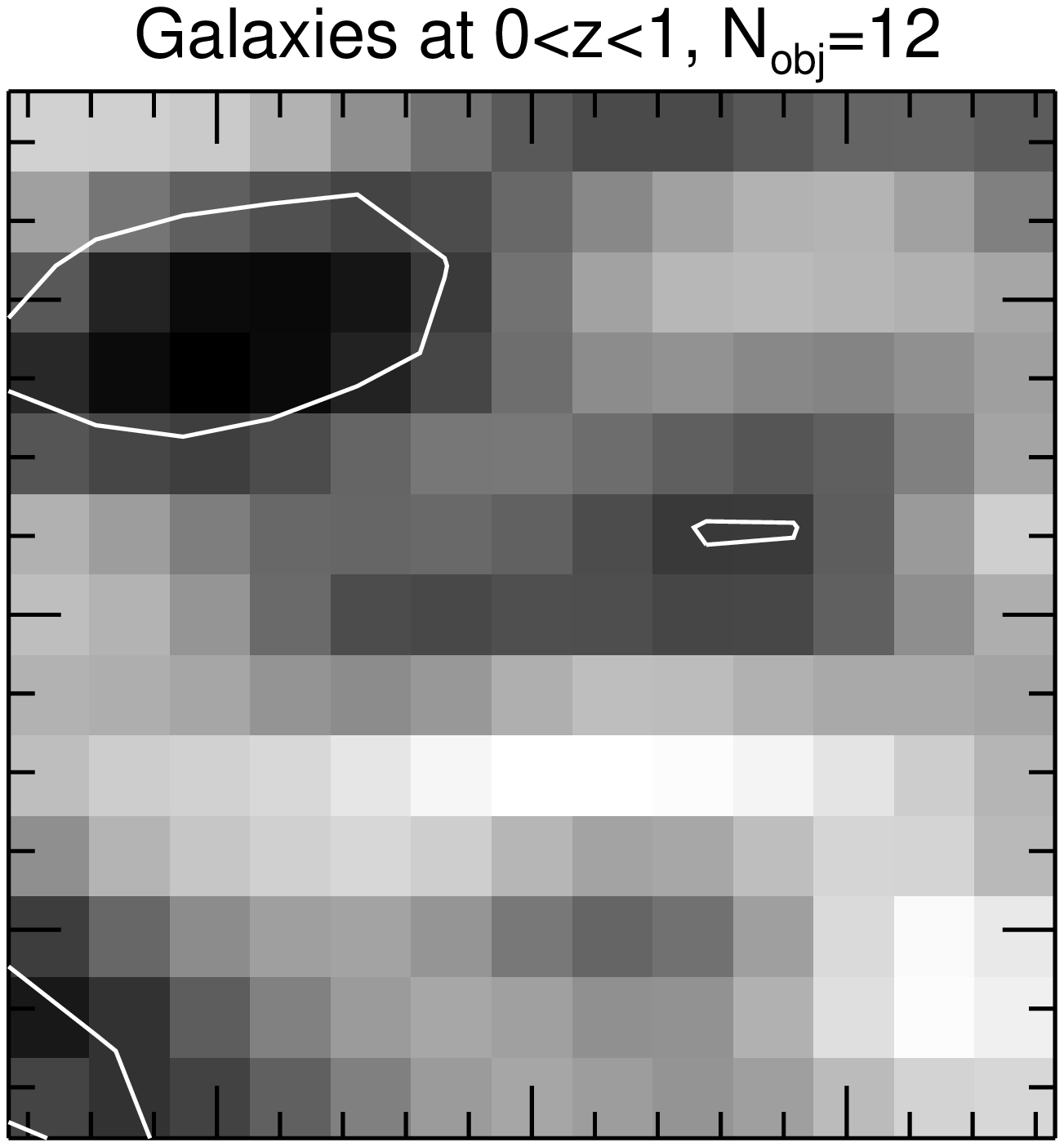}
\includegraphics[scale=0.3]{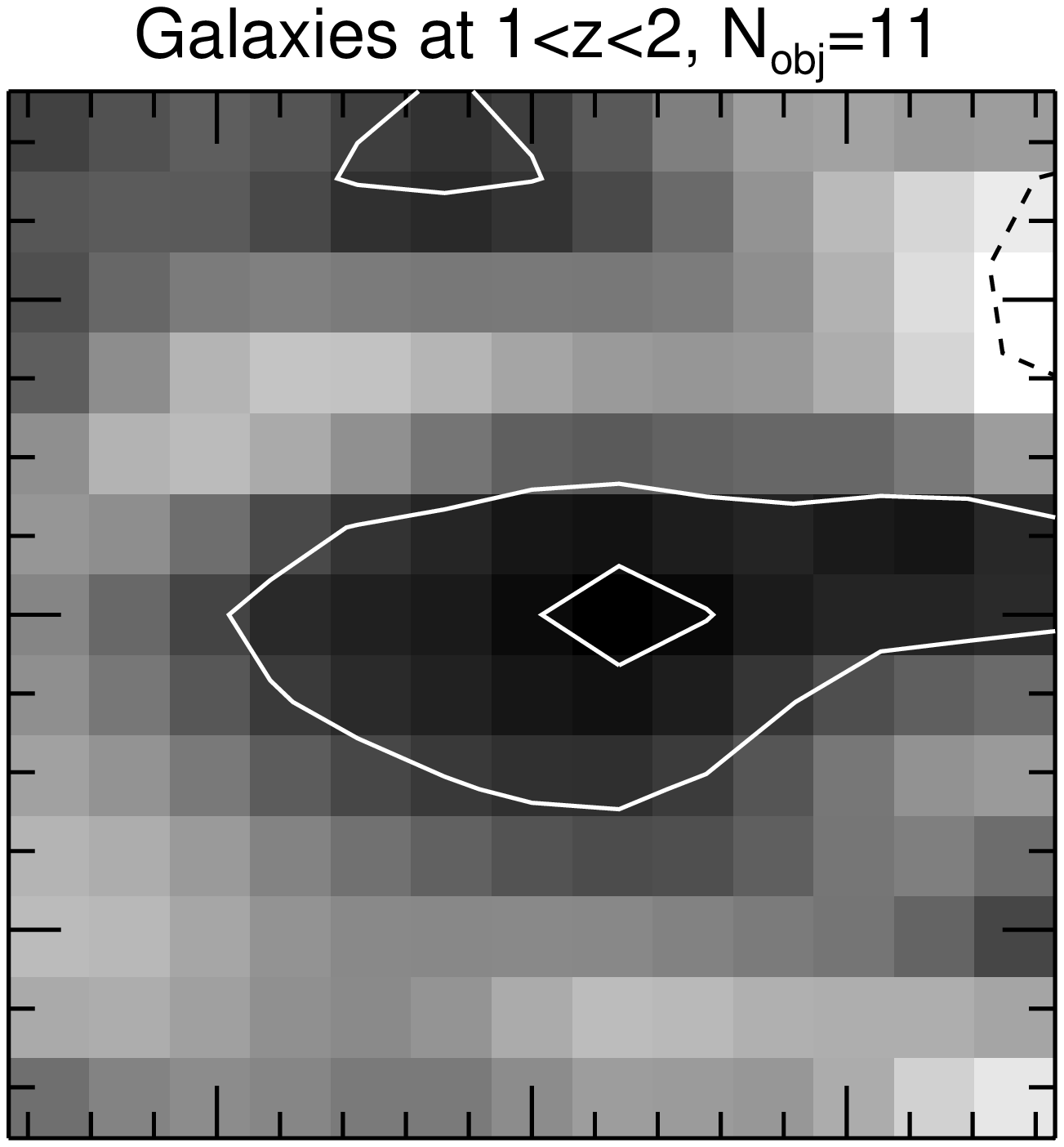}
\includegraphics[scale=0.3]{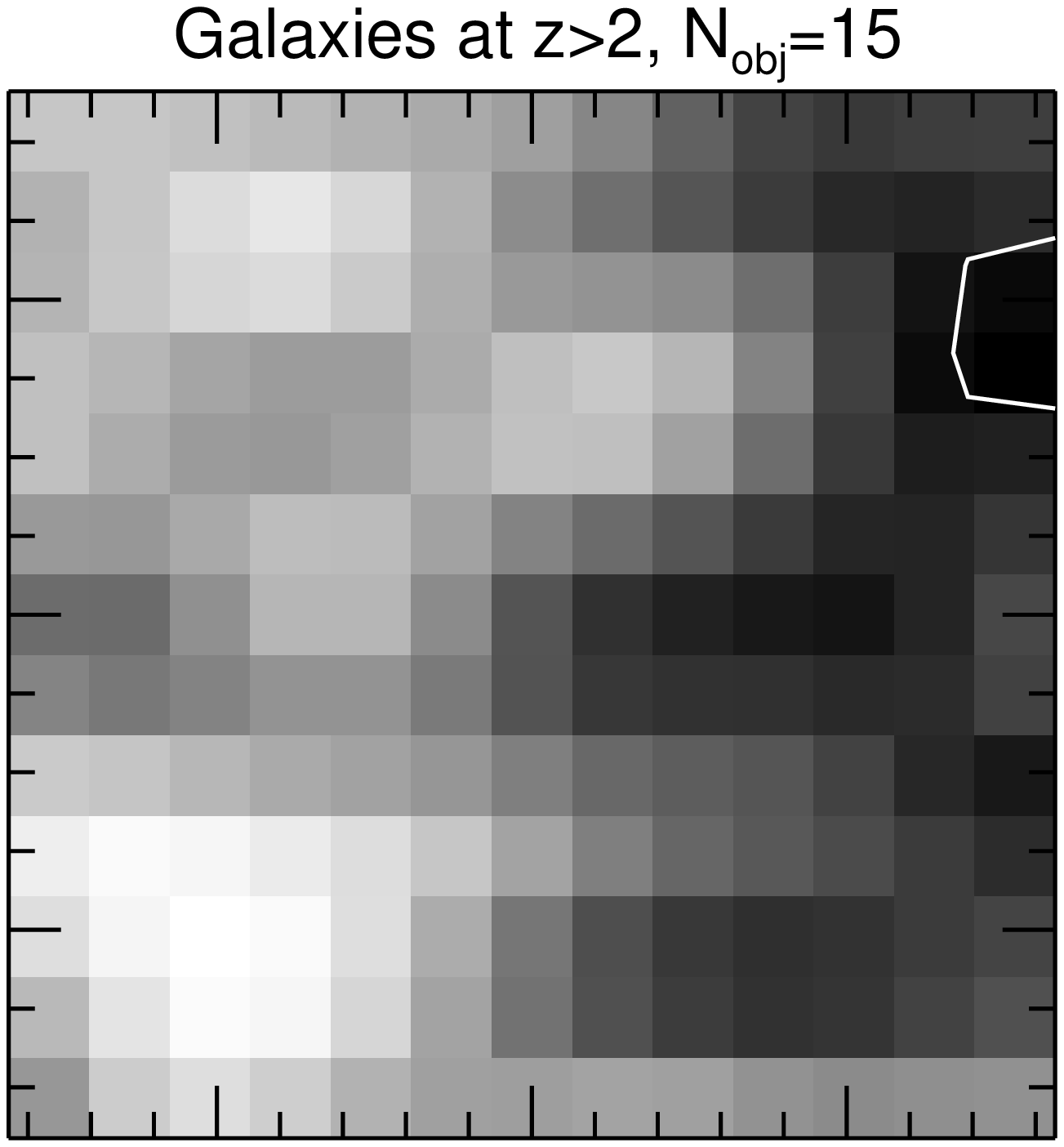}\\

\vspace{3mm}
\includegraphics[scale=0.3]{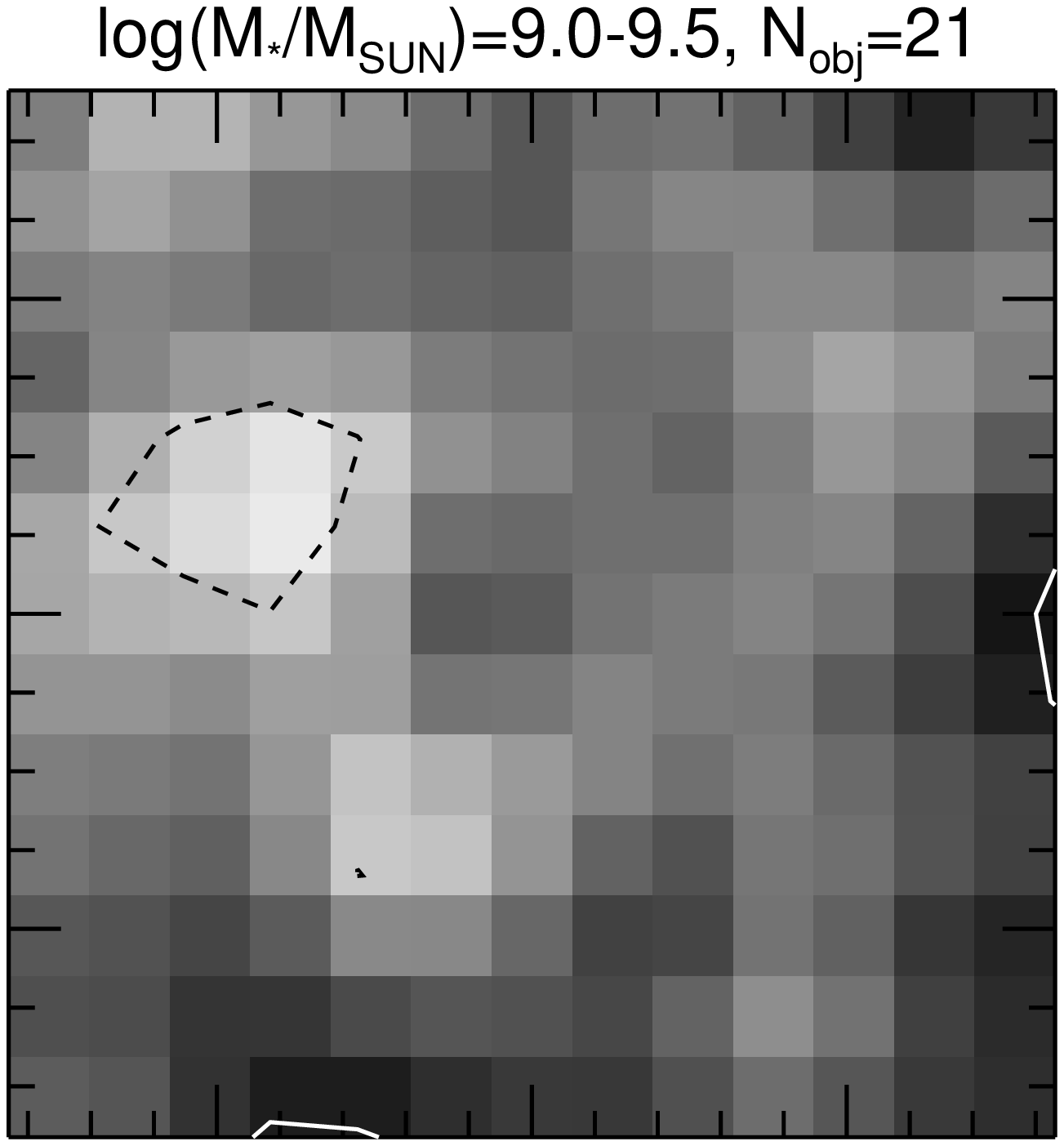}
\includegraphics[scale=0.3]{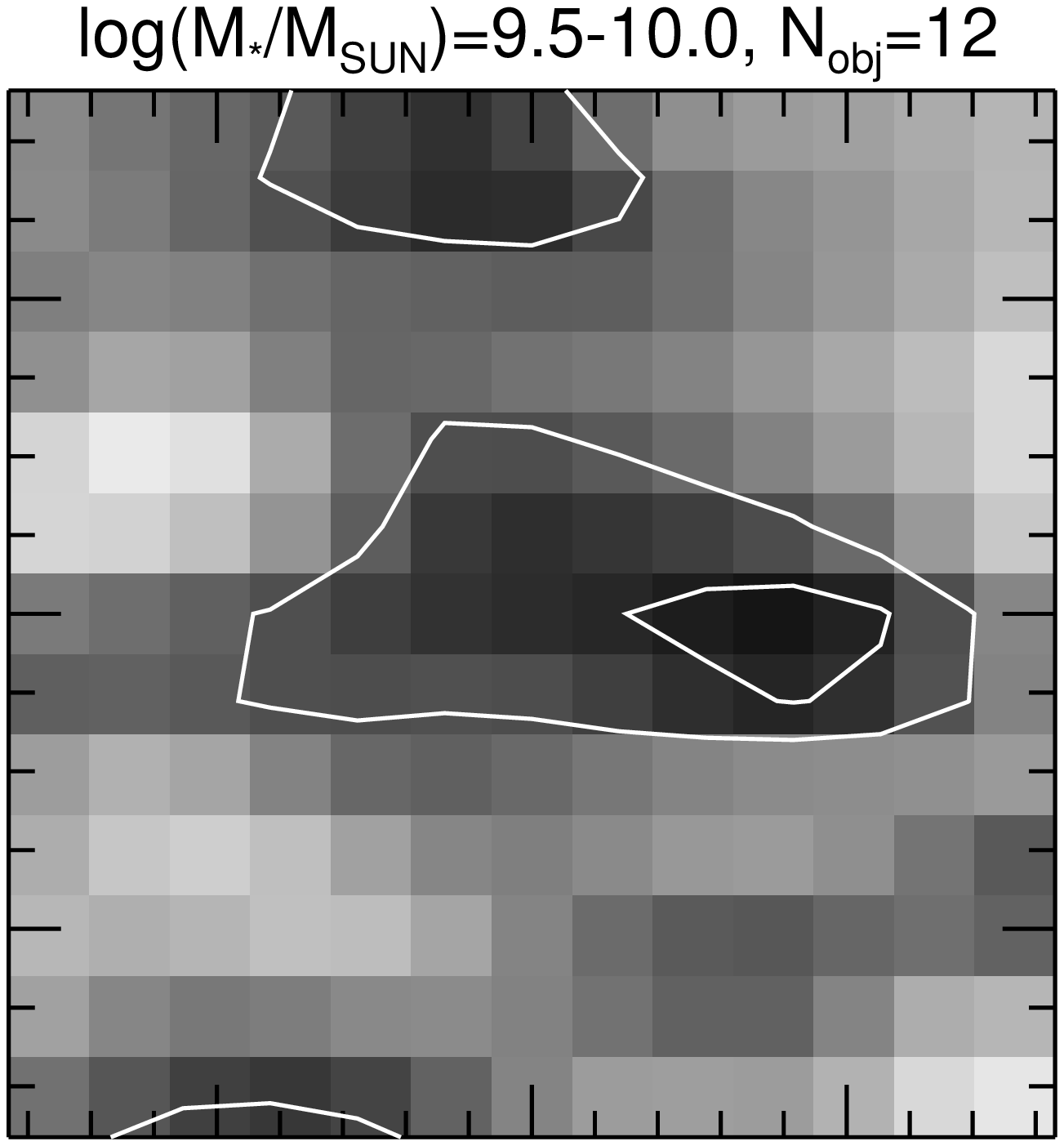}
\includegraphics[scale=0.3]{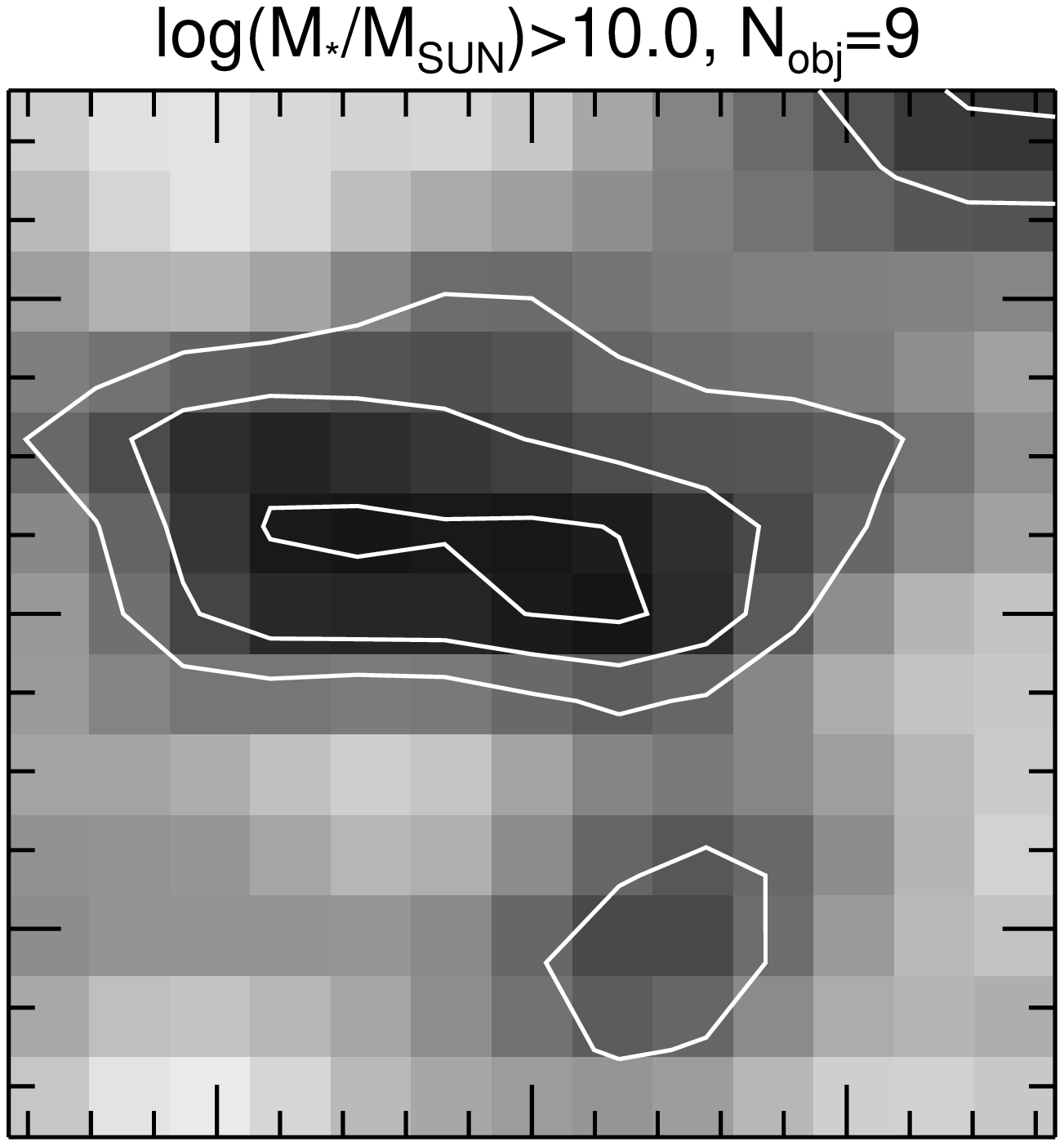}\\
\vspace{3mm}
\includegraphics[scale=0.3]{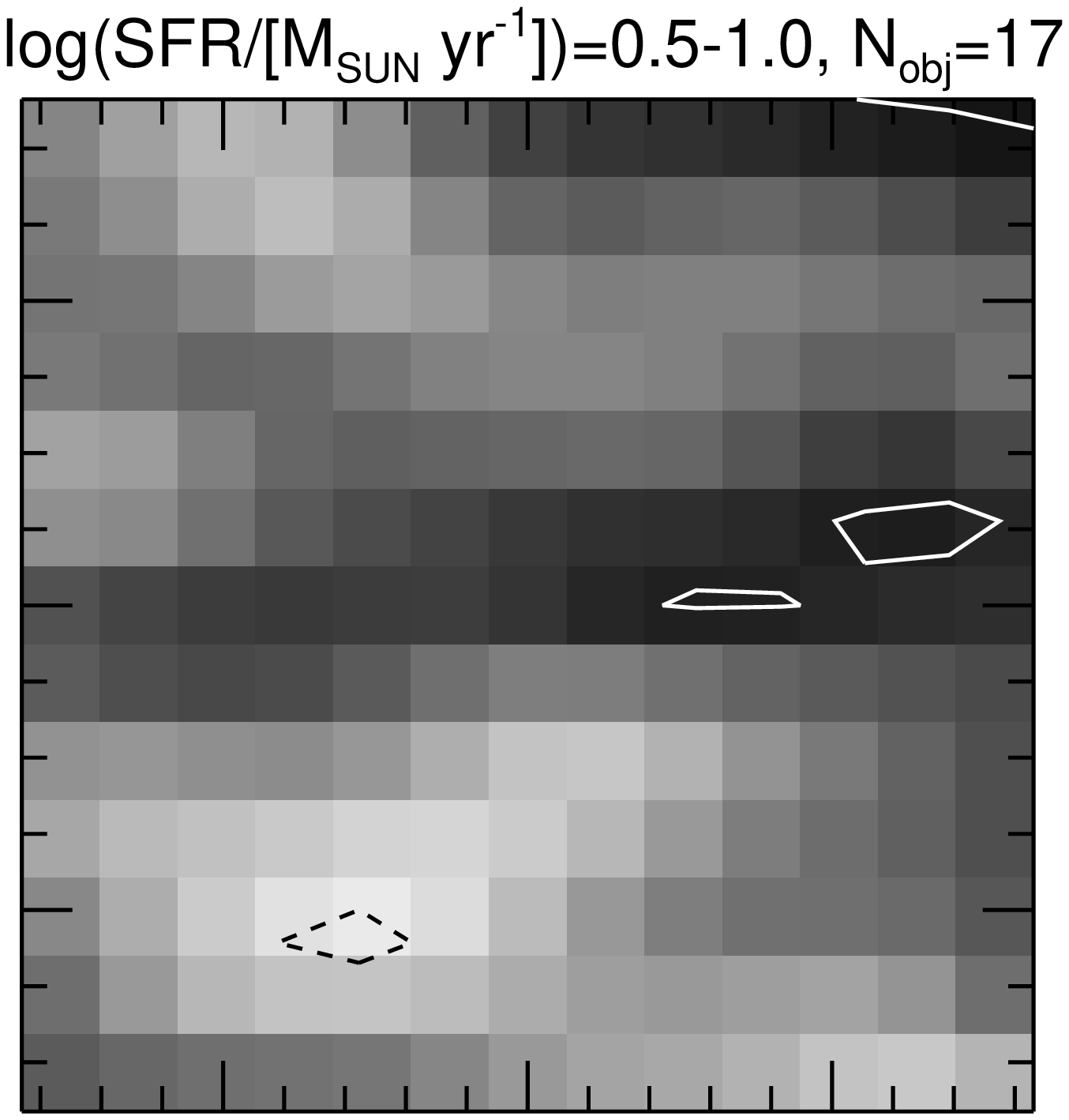}
\includegraphics[scale=0.3]{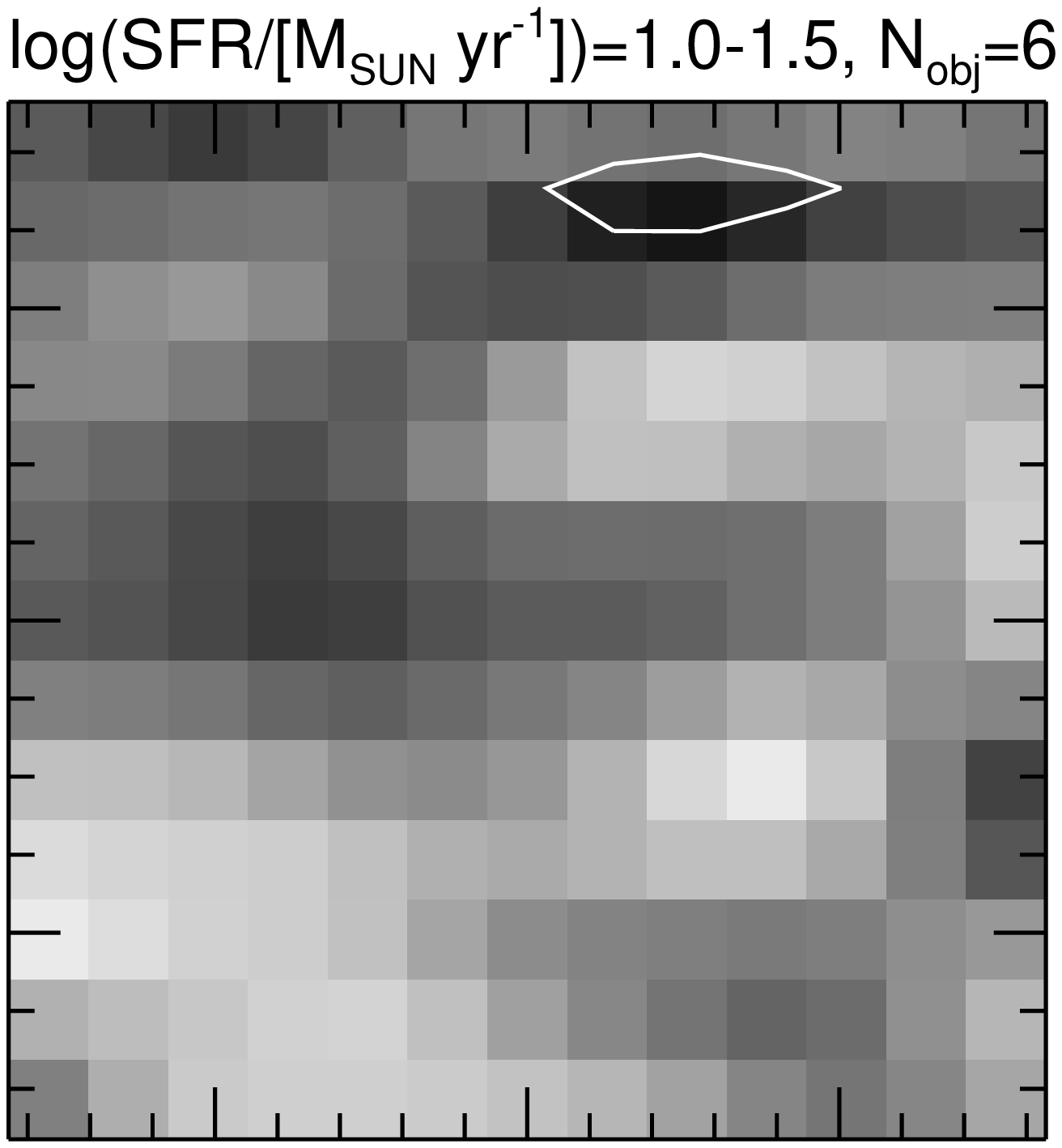}
\includegraphics[scale=0.3]{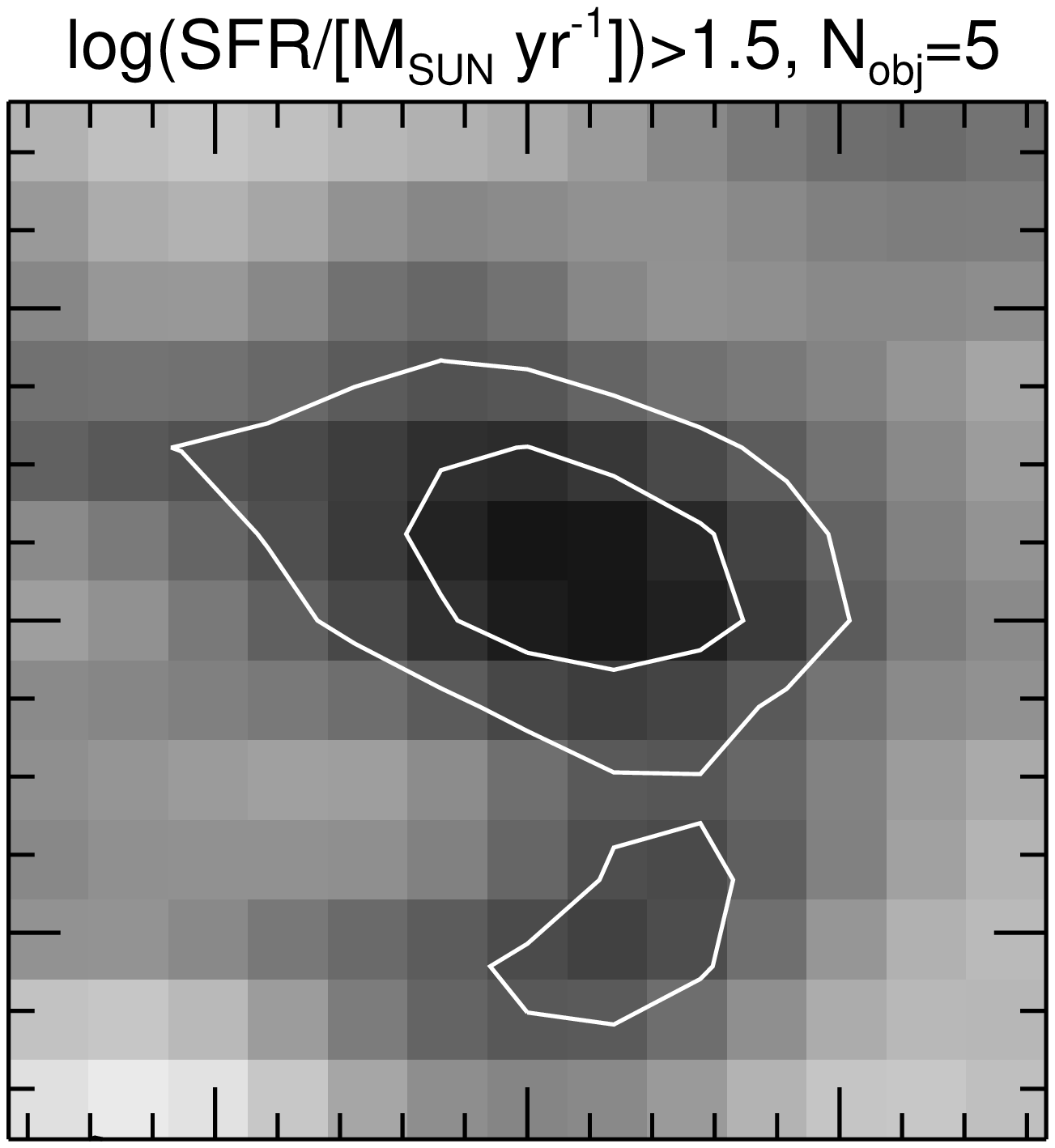}
\caption{Stacked 1.2-mm continuum on the location of galaxies selected as summarised in Table \ref{tab_stack} (see also text): Galaxies selected in the redshift, stellar mass and SFR ranges are shown on the top, middle and bottom panels, respectively. Sources individually detected in the 1.2-mm map at S/N$>3.5$ are not included in the stacks. The images shown are $3.6''\times3.6''$ in size. Solid white and dashed black contours represent the positive and negative signal, respectively. Contours start at $\pm2\sigma$ in steps of $\pm1\sigma$.}\label{fig_stacking12}
\end{figure*}

Recently, \citet{hatsukade15} studied the properties of four 1.3-mm detected sources with fluxes $S_{\rm 1.3mm}>0.2$ mJy (at least two times brighter than our sources). They find that these four galaxies are in the main-sequence, with redshifts $z=1.3-1.6$. However, those sources were selected in fields where these faint millimeter emitters were not the primary target. Most of these continuum sources lie in a dense environment at $z\sim1.3$, and it is thus unclear how representative their redshift and properties is of the field population.

All the sources shown in Fig. \ref{fig_msz} lie within the uniform sensitivity region of our 1.2-mm mosaic, within PB $=0.4$. However, there are a few of them that were not detected in the 1.2-mm continuum even though they have similar SFRs and stellar masses than the detected sources. This could partly be attributed to uncertainties in the SED fitting procedure or to the fact that some galaxies would be located at the very edges of our mosaic. However, it is also possible the non-detection of these sources could also be due to differences in the individual physical properties of these sources. For instance, galaxies with lower dust temperatures or masses would tend to have lower fluxes at 1.2-mm, or they could just be dust poor. In \S \ref{sec_stack} below we address this issue using stacking analysis.

\begin{table*}[ht]
\label{tab_stack}
\caption{Results from the stacking analysis. Columns: (1) Sample name; (2) Selection imposed for this sample. In all cases, we excluded the individually detected sources with $>3.5\sigma$. We limited the samples to have $M_*>10^9\ M_\sun$, to be located within PB=0.4 and to lie $3.5''$ away from the five most significant 1.2-mm continuum detections to avoid contamination. Additionally, in order to reject non-star forming sources in our stacks (i.e. old passive evolving galaxies), we restricted the samples to reside above the main-sequence including its intrinsic scatter at the relevant redshift range (i.e. sources above MS-0.5 dex), using the calibrations from \citet{whitaker14}. Only sources with $m_{\rm F850LP}$ and $m_{\rm 160W}<27.5$ mag AB were included, in order to retain sources with good SED fits; (3) Median redshift of the selected sample; (4) Median SFR obtained from the optical/near-infrared photometry with \texttt{MAGPHYS}; (5) Median stellar mass obtained from the optical/near-infrared photometry with \texttt{MAGPHYS}; (6) Number of objects that entered the stack; (7) Average flux density at 1.2-mm obtained from the stacking procedure.}
\centering
\begin{tabular}{lcccccc}
\hline
Sample $^a$& Selection$^b$ &  $z_{\rm med}$ $^c$ & log$_{10}$(SFR$_{\rm UV, med}$) $^d$ & log$_{10}$($M_{*, med}$)   & $N_{\rm obj}$ $^e$ & $S_{\rm 1.2mm}$ $^f$\\
                   &                         &                        &    ($M_\sun$ yr$^{-1}$)  & ($M_\sun$)   &       &   ($\mu$Jy) \\
      (1) & (2) & (3) & (4) & (5) & (6) & (7) \\
\hline
z1  & $z=0-1$ & $0.76\pm0.19$            &   $0.84\pm0.52$       &    $9.78\pm0.49$       &   12       &  $<13$ \\
z2  & $z=1-2$ & $1.22\pm0.20$            &   $0.52\pm0.63$        &    $9.45\pm0.43$       &   11       & $12\pm4$ \\
z3  & $z=2-4$ & $2.45\pm0.41$            &   $0.75\pm0.36$       &   $9.48\pm0.31$        &   15       &  $<13$ \\

\hline
m1  & log$_{10}(M_*/M_\sun)=9.0-9.5~~$ &   $1.63\pm0.80$         &   $0.46\pm0.35$             &   $9.25\pm0.14$        &  21       & $<8$  \\
m2  & log$_{10}(M_*/M_\sun)=9.5-10.0$ &     $1.29\pm0.95$        &    $0.93\pm0.35$            &   $9.78\pm0.12$        &   12      & $11\pm3.0$  \\%
m3  & log$_{10}(M_*/M_\sun)>10.0$ &   $1.10\pm0.79$          &  $1.00\pm0.51$              &   $10.2\pm0.22$        &   9      & $19\pm5$ \\
\hline
s1  & log$_{10}$(SFR)$={0.5-1.0}\ $M$_\sun$ yr$^{-1}$ &     $1.67\pm0.93$     &    $0.70\pm0.16$           &   $9.58\pm0.36$        &   17      &  $<12$ \\
s2  & log$_{10}$(SFR)$={1.0-1.5}\ $M$_\sun$ yr$^{-1}$&    $2.45\pm0.78$         &     $1.02\pm0.13$          &   $9.81\pm0.27$        &    6     &  $<15$ \\
s3 & log$_{10}$(SFR)$>1.5\ $M$_\sun$ yr$^{-1}$&     $1.05\pm0.48$        &     $1.73\pm0.21$         &    $10.2\pm0.36$       &   5      & $25\pm8$ \\

\hline\hline
\end{tabular}
\end{table*}

\section{Stacking analysis}
\label{sec_stack}

We use the stacking analysis to investigate the nature of the fainter galaxy population not detected at the achieved sensitivity limit of our ALMA 1.2-mm mosaic. To perform the stacking, we extract smaller images, $9''\times9''$ in size, from the final clean ALMA 1.2-mm continuum mosaic, centered at the position of sources that were selected from an independent galaxy catalog (see below). Sub-images of the same size are simultaneously extracted from the PB sensitivity mosaic map. All these sub-images are then combined together, to construct a weighted average using the PB sensitivity map as the weight. The noise in this average image is then obtained from an annulus around the central position with an initial and final radius of 4 and 12 pixels, respectively (1 pixel = $0.3"$). A summary of the stacking analysis results is shown in Fig.~\ref{fig_stacking12}, and listed in Table \ref{tab_stack}. 

\subsection{Nature of undetected galaxies}

Using stacking, we first investigate the emission from galaxies individually undetected at the $3.5\sigma$ level in the ALMA 1.2-mm continuum map as a function of redshift. If these galaxies were to follow a similar redshift distribution as the detected galaxies, then we would expect on average that the galaxies in the $1<z<2$ range would have more 1.2-mm continuum emission than those in other redshift ranges. Figure \ref{fig_stacking12} shows the stacked emission of galaxies in 3 different redshift ranges (samples z1, z2 and z3; see Table \ref{tab_stack}). All samples have been selected to have $M_*>10^9\ M_\sun$ and $z<4$, and sources that enter the stack were required to lie $3.5''$ away from the location of the five most significant individual continuum detections to avoid contamination. The restriction to have a relatively high stellar mass is specifically to not down weight the stack signal. To avoid including passive evolving galaxies with no star formation activity in the stacks, we only select galaxies that are located within and above the main sequence (see Fig. \ref{fig_msz}), taking into account a conservative 0.5 dex of scatter in the main sequence relationship. The main-sequence trends as a function of redshift are taken from \citet{whitaker14}. Additionally, to limit our sample only to galaxies with good measured SED fits, we require that the sample galaxies have magnitudes brighter than 27.5 AB in the F850LP and F160W bands. Galaxies detected at the $>3.5\sigma$ level in the 1.2-mm continuum have been excluded from the stacked samples. Using this selection, we only detect 1.2-mm emission from galaxies at $1<z<2$ (the z2 sample). In all the other redshift samples, we do not find significant emission and thus place $3\sigma$ limits on the 1.2-mm flux densities (see Table \ref{tab_stack}). This implies that most of the underlying millimeter emission that is not directly detected in our ALMA continuum map, comes from galaxies located at similar redshifts as the individually detected galaxies, which have matching redshift distribution with a median $z=1.65$. 

To shed light on whether the most massive or star-forming galaxies could have underlying 1.2-mm emission, we stack on different galaxy samples split in stellar mass and SFR. We use three samples divided by stellar mass and three samples divided by SFR (see Table \ref{tab_stack}). We apply the same restrictions than for the redshift samples, including the limit in stellar mass, the requirement that the galaxies lie within and above the main sequence and the magnitude limit in the optical/near-infrared bands. The galaxies used in these stacks are represented by blue symbols in Fig. \ref{fig_msz} (this Fig. does not show galaxies at $z<1$ and $z>3$).

Figure \ref{fig_stacking12} (middle and bottom panels) shows the results of this exercise. From the three stellar mass samples, only the samples m2 and m3 present a tentative detection of the stacked 1.2-mm emission. For sample m1, we place a $3\sigma$ upper limit. This indicates that less massive galaxies have fainter millimeter continuum emission. Note that the stacked detection for the m2 sample is offset from the center, being unclear the reason for this shift since we are excluding sources near the most significant 1.2-mm sources. It is possible this shift is related to the low S/N of the signal. 

By stacking in samples that were selected based on their UV-SFRs (derived from SED fitting), we find a clear detection for the s3 sample, which includes all galaxies with SFR$_{\rm UV}>30\ M_\odot$ yr$^{-1}$. This is consistent with the detection of emission in the mass-selected samples m2 and m3, which have a concordantly high median UV-derived SFRs. Note that most of the galaxies individually detected at 1.2-mm comply with the s3 sample selection. Thus, the detection of stacked continuum signal in the s3 sample implies that the individually undetected galaxies are just below the detection threshold of our survey, showing on average lower millimeter emission than the individually detected galaxies. The reason for this could be due to uncertainties in the derived stellar masses and SFRs, as well as different physical properties such as lower dust content (lower dust masses).

In summary, we find that most of the millimeter continuum emission of undetected galaxies is produced by galaxies in the redshift range $z = 1-2$ (sample z2). When we make stacks on stellar mass, we obtain detections for the stellar mass ranges $10^{9.5-10.0}\ M_{\sun}$ and $>10^{10}\ M_{\sun}$ (samples m2 and m3). These stellar mass bins have median UV-derived SFRs in the range of $\sim(3-30)\ M_\sun$ yr$^{-1}$. When we explicitly consider galaxy samples with UV-derived SFRs, we only obtain a detection for galaxies with SFRs $>30 \ M_\sun$ yr$^{-1}$ (but not for the $10-30\ M_\sun$ yr$^{-1}$ bin). These stacked detections reach down to 1.2-mm continuum fluxes of $\sim$10 $\mu$Jy.

\subsection{Stacking in the 3-mm continuum}
\label{sec_3mmstack}
\begin{figure}
\centering
\includegraphics[scale=0.28]{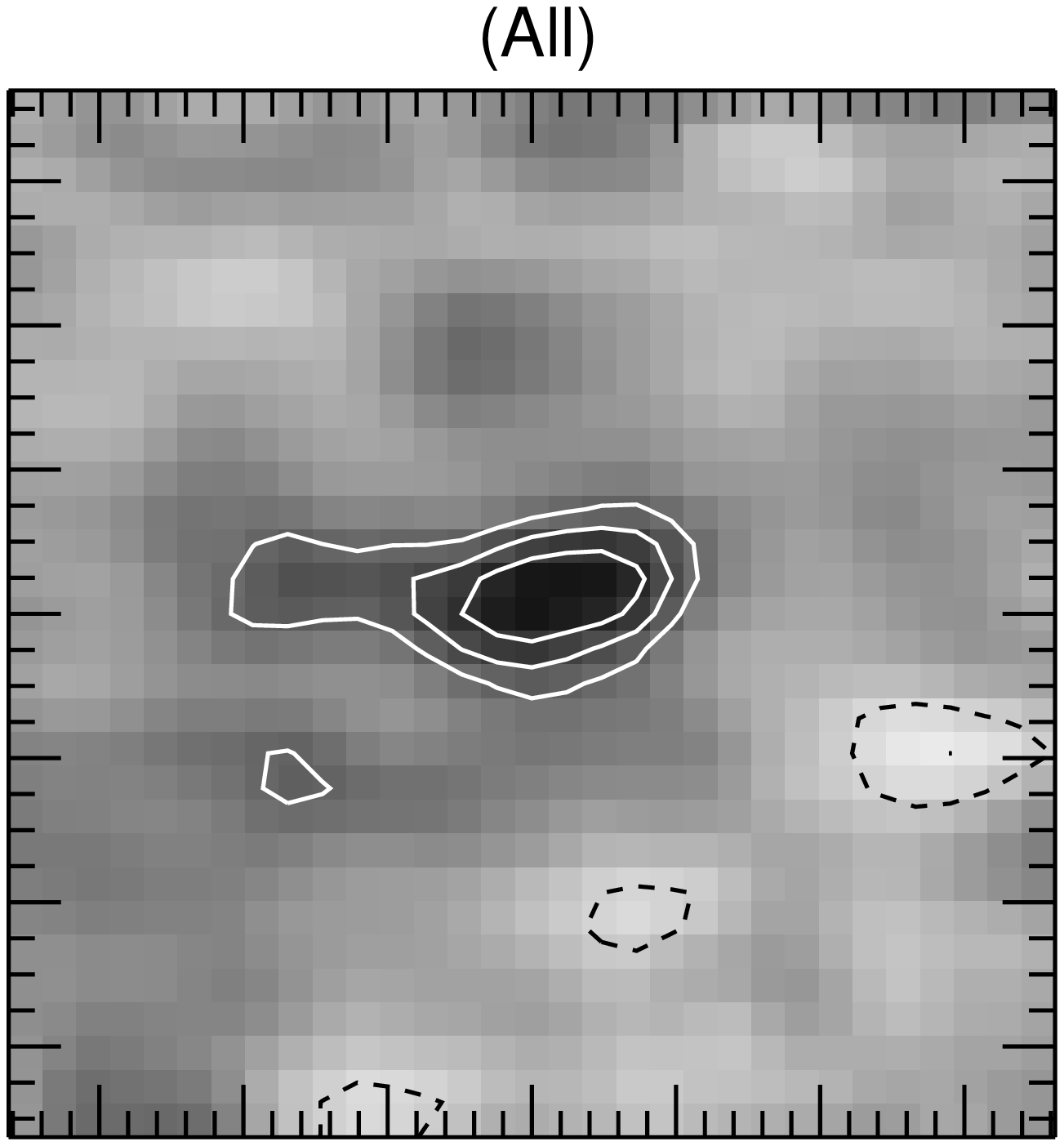}
\includegraphics[scale=0.28]{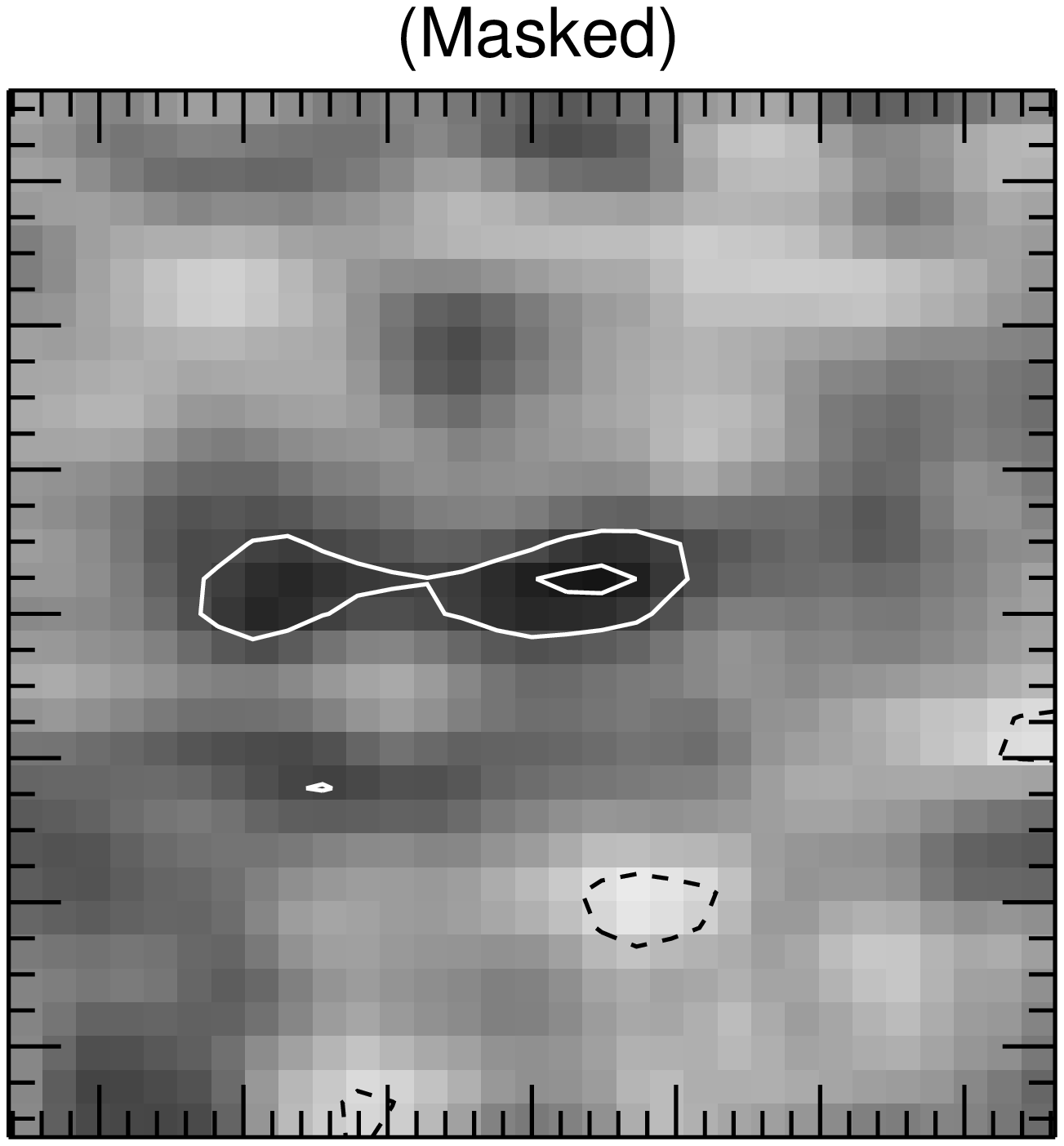}
\caption{Stacked 3-mm emission at the location of the 1.2-mm detected sources ($15"\times15"$ in size). The left panel shows the stacked map when including all sources. The right panel shows the stacked map when including all but the brightest 1.2-mm source, which was also individually detected at 3-mm. White and black contours represent positive and negative emission, respectively. The contours are shown in steps of $\pm1\sigma$ starting at $\pm2\sigma$.}\label{fig_stack3mm}
\end{figure}

Since there is only one significant source in the 3-mm continuum map, we use the stacking analysis to measure the average 3-mm emission from all the sources that were detected at $>3.5\sigma$ in the 1.2-mm map. The result of this procedure is shown in Fig. \ref{fig_stack3mm}. Including all the 1.2-mm sources in the stack, we find an average flux density of $S_{\rm 3mm, all}=12\pm3\ \mu$Jy. Masking the individually detected source in the 3-mm map, we find an average flux density of $S_{\rm 3mm, masked}=9\pm3 \mu$Jy. Using the same stacking procedure and adopting the same samples on the 1.2-mm map (i.e. stacking the 1.2-mm detected sources to obtain the average 1.2-mm flux), we find $S_{\rm 1.2mm, all}=195\pm11\ \mu$Jy and $S_{\rm 1.2mm, masked}=125\pm12\ \mu$Jy, respectively.

The ratio between these measurements can now be used to obtain an estimate of the dust emissivity index $\beta$. We use a single-component modified black body dust model in the optically thin regime of the form $S_\nu\propto (1-e^{-\tau_\nu}) B_\nu(T_{\rm d})$ (see Weiss et al. 2007), where $S_{\nu}$ is the observed flux density, $B_{\nu}$ is the Planck function, and $T_{\rm d}$ is the dust temperature. It can be shown that in the Rayleigh-Jeans (RJ) limit, 

\begin{equation}
\beta = \frac{log(\frac{S(\nu_1)}{S(\nu_2)})}{log(\frac{\nu_1}{\nu_2})} - 2,
\end{equation}

where $S(\nu_1)$ and $S(\nu_2)$ are the flux densities measured at the frequencies $\nu_1$ and $\nu_2$, respectively. Note that at the observed frequencies it is valid to assume the optically thin and RJ approximations.

For the galaxy individually detected in the 1.2-mm and 3-mm maps (ASPECS C1), we find $\beta=1.3\pm0.2$. For the stack sample that includes all the sources, we find $\beta=1.1\pm0.3$. Similarly, for the masked sample we find $\beta=0.9\pm0.4$. This result suggests a significantly lower dust emissivity index for the faint population of DSFGs than what has been typically found in galaxies in the local Universe and the Milky Way, and also at high-redshift, with $\beta$ ranging from 1.5 to 2.0 \citep[e.g.,][]{chapin09,dunne11,draine11, planck11a}. Note that given the relatively small beam size of the 1.2 mm observations, we could be missing flux that could contribute to a larger $\beta$ value. Similarly, the stacked signal detected at 3-mm is weak, and its detection is thus marginal. Both issues could thus be affecting this result. Another possible cause for this low $\beta$ value is the fact that we are tracing fluxes at wavelengths that could receive contribution from free-free emission. This would tend to increase the flux at 3-mm, resulting in larger $\beta$. Finally, it is worth mentioning that due the higher CMB temperature with redshift, we would expect to see an increase in the average $\beta$ value with increasing redshift. Larger samples of faint DSFGs are needed to provide better constraints on this subject. 

\section{ISM properties}

\label{sec_ism}

\subsection{Gas masses from dust, and caveats}

\begin{figure}[t]
\centering
\includegraphics[scale=0.4]{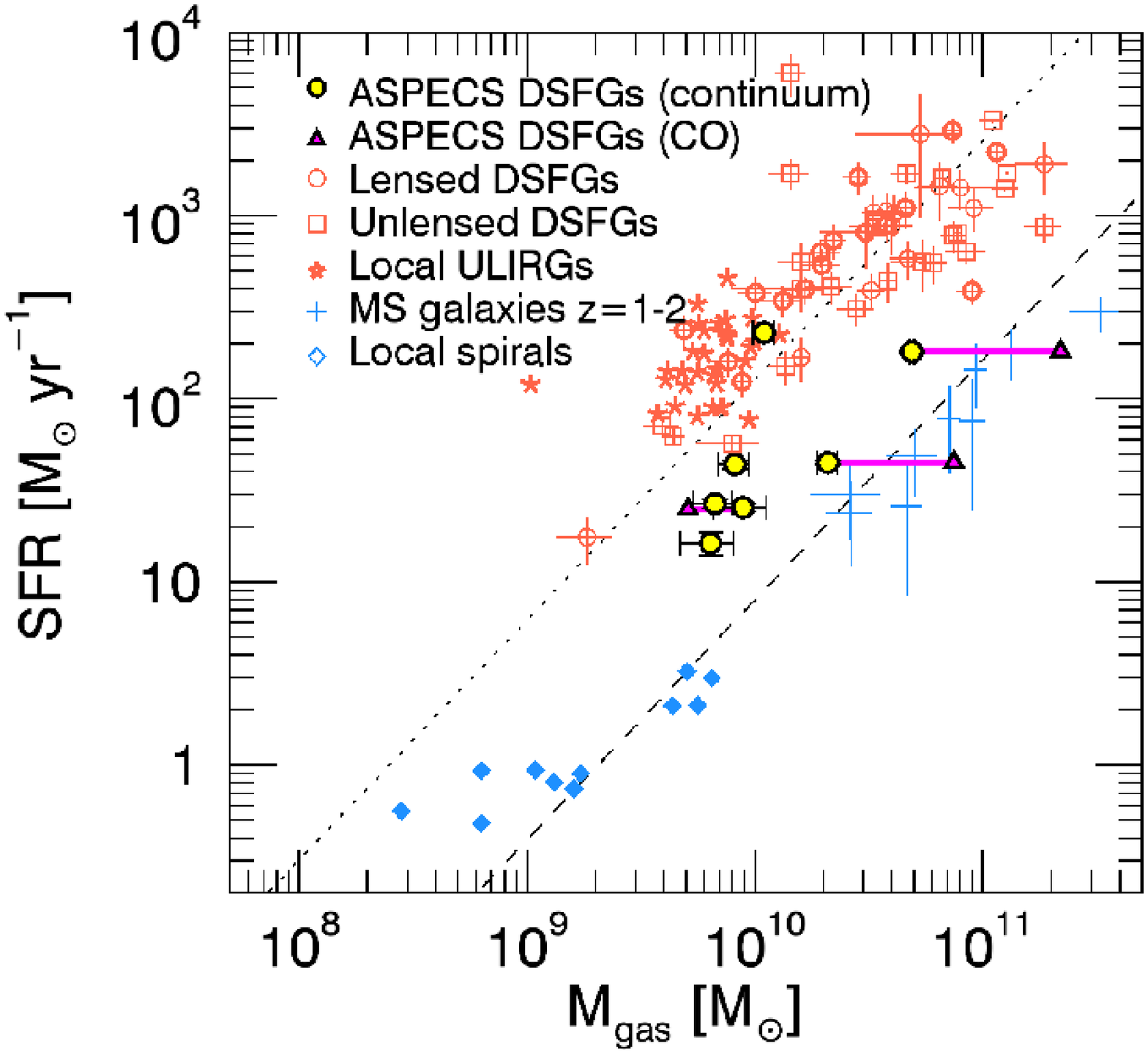}
\caption{ISM mass versus SFR for the ALMA UDF 1.2-mm continuum sources, compared to different galaxy populations that have been detected in CO(1--0) or CO(2--1) from the literature. The ISM mass for the ALMA sources have been computed using the 1.2-mm continuum flux densities following the recipies from \citet{scoville14}. Literature values typically assume a CO luminosity to gas mass conversion factor of 0.8 $M_\odot$ (K km s$^{-1}$ pc$^{2}$)$^{-1}$ for local starburst galaxies and SMGs, and 3.6 or 4.6 (same units) for local spirals and main sequence galaxies at high-redshift. For the CO-based gas mass estimates in the three galaxies detected in CO line emission \citep[see Paper~IV;][]{decarli16b}, we use a conversion factor of 3.6 (same units). For clarity, the magenta lines connect the 1.2-mm continuum and CO-based gas mass estimates. The dashed and dotted lines denote the two sequences of starbursts and main-sequence galaxies defined in \citet{daddi10b}, respectively.}\label{fig_sflaw}
\end{figure}

\begin{figure}[h]
\centering
\includegraphics[scale=0.55]{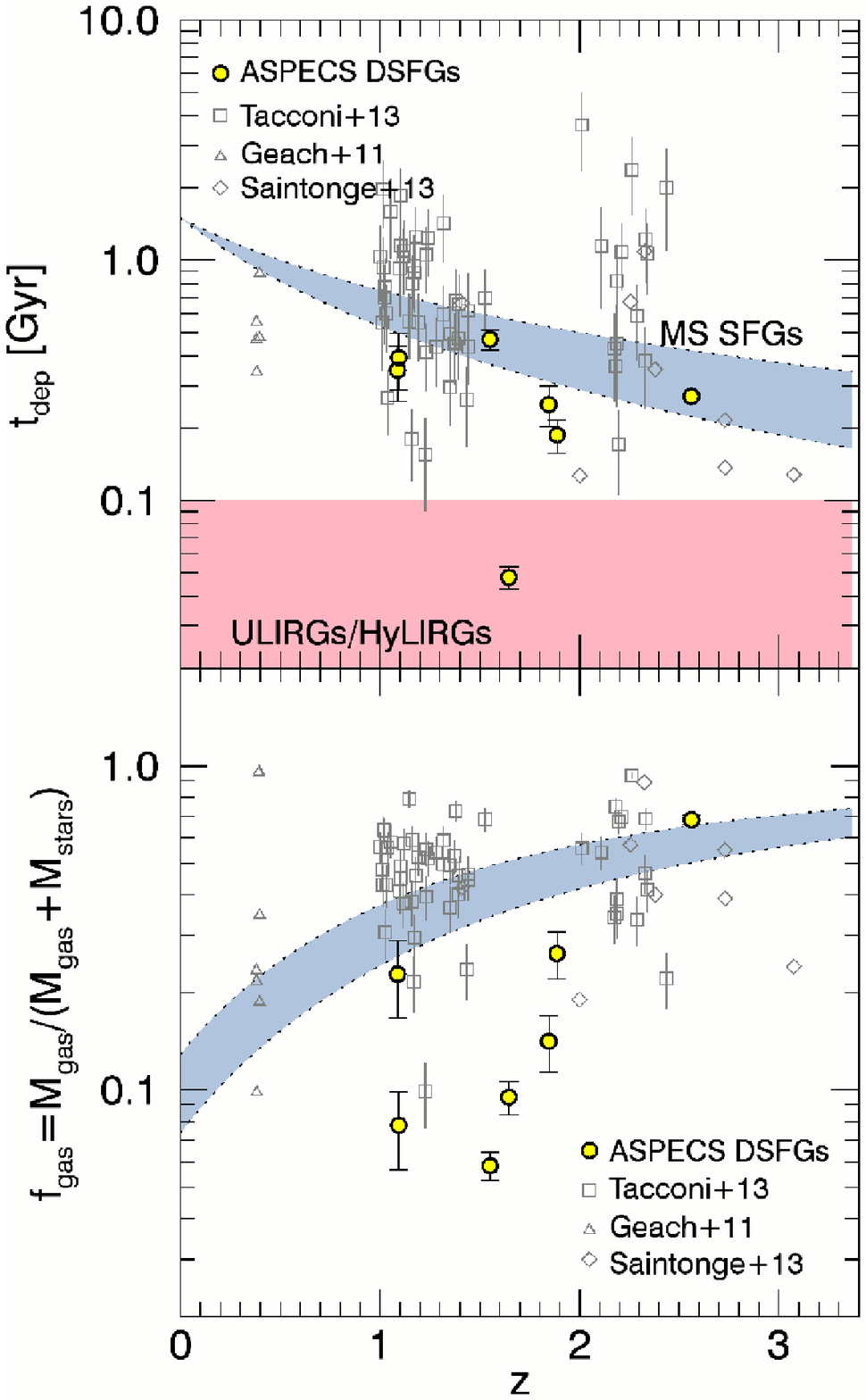}
\caption{Evolution of the gas depletion timescale ($t_{\rm dep}$) and the molecular gas fraction ($f_{\rm gas}$) as a function of redshift for  the ALMA UDF 1.2-mm continuum sources, compared to main sequence galaxies from the literature. Stellar masses and SFRs are computed from SED fitting. The ISM mass for the ALMA sources have been computed using the ALMA 1.2-mm continuum flux density following Scoville et al. (2014). In the top panel, the blue shaded region represents the expected evolution for the gas depletion timescale, $t_{\rm dep}=1.5\times(1+z)^{\gamma}$ with $\gamma=-1.0$ to -1.5, for massive main sequence galaxies \citep{dave12, tacconi13,saintonge13}. The pink region represents the typical gas depletion timescales measured in starburst galaxies \citep[e.g.][]{aravena16a}. In the bottom panel, the blue shaded region represents the evolution of the gas fraction expected for main sequence galaxies with $M_*>10^9\ M_\sun$ following the derivation of \citet{saintonge13}.}\label{fig_tdep}
\end{figure}

A useful method to compute ISM masses in galaxies has been the use of the dust mass as a proxy for the ISM content \citep{leroy11,magdis11,magnelli12,scoville14, genzel15}. Recently, \citet{scoville14} argued that under reasonable assumptions about the dust properties, reliable ISM mass measurements can be made based on flux measurements made in the RJ tail of the dust. The method was calibrated using massive galaxies at low and high redshift and assuming a fixed gas-to-dust ratio, which is expected to be fairly constant for a relatively ample range in properties \citep[see ][ for details]{scoville14}, and assumes a fixed dust temperature of $T_{\rm d}=25$ K. Note that there is a weak dependance of this method on $T_{\rm d}$, since we are probing the RJ part of the spectrum. Following \citet{scoville14}, we compute the ISM mass in units of $10^{10} M_\sun$ as:

\begin{equation}
M_{\rm ISM}=1.2 (1+z)^{-4.8} (\frac{\nu_{\rm obs}}{350})^{-3.8} \frac{\Gamma_0}{\Gamma_{\rm RJ}} S_\nu D_{\rm L}^2,
\end{equation}
where  $D_{\rm L}$ is the luminosity distance in Gpc at redshift $z$, and $S_\nu$ is the measured flux density in mJy at the observing frequency $\nu_{\rm obs}$ (in GHz). $\Gamma_{\rm RJ}$ is a correction factor that takes into account the deviation from the RJ limit as we approach higher redshifts. This factor depends on $z$, $T_{\rm d}$ and $\nu_{\rm obs}$, and becomes $\Gamma_0=0.76$ at $z=0$ for $\nu_{\rm obs}=$ 242 GHz and $T_{\rm d}=25$ K. This method to compute ISM masses assumes a dust emissivity index $\beta=1.8$, which we use throughout for consistency with other studies. 

\texttt{MAGPHYS} also delivers an estimate of the dust mass ($M_{\rm d}$) using the median of the dust mass posterior probability when fitting the available photometry. From this dust mass estimate, and under the assumption of a fixed gas-to-dust ratio $(\delta_{\rm GDR})$ and that the ISM is mostly molecular, one can obtain a measurement of the gas mass as $M_{\rm gas}=\delta_{\rm GDR} M_{\rm d}$. For local galaxies it has been found that typically, $\delta_{\rm GDR}\sim72$ \citep{sandstrom13}, however metallicity-dependent variations are likely to play a significant role \citep[e.g., ][]{remyruyer14}. For the typical stellar masses of our sources ($\sim10^{10-11} M_\odot$) and assuming that local calibrations apply, we would expect metallicities close to the solar value, 12+log(O/H) $\sim9$ \citep{tremonti04}. However, since the metallicities are lower at high redshift, the typical stellar masses of our sample imply metallicities of $\sim8.4$ at $z\sim1.5$ \citep{yabe14, zahid14}. This metallicity value would translate into  $\delta_{\rm GDR}\sim200$ \citep{remyruyer14}. Hence, we adopt this value to convert the dust masses obtained with \texttt{MAGPHYS} into gas mass estimates.  

\citet[][; Paper~IV]{decarli16b} provide a detailed discussion of the different available methods to compute the gas masses, based on the CO measurements for four sources in the ASPECS field. From Table \ref{tab_prop}, we find that the gas masses obtained using \texttt{MAGPHYS} SED fitting are consistent with the ISM estimates from the Scoville et al. method for the assumed $\delta_{\rm GDR}$. Decarli et al. (2016b; Paper~IV) finds that the gas estimates following Scoville et al. and the \texttt{MAGPHYS} SED fitting methods under-predict the gas masses by a factor of $\sim3-4$ compared to the CO based estimates. There are several reasons that could explain this discrepancy, including (i) a combination of high excitation and low $\alpha_{\rm CO}$ values in the CO measurements, (ii) systematics in the calibration of the dust-based measurements, and (iii) different spatial distributions of dust and molecular gas within individual galaxies (see Paper~IV for details). Another important issue is that the \citet{scoville14} calibration uses a fixed $\delta_{\rm GDR}$ value assuming solar metallicity. This assumption is reasonable for massive galaxies ($\sim10^{11}\ M_\sun$) as applied in their study, however, it may potentially underestimate the gas masses for less massive, lower metallicity galaxies, for which a higher $\delta_{\rm GDR}$ should be used.

Most importantly, perhaps, is the fact that the \citet{scoville14} calibration uses a gas to dust ratio fixed value for a solar metallicity. This assumption is reasonable for massive galaxies as applied in their study ($\sim10^{11}\ M_\sun$), however, it will likely result in lower gas masses for less massive, lower metallicity galaxies for which a higher $\delta_{\rm GDR}$ should be used.

Despite these uncertainties, the dust-based estimates constitute the only means to provide a measurement of the gas masses in our 1.2-mm continuum detected sources, given that most of them do not have CO line detections. Table \ref{tab_prop} lists the gas masses obtained using both the Scoville et al. and the \texttt{MAGPHYS} SED fitting method. In what follows we only use the ISM masses obtained with the Scoville et al. method as a measure of the total molecular gas mass, under the assumption that most of the ISM of high-redshift galaxies is in the form of molecular gas.

\subsection{Gas depletion timescales and fractions}

Figure \ref{fig_sflaw} shows the ISM mass (using Scoville et al. method) versus SFR (derived using SED fitting) for the galaxies detected at 1.2-mm continuum emission in our survey. For comparison, we also show the gas masses and SFRs of literature sources that have been detected in CO emission. To avoid uncertainties due to gas excitation, we only chose literature sources with low-$J$ CO measurements. We use a $^{12}$CO to gas mass conversion factor $\alpha_{\rm CO}=0.8$ K km s$^{-1}$ pc$^2$ for the samples of ultra-luminous IR galaxies (ULIRGs; Solomon et al. 1997) and both unlensed \citep{riechers11b, riechers11c,ivison11,ivison13,frayer08,thomson12,carilli11,hodge13,bothwell13,walter12,combes12,coppin10,debreuck14} and lensed DSFGs \citep{ivison10,lestrade11,swinbank10,harris10,decarli12,harris12,fu12,aravena16a}. For the samples of local spirals (Leroy et al. 2008) and main sequence galaxies \citep{daddi10a,magdis11,magnelli12}, we use $\alpha_{\rm CO}=4.6$ and 3.6 K km s$^{-1}$ pc$^2$, respectively. For reference, we also show the available CO-based gas mass estimates for the three 1.2-mm continuum sources in our sample that were detected in CO line emission \citep[C1, C2 and C6;][; Paper~IV]{decarli16b}. For these, a conversion factor of 3.6 K km s$^{-1}$ pc$^2$ has been used.

Our galaxies seem to span a significant range in ISM masses and SFRs. Two of our ALMA 1.2-mm sources appear to be aligned with the sequence formed by the local spirals and main-sequence galaxies at $z=1-2$ defined by the dashed line \citep{daddi10b}. This includes two of the CO detected galaxies, which are also detected in continuum. In particular, the 1.2-mm brightest galaxy in our sample falls into the group of main-sequence galaxies, supporting the identification of this galaxy as main sequence based on SFR--$M_*$. Only one galaxy, the third brightest in our continuum sample, is clearly located in the starburst regime. Four other sources appear to lie in between the trends of starburst or main-sequence galaxies. We remark that the gas mass values derived from the 1.2-mm fluxes could be underestimated as discussed in the previous section. This would thus imply that these four sources in our sample could belong to the trend of main sequence galaxies.

We note that the fact that the starburst and main-sequence galaxy trends in this SFR--$M_{\rm gas}$ plane appear to be well separated from each other, with virtually no source lying in between, partly relies on the use of fixed $\alpha_{\rm CO}$ factors for each particular sample.  While in several cases, the $\alpha_{\rm CO}$ conversion factor has been measured directly for the literature sources, we caution that the use of a binary set of values for this parameter may artificially lead to different star formation laws for starbursts and main-sequence galaxies \citep{ivison11}. The $\alpha_{\rm CO}$ factor depends on several parameters including metallicity, gas temperature and velocity dispersion and should depend on individual galaxy properties such as the gas or SFR surface density \citep[see ][]{casey14}. Furthermore, the bi-modality might be in part caused by the pre-selection of individual sources for CO follow-up which biases the range of properties covered by targeted current observations. However, it should be pointed out that this separation is already seen when comparing the direct observables $L'_{\rm CO}$ and $L_{\rm IR}$ \citep[e.g.,][]{daddi10a,genzel10,aravena16a}. 

Figure \ref{fig_tdep} shows the implied gas depletion timescales ($t_{\rm dep}$) and gas fractions ($f_{\rm gas}$) as a function of redshift for our ALMA 1.2-mm continuum sources, compared to recent measurements of main-sequence galaxies at $z=0.5-3.0$ \citep{geach11,tacconi13,saintonge13}. Observations of massive main-sequence galaxies ($M_*>10^{10}\ M_\sun$) have shown evidence for a significant dependency of $t_{\rm dep}$ out to $z=3$ \citep{tacconi13,saintonge13,genzel15}, consistent with models of galaxy formation. These studies show a dependency of $t_{\rm dep}$ with redshift with the form $(1+z)^\gamma$, with $\gamma$ varying between -1.5 to -1.0 \citep{tacconi13}, as shown in Fig. \ref{fig_tdep}. Recent studies, however, show that $\gamma$ can be as low as -0.3 \citep{genzel15}. Similarly, as shown in the bottom panel of Fig. \ref{fig_tdep}, $f_{\rm gas}$ shows a significant dependency with redshift, which appears to flatten at $z>3$ \citep{saintonge13}. 

The gas depletion timescales for our faint 1.2-mm sources is consistent with the ranges found for main sequence galaxies at similar redshifts. Only one galaxy has a $t_{\rm dep}$ value that puts it clearly in the range occupied by starburst galaxies. However, our galaxies present gas fractions ranging from $0.06-0.2$ for the $z\sim1.5$ sample, which significantly lower than other main sequence galaxies at similar redshifts. Only the higher redshift galaxy in our sample, ASPECS C1 at $z=2.5$, has a value of $f_{\rm gas}$ comparable to literature sources at its redshift. This implies that while most of our galaxies have measured gas depletion timescales that agree with previous studies for main sequence galaxies, they have gas fractions that are much lower than the those found for same comparison samples. 

Several factors could affect the measured $t_{\rm dep}$ and $f_{\rm gas}$. This can partly be attributed to uncertainties in the derived parameters through SED fitting. However, we are using very deep multi-wavelength photometry, and thus the derived SFRs and stellar masses should be as accurate as in previous studies. This is indicated by the fact that the ranges for the location of the main sequence at different redshifts in Fig. \ref{fig_msz} are consistent with those from the literature \citep{whitaker14}. Another possible explanation is that the gas masses computed using the 1.2-mm flux densities are being underestimated. A factor of $\sim2-3$ higher gas masses, as those derived from CO \citep[see ][; Paper~IV]{decarli16b}, would place the measured gas fractions more in line with the expected values for main sequence galaxies, while retaining high gas depletion timescales.

Additionally, our sample presents significant scatter in both plots. This scatter is unlikely caused by the possible underestimation of the gas masses where we would expect a more systematic effect. In this case, our sources present a scatter that is consistent with the typical one found in other samples studied in CO emission \citep{geach11, tacconi13, saintonge13}. Because of this scatter and the relatively narrow redshift range covered by our ALMA detections, it is hard to establish any evolutionary trend with the available data.

\section{Contribution to the EBL at 1.2-mm}

\label{sec_cib}

\subsection{Integrated intensity and fraction of the EBL}

We use the number counts at 1.2-mm derived in Section \ref{sec_counts} to calculate the contribution to the EBL at 1.2-mm. Although our source number counts are derived from a small area of the sky, they are based in a deep contiguous blank field. 

To calculate the contribution to the cosmic background at 1.2-mm from our measurements, we directly integrate the number counts, corrected for fidelity and completeness, down to the faintest flux bin ($S_{\rm 1.2mm}\sim37\mu$Jy).  We obtain an integrated intensity of $7.8\pm0.4$ Jy deg$^{-2}$. The uncertainty is derived from the sum of the uncertainties of the individual detections, corrected for fidelity and completeness. However, our number counts do not extend to fluxes above 0.6 mJy. To estimate the contribution of the bright-end of the number counts, which are not traced by our survey, we use the results from from \citet{karim13} and \citet{oteo15}. While the \citet{karim13} results are measured at 870$\mu$m, we chose them since they are based on ALMA high resolution observations and thus take better into account the multiplicity and false detection rate issues seen in single-dish telescope bolometer surveys. It is a well known result from their study that bolometer surveys overpredict the number counts at the bright end (above $S_{870\mu{\rm m}}>6$ mJy). We convert their counts from 870$\mu$m to 1.2-mm using $S_{\rm 1.2mm}=0.4\times S_{\rm 870}\mu{\rm m}$, and add their contribution by integrating the values in their Table 1. Similarly, we use the \citet{oteo15} results to account for the contribution to the integrated intensity between 1.2-mm fluxes of 0.6 to 1.9 mJy, which are not covered by either the Karim et al. or our measurements. To fill this gap, we extrapolate the Oteo et al. number counts (in log-log space). By adding up the contribution of all galaxies starting at our faintest flux bin, we find that an integrated intensity of $8.6\pm0.7$ Jy deg$^{-2}$.

To compute the CIB at the frequency of our observations, we make use of the latest values derived by \citet{planck14}. By interpolating the {\it Planck} measurements (see their Table 10) over the frequency range of our observations (212-272 GHz), we find an EBL at $\sim242$ GHz of $14.2\pm0.6$ Jy deg$^{-2}$. From this, we find that our number counts recover $\sim60\pm6\%$ of the EBL at 242 GHz. Note that the EBL value at 242 GHz measured by {\it Planck} is much more precise than that measured by COBE 20 years ago, and we thus adopt this value.

In order to account for the missing contribution to the EBL, we use stacking analysis. We follow the procedure explained in \S \ref{sec_stack}. We select the same samples (see Table \ref{tab_stack}), but in this case we limit them to exclude all sources with a detection at the $>3\sigma$ level in order to be consistent with the faintest flux level taken into account to derive the number counts. In all cases, the samples differ by at most two sources with respect to those listed in Table \ref{tab_stack}. Hence, we find similar results than those presented in \S \ref{sec_stack}. We thus use the fluxes and number of objects for the m2 and m3 samples to compute the integrated intensity from the faintest, undetected sources. We find an extra contribution of $2.8\pm0.5$ Jy deg$^{-2}$ or $\sim20\pm4\%$ of the EBL at 242 GHz. Combining this to our measurement from the number counts, implies a total intensity of $11.4\pm0.8$ Jy deg$^{-2}$, which makes up $80\pm7\%$ ($\sim77--84\%$) of the EBL at 242 GHz measured by {\it Planck}.

\subsection{Nature of the sources that make up the EBL} 
 
A critical result from this study corresponds to the properties of the galaxies that contribute to the EBL at 242 GHz. Based on our number count measurements only, we obtained an integrated intensity of $7.8\pm0.4$ Jy deg$^{-2}$. This makes up $55\pm4\%$ of the EBL measured by {\it Planck} at 242 GHz, implying that the population of galaxies that dominates this background is composed by the galaxies individually resolved by our ASPECS survey. From \S \ref{sec_prop}, we determined that these galaxies have typical stellar masses of $\sim4\times10^{10}\ M_\sun$, SFRs of $\sim40~M_\sun$ yr$^{-1}$ at $z\sim1.7$, which corresponds to the main sequence at this redshift. This is supported by the ISM masses of these galaxies, which places them in the star-forming sequence in the $M_{\rm ISM}$ vs SFR plane. By using stacking, we find that on average the galaxies that make up another 20\% of the EBL at 242 GHz, at the faintest end, is composed by slightly less massive galaxies ($\sim(0.5-1.5)\times10^{10}\ M_\sun$) and low SFRs ($10-20\ M_\sun$ yr$^{-1}$) at similar redshifts. These findings imply that the bulk of galaxies that make up the CIB consists of faint, main-sequence galaxies at $z\sim1.7$.

Our measurements indicate that $\sim77-84\%$ of the EBL at 242 GHz can be resolved by individually detected galaxies, by those identified by stacking (in the m2+m3 samples). If we use the upper limit in the mass bin m1, we find that these galaxies could contribute up $6\%$ of the EBL at 242 GHz ($3\sigma$). This implies that up to $84\%+6\% = 90\%$ of the EBL could be identified by our observations (plus literature for the bright end), and hence only about 10\% of the EBL measured by {\it Planck} at this frequency is left unresolved. Since we have included the most massive samples in our stacking, $M_{\star}>10^9 M_\sun$, the remainder of the EBL at these frequencies would likely come from less massive galaxies ($M_{\star}<10^9 M_\sun$).

\subsection{The effect of cosmic variance} 
 
A number of recent studies have used the archival ALMA 1.2-mm data to provide constraints on the EBL at 1.2-mm. These studies measure significantly higher integrated intensities at 1.2-mm compared to our estimates: \citet{fujimoto16} measure the number counts down to a flux limit of 15 $\mu$Jy, just below our ALMA UDF flux limit, with an integrated intensity of $\sim22$ Jy deg$^{-2}$; \citet{hatsukade13} integrated their number counts down to 0.15 mJy, obtaining an intensity of $\sim16.9$ Jy deg$^{-2}$ (converting their measurement from 1.3-mm to 1.2-mm); \citet{ono14} measures $\sim11$ Jy deg$^{-2}$ down to 0.1 mJy; similarly, \citet{carniani15} measures $\sim17$ Jy deg$^{-2}$ down to 0.1 mJy at 1.2-mm. To derive the fraction of the EBL at 1.2-mm resolved, most of these literature results use early measurements from the Far Infrared Absolute Spectrophotometer (FIRAS) on board of the COBE satellite \citep{fixsen98}, which measures an integrated intensity of $22_{-8}^{+14}$ Jy deg$^{-2}$ at this wavelength. However, the COBE spectrum of the IR background becomes highly uncertain at frequencies below 350 GHz (see Fig. 4 of Fixsen et al.), mostly due to Galactic contamination. The newer measurement from the {\it Planck} satellite has much better precision and is within the uncertainties of the COBE measurement. As such, the recent measurements from the literature imply very high resolved fractions of the EBL, in some cases even exceeding the {\it Planck} measurements at 242 GHz. We note that the EBL is a grand average of the extragalactic emission over the whole sky. Therefore measurements covering $\sim$1 arcmin$^{-2}$ or less of the sky, aiming to resolve the sources contributing to this background will be most likely highly affected by cosmic variance. If the observations were pointed to an overdense region of the sky, this will translate into a higher number of sources and higher resolved fraction of the EBL. In particular, Fig. \ref{fig_ncounts} shows that for the flux range $0.08-0.6$ mJy our cumulative number counts are significantly below, by a factor of $\sim2$, with respect to the values derived by \citet{hatsukade13} and \citet{fujimoto16}\footnote{Over this flux range, the Fujimoto et al. results fully rely on the observations analysed by Hatsukade et al. Thus, these studies measure effectively the same number of sources.}, yet more consistent with the counts derived by \citet{oteo15} and \citet{carniani15}. This substantial difference in the number counts, possibly due to the small areas covered but also to the fact that these studies are not ``blank-field'', would explain the differences in the measured intensities and resolved fraction of the EBL between different studies. As shown in \citet{scoville13}, small scale source density variations can cover significant fractions of the sky (see their Figs. 9-11). As explained in \S \ref{sec_counts}, the number count differences might also be due to different methods and analysis tools used. In any case, measurements on larger fields will help to elucidate the effect of small scale structure on the EBL at millimeter wavelengths.

\begin{figure}
\centering
\includegraphics[scale=0.4]{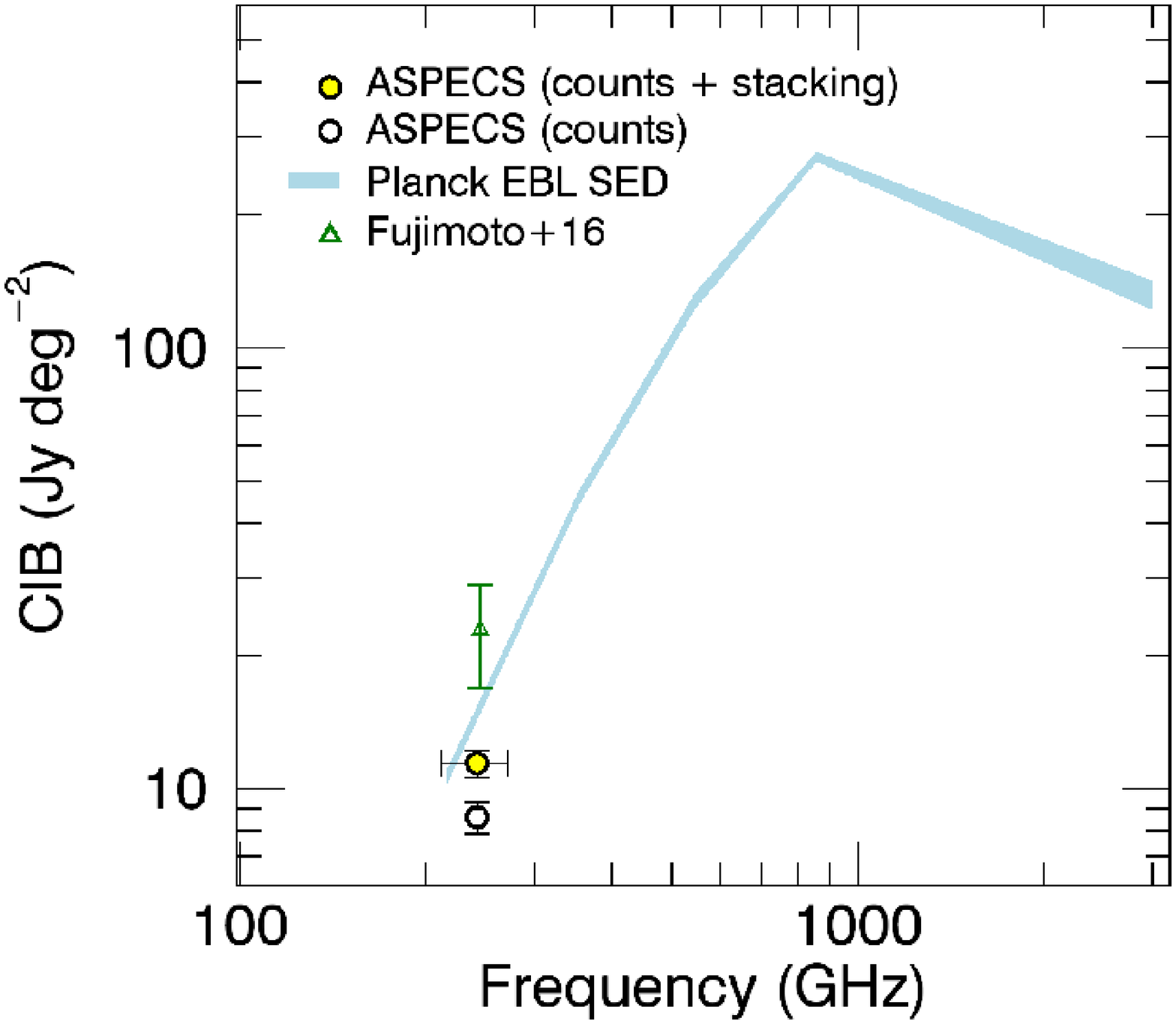}
\caption{Extragalactic infrared background spectral energy distribution compared to the amount of intensity resolved by our ALMA UDF observations. The shaded blue area represents the cosmic IR background revealed by the {\it Planck} satellite observations and the range on uncertainties in the measured data. Note that the uncertainty is so small that the shaded area resembles a thick line. The yellow circle shows the integrated intensity of our ASPECS observations at 242 GHz (11.4 Jy deg$^{-2}$) including both the measurement based on the number counts and the stacking analysis. The open circle shows the intensity recovered by the number count measurements only (without stacking). The green triangle shows the measurement made by \citet{fujimoto16} based on archival 1.2-mm data.}\label{fig_cib}
\end{figure}

\section{Conclusions/summary}\label{sec_concl}

Using ALMA in cycle-2, we have conducted a millimeter spectroscopic survey by scanning the full 3-mm and 1.2-mm bands over a region in the {\it Hubble} UDF. The collapsed cubes constitute the deepest continuum images ever obtained over an 1 arcmin$^2$ contiguous area of the sky. The main results of our continuum measurements can be summarised as follows:

\begin{itemize}

\item We detect nine sources with significances $>3.5\sigma$ at 1.2-mm and only one source at 3-mm. From these detections, we measure the 1.2-mm number counts over the flux density range $S_{\rm 1.2mm}=0.036-0.57$ mJy. Our number counts are similar to previous measurements, with differences within a factor of $\sim2$.

\item We measure the properties of the individually detected galaxies at S/N$>3.5$. We find that there is a large spread in stellar masses and SFRs, with median values of $4\times10^{10}\ M_\sun$ and $\sim40~M_\sun$ yr$^{-1}$, much lower than found in brighter SMGs. We find that these faint DSFGs are systematically located at lower redshifts than millimeter-selected SMGs, with a median redshift of $z=1.7$. All galaxies are consistent with being close to the main sequence at their respective redshift.

\item We use stacking analysis to estimate the average emission from samples of galaxies selected by redshift, stellar mass and SFRs. We only find detections in samples selected in the redshift range $1<z<2$, as well as in the stellar mass ranges log$(M_*/M_\sun)=9.5-10.0$ and log$(M_*/M_\sun)=10.0-10.5$, with typical SFRs of $3-10\ M_\sun$ yr$^{-1}$ . This suggests that the rest of the emission, not individually detected in our survey, comes from galaxies less massive, with lower SFRs, but at a similar redshift than the detected sources.

\item We use the 1.2-mm flux as a proxy for the ISM masses in our individually detected galaxies. We find that most of our sources are located in the star-forming trend occupied by main-sequence galaxies and local spirals, implying relatively large gas time depletion timescales, typically above 300 Myr, and a large spread in the molecular gas fractions ranging from 0.1 to 1.0. We compare these results to ISM mass estimates using CO as a tracer in \citet[][; Paper~IV]{decarli16b}.

\item Our individual detections alone are able to resolve $55\pm4\%$ of the EBL at 242 GHz measured by the {\it Planck} satellite. By adding up the integrated intensity from our number counts, to the contribution from the bright end of the number counts -- mostly composed by SMGs -- and the contribution of faint galaxies detected using stacking, we are able to resolve between 77--84\% of the CIB at 242 GHz. The typical properties of the population that makes up most of the EBL at these frequencies corresponds to that of the galaxies described in this work.

\end{itemize}

\acknowledgements
We thank the anonymous referee for her/his positive feedback and useful comments. M.A.~acknowledges partial support from FONDECYT through grant 1140099. FW, IRS, and RJI acknowledge support through ERC grants COSMIC--DAWN, DUSTYGAL, and COSMICISM, respectively. FEB and LI acknowledge Conicyt grants Basal-CATA PFB--06/2007 and Anilo ACT1417. FEB also acknowledge support from FONDECYT Regular 1141218 (FEB), and the Ministry of Economy, Development, and Tourism's Millennium Science Initiative through grant IC120009, awarded to The Millennium Institute of Astrophysics, MAS. EdC gratefully acknowledges the Australian Research Council as the recipient of a Future Fellowship (project FT150100079). DR acknowledges support from the National Science Foundation under grant number AST-1614213 to Cornell University. IRS also acknowledges support from STFC (ST/L00075X/1) and a Royal Society / Wolfson Merit award. Support for RD and BM was provided by the DFG priority program 1573 `The physics of the interstellar medium'.  AK and FB acknowledge support by the Collaborative Research Council 956, sub-project A1, funded by the Deutsche Forschungsgemeinschaft (DFG). PI acknowledges Conict grants Basal-CATA PFB--06/2007 and Anilo ACT1417. RJA was supported by FONDECYT grant number 1151408. This paper makes use of the following ALMA data: ADS/JAO.ALMA\#2013.1.00146.S and ADS/JAO.ALMA\#2013.1.00718.S. ALMA is a partnership of ESO (representing its member states), NSF (USA) and NINS (Japan), together with NRC (Canada), NSC and ASIAA (Taiwan), and KASI (Republic of Korea), in cooperation with the Republic of Chile. The Joint ALMA Observatory is operated by ESO, AUI/NRAO and NAOJ. The 3mm-part of ALMA project had been supported by the German ARC.

\bibliographystyle{apj}
\bibliography{alma_udf_continuum_MA}

\begin{thebibliography}{}
\expandafter\ifx\csname natexlab\endcsname\relax\def\natexlab#1{#1}\fi

\bibitem[{{Aravena} {et~al.}(2016{\natexlab{a}}){Aravena}, {Spilker},
  {Bethermin}, {Bothwell}, {Chapman}, {de Breuck}, {Furstenau},
  {G{\'o}nzalez-L{\'o}pez}, {Greve}, {Litke}, {Ma}, {Malkan}, {Marrone},
  {Murphy}, {Stark}, {Strandet}, {Vieira}, {Weiss}, {Welikala}, {Wong}, \&
  {Collier}}]{aravena16a}
{Aravena}, M., {Spilker}, J.~S., {Bethermin}, M., {et~al.} 2016{\natexlab{a}},
  \mnras, 457, 4406

\bibitem[{{Aravena} {et~al.}(2016{\natexlab{b}}){Aravena}, {Decarli}, {Walter},
  {Bouwens}, {Oesch}, {Carilli}, {Bauer}, {Da Cunha}, {Daddi},
  {G{\'o}nzalez-L{\'o}pez}, {Ivison}, {Riechers}, {Smail}, {Swinbank}, {Weiss},
  {Anguita}, {Bacon}, {Bell}, {Bertoldi}, {Cortes}, {Cox}, {Hodge}, {Ibar},
  {Inami}, {Infante}, {Karim}, {Magnelli}, {Ota}, {Popping}, {van der Werf},
  {Wagg}, \& {Fudamoto}}]{aravena16b}
{Aravena}, M., {Decarli}, R., {Walter}, F., {et~al.} 2016{\natexlab{b}}, ArXiv
  e-prints, arXiv:1607.06772

\bibitem[{{Aretxaga} {et~al.}(2011){Aretxaga}, {Wilson}, {Aguilar}, {Alberts},
  {Scott}, {Scoville}, {Yun}, {Austermann}, {Downes}, {Ezawa}, {Hatsukade},
  {Hughes}, {Kawabe}, {Kohno}, {Oshima}, {Perera}, {Tamura}, \&
  {Zeballos}}]{aretxaga11}
{Aretxaga}, I., {Wilson}, G.~W., {Aguilar}, E., {et~al.} 2011, \mnras, 415,
  3831

\bibitem[{{Austermann} {et~al.}(2010){Austermann}, {Dunlop}, {Perera}, {Scott},
  {Wilson}, {Aretxaga}, {Hughes}, {Almaini}, {Chapin}, {Chapman}, {Cirasuolo},
  {Clements}, {Coppin}, {Dunne}, {Dye}, {Eales}, {Egami}, {Farrah}, {Ferrusca},
  {Flynn}, {Haig}, {Halpern}, {Ibar}, {Ivison}, {van Kampen}, {Kang}, {Kim},
  {Lacey}, {Lowenthal}, {Mauskopf}, {McLure}, {Mortier}, {Negrello}, {Oliver},
  {Peacock}, {Pope}, {Rawlings}, {Rieke}, {Roseboom}, {Rowan-Robinson},
  {Scott}, {Serjeant}, {Smail}, {Swinbank}, {Stevens}, {Velazquez}, {Wagg}, \&
  {Yun}}]{austermann10}
{Austermann}, J.~E., {Dunlop}, J.~S., {Perera}, T.~A., {et~al.} 2010, \mnras,
  401, 160

\bibitem[{{Barger} {et~al.}(1999){Barger}, {Cowie}, \& {Sanders}}]{barger99}
{Barger}, A.~J., {Cowie}, L.~L., \& {Sanders}, D.~B. 1999, \apjl, 518, L5

\bibitem[{{Barger} {et~al.}(1998){Barger}, {Cowie}, {Sanders}, {Fulton},
  {Taniguchi}, {Sato}, {Kawara}, \& {Okuda}}]{barger98}
{Barger}, A.~J., {Cowie}, L.~L., {Sanders}, D.~B., {et~al.} 1998, \nat, 394,
  248

\bibitem[{{Bertin} \& {Arnouts}(1996)}]{bertin96}
{Bertin}, E., \& {Arnouts}, S. 1996, \aaps, 117, 393

\bibitem[{{Bertoldi} {et~al.}(2000){Bertoldi}, {Carilli}, {Menten}, {Owen},
  {Dey}, {Gueth}, {Graham}, {Kreysa}, {Ledlow}, {Liu}, {Motte}, {Reichertz},
  {Schilke}, \& {Zylka}}]{bertoldi00}
{Bertoldi}, F., {Carilli}, C.~L., {Menten}, K.~M., {et~al.} 2000, \aap, 360, 92

\bibitem[{{Bertoldi} {et~al.}(2007){Bertoldi}, {Carilli}, {Aravena},
  {Schinnerer}, {Voss}, {Smolcic}, {Jahnke}, {Scoville}, {Blain}, {Menten},
  {Lutz}, {Brusa}, {Taniguchi}, {Capak}, {Mobasher}, {Lilly}, {Thompson},
  {Aussel}, {Kreysa}, {Hasinger}, {Aguirre}, {Schlaerth}, \&
  {Koekemoer}}]{bertoldi07}
{Bertoldi}, F., {Carilli}, C., {Aravena}, M., {et~al.} 2007, \apjs, 172, 132

\bibitem[{{B{\'e}thermin} {et~al.}(2015){B{\'e}thermin}, {De Breuck},
  {Sargent}, \& {Daddi}}]{bethermin15b}
{B{\'e}thermin}, M., {De Breuck}, C., {Sargent}, M., \& {Daddi}, E. 2015, \aap,
  576, L9

\bibitem[{{Blanc} {et~al.}(2008){Blanc}, {Lira}, {Barrientos}, {Aguirre},
  {Francke}, {Taylor}, {Quadri}, {Marchesini}, {Infante}, {Gawiser}, {Hall},
  {Willis}, {Herrera}, {Maza}, \& {MUSYC Collaboration}}]{blanc08}
{Blanc}, G.~A., {Lira}, P., {Barrientos}, L.~F., {et~al.} 2008, \apj, 681, 1099

\bibitem[{{Bothwell} {et~al.}(2013){Bothwell}, {Smail}, {Chapman}, {Genzel},
  {Ivison}, {Tacconi}, {Alaghband-Zadeh}, {Bertoldi}, {Blain}, {Casey}, {Cox},
  {Greve}, {Lutz}, {Neri}, {Omont}, \& {Swinbank}}]{bothwell13}
{Bothwell}, M.~S., {Smail}, I., {Chapman}, S.~C., {et~al.} 2013, \mnras, 429,
  3047

\bibitem[{{Bouwens} {et~al.}(2014){Bouwens}, {Bradley}, {Zitrin}, {Coe},
  {Franx}, {Zheng}, {Smit}, {Host}, {Postman}, {Moustakas}, {Labb{\'e}},
  {Carrasco}, {Molino}, {Donahue}, {Kelson}, {Meneghetti}, {Ben{\'{\i}}tez},
  {Lemze}, {Umetsu}, {Broadhurst}, {Moustakas}, {Rosati}, {Jouvel},
  {Bartelmann}, {Ford}, {Graves}, {Grillo}, {Infante}, {Jimenez-Teja}, {Lahav},
  {Maoz}, {Medezinski}, {Melchior}, {Merten}, {Nonino}, {Ogaz}, \&
  {Seitz}}]{bouwens14}
{Bouwens}, R.~J., {Bradley}, L., {Zitrin}, A., {et~al.} 2014, \apj, 795, 126

\bibitem[{{Brinchmann} {et~al.}(2004){Brinchmann}, {Charlot}, {White},
  {Tremonti}, {Kauffmann}, {Heckman}, \& {Brinkmann}}]{brinchmann04}
{Brinchmann}, J., {Charlot}, S., {White}, S.~D.~M., {et~al.} 2004, \mnras, 351,
  1151

\bibitem[{{Carilli} {et~al.}(2011){Carilli}, {Hodge}, {Walter}, {Riechers},
  {Daddi}, {Dannerbauer}, \& {Morrison}}]{carilli11}
{Carilli}, C.~L., {Hodge}, J., {Walter}, F., {et~al.} 2011, \apjl, 739, L33

\bibitem[{{Carniani} {et~al.}(2015){Carniani}, {Maiolino}, {De Zotti},
  {Negrello}, {Marconi}, {Bothwell}, {Capak}, {Carilli}, {Castellano},
  {Cristiani}, {Ferrara}, {Fontana}, {Gallerani}, {Jones}, {Ohta}, {Ota},
  {Pentericci}, {Santini}, {Sheth}, {Vallini}, {Vanzella}, {Wagg}, \&
  {Williams}}]{carniani15}
{Carniani}, S., {Maiolino}, R., {De Zotti}, G., {et~al.} 2015, \aap, 584, A78

\bibitem[{{Casey} {et~al.}(2014){Casey}, {Narayanan}, \& {Cooray}}]{casey14}
{Casey}, C.~M., {Narayanan}, D., \& {Cooray}, A. 2014, \physrep, 541, 45

\bibitem[{{Chapin} {et~al.}(2009){Chapin}, {Pope}, {Scott}, {Aretxaga},
  {Austermann}, {Chary}, {Coppin}, {Halpern}, {Hughes}, {Lowenthal},
  {Morrison}, {Perera}, {Scott}, {Wilson}, \& {Yun}}]{chapin09}
{Chapin}, E.~L., {Pope}, A., {Scott}, D., {et~al.} 2009, \mnras, 398, 1793

\bibitem[{{Chapman} {et~al.}(2005){Chapman}, {Blain}, {Smail}, \&
  {Ivison}}]{chapman05}
{Chapman}, S.~C., {Blain}, A.~W., {Smail}, I., \& {Ivison}, R.~J. 2005, \apj,
  622, 772

\bibitem[{{Chen} {et~al.}(2013){Chen}, {Cowie}, {Barger}, {Casey}, {Lee},
  {Sanders}, {Wang}, \& {Williams}}]{chen13}
{Chen}, C.-C., {Cowie}, L.~L., {Barger}, A.~J., {et~al.} 2013, \apj, 776, 131

\bibitem[{{Coe} {et~al.}(2006){Coe}, {Ben{\'{\i}}tez}, {S{\'a}nchez}, {Jee},
  {Bouwens}, \& {Ford}}]{coe06}
{Coe}, D., {Ben{\'{\i}}tez}, N., {S{\'a}nchez}, S.~F., {et~al.} 2006, \aj, 132,
  926

\bibitem[{{Combes} {et~al.}(2012){Combes}, {Rex}, {Rawle}, {Egami}, {Boone},
  {Smail}, {Richard}, {Ivison}, {Gurwell}, {Casey}, {Omont}, {Berciano Alba},
  {Dessauges-Zavadsky}, {Edge}, {Fazio}, {Kneib}, {Okabe}, {Pell{\'o}},
  {P{\'e}rez-Gonz{\'a}lez}, {Schaerer}, {Smith}, {Swinbank}, \& {van der
  Werf}}]{combes12}
{Combes}, F., {Rex}, M., {Rawle}, T.~D., {et~al.} 2012, \aap, 538, L4

\bibitem[{{Condon}(2007)}]{condon07}
{Condon}, J.~J. 2007, in Astronomical Society of the Pacific Conference Series,
  Vol. 380, Deepest Astronomical Surveys, ed. J.~{Afonso}, H.~C. {Ferguson},
  B.~{Mobasher}, \& R.~{Norris}, 189

\bibitem[{{Coppin} {et~al.}(2006){Coppin}, {Chapin}, {Mortier}, {Scott},
  {Borys}, {Dunlop}, {Halpern}, {Hughes}, {Pope}, {Scott}, {Serjeant}, {Wagg},
  {Alexander}, {Almaini}, {Aretxaga}, {Babbedge}, {Best}, {Blain}, {Chapman},
  {Clements}, {Crawford}, {Dunne}, {Eales}, {Edge}, {Farrah}, {Gazta{\~n}aga},
  {Gear}, {Granato}, {Greve}, {Fox}, {Ivison}, {Jarvis}, {Jenness}, {Lacey},
  {Lepage}, {Mann}, {Marsden}, {Martinez-Sansigre}, {Oliver}, {Page},
  {Peacock}, {Pearson}, {Percival}, {Priddey}, {Rawlings}, {Rowan-Robinson},
  {Savage}, {Seigar}, {Sekiguchi}, {Silva}, {Simpson}, {Smail}, {Stevens},
  {Takagi}, {Vaccari}, {van Kampen}, \& {Willott}}]{coppin06}
{Coppin}, K., {Chapin}, E.~L., {Mortier}, A.~M.~J., {et~al.} 2006, \mnras, 372,
  1621

\bibitem[{{Coppin} {et~al.}(2010){Coppin}, {Chapman}, {Smail}, {Swinbank},
  {Walter}, {Wardlow}, {Weiss}, {Alexander}, {Brandt}, {Dannerbauer}, {De
  Breuck}, {Dickinson}, {Dunlop}, {Edge}, {Emonts}, {Greve}, {Huynh}, {Ivison},
  {Knudsen}, {Menten}, {Schinnerer}, \& {van der Werf}}]{coppin10}
{Coppin}, K.~E.~K., {Chapman}, S.~C., {Smail}, I., {et~al.} 2010, \mnras, 407,
  L103

\bibitem[{{Cowie} {et~al.}(2002){Cowie}, {Barger}, \& {Kneib}}]{cowie02}
{Cowie}, L.~L., {Barger}, A.~J., \& {Kneib}, J. 2002, \aj, 123, 2197

\bibitem[{{da~Cunha} {et~al.}(2008){da~Cunha}, {Charlot}, \&
  {Elbaz}}]{dacunha08}
{da~Cunha}, E., {Charlot}, S., \& {Elbaz}, D. 2008, \mnras, 388, 1595

\bibitem[{{da Cunha} {et~al.}(2015){da Cunha}, {Walter}, {Smail}, {Swinbank},
  {Simpson}, {Decarli}, {Hodge}, {Weiss}, {van der Werf}, {Bertoldi},
  {Chapman}, {Cox}, {Danielson}, {Dannerbauer}, {Greve}, {Ivison}, {Karim}, \&
  {Thomson}}]{dacunha15}
{da Cunha}, E., {Walter}, F., {Smail}, I.~R., {et~al.} 2015, \apj, 806, 110

\bibitem[{{Daddi} {et~al.}(2007){Daddi}, {Dickinson}, {Morrison}, {Chary},
  {Cimatti}, {Elbaz}, {Frayer}, {Renzini}, {Pope}, {Alexander}, {Bauer},
  {Giavalisco}, {Huynh}, {Kurk}, \& {Mignoli}}]{daddi07}
{Daddi}, E., {Dickinson}, M., {Morrison}, G., {et~al.} 2007, \apj, 670, 156

\bibitem[{{Daddi} {et~al.}(2010{\natexlab{a}}){Daddi}, {Elbaz}, {Walter},
  {Bournaud}, {Salmi}, {Carilli}, {Dannerbauer}, {Dickinson}, {Monaco}, \&
  {Riechers}}]{daddi10b}
{Daddi}, E., {Elbaz}, D., {Walter}, F., {et~al.} 2010{\natexlab{a}}, \apjl,
  714, L118

\bibitem[{{Daddi} {et~al.}(2010{\natexlab{b}}){Daddi}, {Bournaud}, {Walter},
  {Dannerbauer}, {Carilli}, {Dickinson}, {Elbaz}, {Morrison}, {Riechers},
  {Onodera}, {Salmi}, {Krips}, \& {Stern}}]{daddi10a}
{Daddi}, E., {Bournaud}, F., {Walter}, F., {et~al.} 2010{\natexlab{b}}, \apj,
  713, 686

\bibitem[{{Dav{\'e}} {et~al.}(2012){Dav{\'e}}, {Finlator}, \&
  {Oppenheimer}}]{dave12}
{Dav{\'e}}, R., {Finlator}, K., \& {Oppenheimer}, B.~D. 2012, \mnras, 421, 98

\bibitem[{{De Breuck} {et~al.}(2014){De Breuck}, {Williams}, {Swinbank},
  {Caselli}, {Coppin}, {Davis}, {Maiolino}, {Nagao}, {Smail}, {Walter},
  {Wei{\ss}}, \& {Zwaan}}]{debreuck14}
{De Breuck}, C., {Williams}, R.~J., {Swinbank}, M., {et~al.} 2014, \aap, 565,
  A59

\bibitem[{{Decarli} {et~al.}(2012){Decarli}, {Walter}, {Neri}, {Bertoldi},
  {Carilli}, {Cox}, {Kneib}, {Lestrade}, {Maiolino}, {Omont}, {Richard},
  {Riechers}, {Thanjavur}, \& {Weiss}}]{decarli12}
{Decarli}, R., {Walter}, F., {Neri}, R., {et~al.} 2012, \apj, 752, 2

\bibitem[{{Decarli} {et~al.}(2014){Decarli}, {Walter}, {Carilli}, {Bertoldi},
  {Cox}, {Ferkinhoff}, {Groves}, {Maiolino}, {Neri}, {Riechers}, \&
  {Weiss}}]{decarli14}
{Decarli}, R., {Walter}, F., {Carilli}, C., {et~al.} 2014, \apjl, 782, L17

\bibitem[{{Decarli} {et~al.}(2016{\natexlab{a}}){Decarli}, {Walter}, {Aravena},
  {Carilli}, {Bouwens}, {da Cunha}, {Daddi}, {Ivison}, {Popping}, {Riechers},
  {Smail}, {Swinbank}, {Weiss}, {Anguita}, {Assef}, {Bauer}, {Bell},
  {Bertoldi}, {Chapman}, {Colina}, {Cortes}, {Cox}, {Dickinson}, {Elbaz},
  {G{\'o}nzalez-L{\'o}pez}, {Ibar}, {Infante}, {Hodge}, {Karim}, {Le Fevre},
  {Magnelli}, {Neri}, {Oesch}, {Ota}, {Rix}, {Sargent}, {Sheth}, {van der Wel},
  {van der Werf}, \& {Wagg}}]{decarli16a}
{Decarli}, R., {Walter}, F., {Aravena}, M., {et~al.} 2016{\natexlab{a}}, ArXiv
  e-prints, arXiv:1607.06770

\bibitem[{{Decarli} {et~al.}(2016{\natexlab{b}}){Decarli}, {Walter}, {Aravena},
  {Carilli}, {Bouwens}, {da Cunha}, {Daddi}, {Elbaz}, {Riechers}, {Smail},
  {Swinbank}, {Weiss}, {Bacon}, {Bauer}, {Bell}, {Bertoldi}, {Chapman},
  {Colina}, {Cortes}, {Cox}, {G{\'o}nzalez-L{\'o}pez}, {Inami}, {Ivison},
  {Hodge}, {Karim}, {Magnelli}, {Ota}, {Popping}, {Rix}, {Sargent}, {van der
  Wel}, \& {van der Werf}}]{decarli16b}
---. 2016{\natexlab{b}}, ArXiv e-prints, arXiv:1607.06771

\bibitem[{{Draine}(2011)}]{draine11}
{Draine}, B.~T. 2011, \apj, 732, 100

\bibitem[{{Dunlop} {et~al.}(2016){Dunlop}, {McLure}, {Biggs}, {Geach},
  {Michalowski}, {Ivison}, {Rujopakarn}, {van Kampen}, {Kirkpatrick}, {Pope},
  {Scott}, {Swinbank}, {Targett}, {Aretxaga}, {Austermann}, {Best}, {Bruce},
  {Chapin}, {Charlot}, {Cirasuolo}, {Coppin}, {Ellis}, {Finkelstein},
  {Hayward}, {Hughes}, {Ibar}, {Khochfar}, {Koprowski}, {Narayanan},
  {Papovich}, {Peacock}, {Robertson}, {Vernstrom}, {van der Werf}, {Wilson}, \&
  {Yun}}]{dunlop16}
{Dunlop}, J.~S., {McLure}, R.~J., {Biggs}, A.~D., {et~al.} 2016, ArXiv
  e-prints, arXiv:1606.00227

\bibitem[{{Dunne} {et~al.}(2011){Dunne}, {Gomez}, {da Cunha}, {Charlot}, {Dye},
  {Eales}, {Maddox}, {Rowlands}, {Smith}, {Auld}, {Baes}, {Bonfield}, {Bourne},
  {Buttiglione}, {Cava}, {Clements}, {Coppin}, {Cooray}, {Dariush}, {de Zotti},
  {Driver}, {Fritz}, {Geach}, {Hopwood}, {Ibar}, {Ivison}, {Jarvis}, {Kelvin},
  {Pascale}, {Pohlen}, {Popescu}, {Rigby}, {Robotham}, {Rodighiero}, {Sansom},
  {Serjeant}, {Temi}, {Thompson}, {Tuffs}, {van der Werf}, \&
  {Vlahakis}}]{dunne11}
{Dunne}, L., {Gomez}, H.~L., {da Cunha}, E., {et~al.} 2011, \mnras, 417, 1510

\bibitem[{{Eales} {et~al.}(1999){Eales}, {Lilly}, {Gear}, {Dunne}, {Bond},
  {Hammer}, {Le F{\`e}vre}, \& {Crampton}}]{eales99}
{Eales}, S., {Lilly}, S., {Gear}, W., {et~al.} 1999, \apj, 515, 518

\bibitem[{{Eales} {et~al.}(2000){Eales}, {Lilly}, {Webb}, {Dunne}, {Gear},
  {Clements}, \& {Yun}}]{eales00}
{Eales}, S., {Lilly}, S., {Webb}, T., {et~al.} 2000, \aj, 120, 2244

\bibitem[{{Elbaz} {et~al.}(2007){Elbaz}, {Daddi}, {Le Borgne}, {Dickinson},
  {Alexander}, {Chary}, {Starck}, {Brandt}, {Kitzbichler}, {MacDonald},
  {Nonino}, {Popesso}, {Stern}, \& {Vanzella}}]{elbaz07}
{Elbaz}, D., {Daddi}, E., {Le Borgne}, D., {et~al.} 2007, \aap, 468, 33

\bibitem[{{Elbaz} {et~al.}(2011){Elbaz}, {Dickinson}, {Hwang},
  {D{\'{\i}}az-Santos}, {Magdis}, {Magnelli}, {Le Borgne}, {Galliano},
  {Pannella}, {Chanial}, {Armus}, {Charmandaris}, {Daddi}, {Aussel}, {Popesso},
  {Kartaltepe}, {Altieri}, {Valtchanov}, {Coia}, {Dannerbauer}, {Dasyra},
  {Leiton}, {Mazzarella}, {Alexander}, {Buat}, {Burgarella}, {Chary}, {Gilli},
  {Ivison}, {Juneau}, {Le Floc'h}, {Lutz}, {Morrison}, {Mullaney}, {Murphy},
  {Pope}, {Scott}, {Brodwin}, {Calzetti}, {Cesarsky}, {Charlot}, {Dole},
  {Eisenhardt}, {Ferguson}, {F{\"o}rster Schreiber}, {Frayer}, {Giavalisco},
  {Huynh}, {Koekemoer}, {Papovich}, {Reddy}, {Surace}, {Teplitz}, {Yun}, \&
  {Wilson}}]{elbaz11}
{Elbaz}, D., {Dickinson}, M., {Hwang}, H.~S., {et~al.} 2011, \aap, 533, A119

\bibitem[{{Engel} {et~al.}(2010){Engel}, {Tacconi}, {Davies}, {Neri}, {Smail},
  {Chapman}, {Genzel}, {Cox}, {Greve}, {Ivison}, {Blain}, {Bertoldi}, \&
  {Omont}}]{engel10}
{Engel}, H., {Tacconi}, L.~J., {Davies}, R.~I., {et~al.} 2010, \apj, 724, 233

\bibitem[{{Fixsen} {et~al.}(1998){Fixsen}, {Dwek}, {Mather}, {Bennett}, \&
  {Shafer}}]{fixsen98}
{Fixsen}, D.~J., {Dwek}, E., {Mather}, J.~C., {Bennett}, C.~L., \& {Shafer},
  R.~A. 1998, \apj, 508, 123

\bibitem[{{Frayer} {et~al.}(2008){Frayer}, {Koda}, {Pope}, {Huynh}, {Chary},
  {Scott}, {Dickinson}, {Bock}, {Carpenter}, {Hawkins}, {Hodges}, {Lamb},
  {Plambeck}, {Pound}, {Scott}, {Scoville}, \& {Woody}}]{frayer08}
{Frayer}, D.~T., {Koda}, J., {Pope}, A., {et~al.} 2008, \apjl, 680, L21

\bibitem[{{Fu} {et~al.}(2012){Fu}, {Jullo}, {Cooray}, {Bussmann}, {Ivison},
  {P{\'e}rez-Fournon}, {Djorgovski}, {Scoville}, {Yan}, {Riechers}, {Aguirre},
  {Auld}, {Baes}, {Baker}, {Bradford}, {Cava}, {Clements}, {Dannerbauer},
  {Dariush}, {De Zotti}, {Dole}, {Dunne}, {Dye}, {Eales}, {Frayer}, {Gavazzi},
  {Gurwell}, {Harris}, {Herranz}, {Hopwood}, {Hoyos}, {Ibar}, {Jarvis}, {Kim},
  {Leeuw}, {Lupu}, {Maddox}, {Mart{\'{\i}}nez-Navajas}, {Micha{\l}owski},
  {Negrello}, {Omont}, {Rosenman}, {Scott}, {Serjeant}, {Smail}, {Swinbank},
  {Valiante}, {Verma}, {Vieira}, {Wardlow}, \& {van der Werf}}]{fu12}
{Fu}, H., {Jullo}, E., {Cooray}, A., {et~al.} 2012, \apj, 753, 134

\bibitem[{{Fujimoto} {et~al.}(2016){Fujimoto}, {Ouchi}, {Ono}, {Shibuya},
  {Ishigaki}, {Nagai}, \& {Momose}}]{fujimoto16}
{Fujimoto}, S., {Ouchi}, M., {Ono}, Y., {et~al.} 2016, \apjs, 222, 1

\bibitem[{{Geach} {et~al.}(2011){Geach}, {Smail}, {Moran}, {MacArthur},
  {Lagos}, \& {Edge}}]{geach11}
{Geach}, J.~E., {Smail}, I., {Moran}, S.~M., {et~al.} 2011, \apjl, 730, L19

\bibitem[{{Genzel} {et~al.}(2010){Genzel}, {Tacconi}, {Gracia-Carpio},
  {Sternberg}, {Cooper}, {Shapiro}, {Bolatto}, {Bouch{\'e}}, {Bournaud},
  {Burkert}, {Combes}, {Comerford}, {Cox}, {Davis}, {Schreiber},
  {Garcia-Burillo}, {Lutz}, {Naab}, {Neri}, {Omont}, {Shapley}, \&
  {Weiner}}]{genzel10}
{Genzel}, R., {Tacconi}, L.~J., {Gracia-Carpio}, J., {et~al.} 2010, \mnras,
  407, 2091

\bibitem[{{Genzel} {et~al.}(2015){Genzel}, {Tacconi}, {Lutz}, {Saintonge},
  {Berta}, {Magnelli}, {Combes}, {Garc{\'{\i}}a-Burillo}, {Neri}, {Bolatto},
  {Contini}, {Lilly}, {Boissier}, {Boone}, {Bouch{\'e}}, {Bournaud}, {Burkert},
  {Carollo}, {Colina}, {Cooper}, {Cox}, {Feruglio}, {F{\"o}rster Schreiber},
  {Freundlich}, {Gracia-Carpio}, {Juneau}, {Kovac}, {Lippa}, {Naab}, {Salome},
  {Renzini}, {Sternberg}, {Walter}, {Weiner}, {Weiss}, \& {Wuyts}}]{genzel15}
{Genzel}, R., {Tacconi}, L.~J., {Lutz}, D., {et~al.} 2015, \apj, 800, 20

\bibitem[{{Greve} {et~al.}(2008){Greve}, {Pope}, {Scott}, {Ivison}, {Borys},
  {Conselice}, \& {Bertoldi}}]{greve08}
{Greve}, T.~R., {Pope}, A., {Scott}, D., {et~al.} 2008, \mnras, 389, 1489

\bibitem[{{Greve} {et~al.}(2010){Greve}, {Wei{$\beta$}}, {Walter}, {Smail},
  {Zheng}, {Knudsen}, {Coppin}, {Kov{\'a}cs}, {Bell}, {de Breuck},
  {Dannerbauer}, {Dickinson}, {Gawiser}, {Lutz}, {Rix}, {Schinnerer},
  {Alexander}, {Bertoldi}, {Brandt}, {Chapman}, {Ivison}, {Koekemoer},
  {Kreysa}, {Kurczynski}, {Menten}, {Siringo}, {Swinbank}, \& {van der
  Werf}}]{greve10}
{Greve}, T.~R., {Wei{$\beta$}}, A., {Walter}, F., {et~al.} 2010, \apj, 719, 483

\bibitem[{{Hainline} {et~al.}(2011){Hainline}, {Blain}, {Smail}, {Alexander},
  {Armus}, {Chapman}, \& {Ivison}}]{hainline11}
{Hainline}, L.~J., {Blain}, A.~W., {Smail}, I., {et~al.} 2011, \apj, 740, 96

\bibitem[{{Harris} {et~al.}(2010){Harris}, {Baker}, {Zonak}, {Sharon},
  {Genzel}, {Rauch}, {Watts}, \& {Creager}}]{harris10}
{Harris}, A.~I., {Baker}, A.~J., {Zonak}, S.~G., {et~al.} 2010, \apj, 723, 1139

\bibitem[{{Harris} {et~al.}(2012){Harris}, {Baker}, {Frayer}, {Smail},
  {Swinbank}, {Riechers}, {van der Werf}, {Auld}, {Baes}, {Bussmann},
  {Buttiglione}, {Cava}, {Clements}, {Cooray}, {Dannerbauer}, {Dariush},
  {DeZotti}, {Dunne}, {Dye}, {Eales}, {Fritz}, {Gonzalez-Nuevo}, {Hopwood},
  {Ibar}, {Ivison}, {Jarvis}, {Maddox}, {Negrello}, {Rigby}, {Smith}, {Temi},
  \& {Wardlow}}]{harris12}
{Harris}, A.~I., {Baker}, A.~J., {Frayer}, D.~T., {et~al.} 2012, ArXiv
  e-prints, arXiv:1204.4706

\bibitem[{{Hatsukade} {et~al.}(2013){Hatsukade}, {Ohta}, {Seko}, {Yabe}, \&
  {Akiyama}}]{hatsukade13}
{Hatsukade}, B., {Ohta}, K., {Seko}, A., {Yabe}, K., \& {Akiyama}, M. 2013,
  \apjl, 769, L27

\bibitem[{{Hatsukade} {et~al.}(2015){Hatsukade}, {Ohta}, {Yabe}, {Seko},
  {Makiya}, \& {Akiyama}}]{hatsukade15}
{Hatsukade}, B., {Ohta}, K., {Yabe}, K., {et~al.} 2015, \apj, 810, 91

\bibitem[{{Hatsukade} {et~al.}(2011){Hatsukade}, {Kohno}, {Aretxaga},
  {Austermann}, {Ezawa}, {Hughes}, {Ikarashi}, {Iono}, {Kawabe}, {Khan},
  {Matsuo}, {Matsuura}, {Nakanishi}, {Oshima}, {Perera}, {Scott}, {Shirahata},
  {Takeuchi}, {Tamura}, {Tanaka}, {Tosaki}, {Wilson}, \& {Yun}}]{hatsukade11}
{Hatsukade}, B., {Kohno}, K., {Aretxaga}, I., {et~al.} 2011, \mnras, 411, 102

\bibitem[{{Hatsukade} {et~al.}(2016){Hatsukade}, {Kohno}, {Umehata},
  {Aretxaga}, {Caputi}, {Dunlop}, {Ikarashi}, {Iono}, {Ivison}, {Lee},
  {Makiya}, {Matsuda}, {Motohara}, {Nakanishi}, {Ohta}, {Tadaki}, {Tamura},
  {Wang}, {Wilson}, {Yamaguchi}, \& {Yun}}]{hatsukade16}
{Hatsukade}, B., {Kohno}, K., {Umehata}, H., {et~al.} 2016, ArXiv e-prints,
  arXiv:1602.08167

\bibitem[{{Helou} \& {Beichman}(1990)}]{helou90}
{Helou}, G., \& {Beichman}, C.~A. 1990, in Liege International Astrophysical
  Colloquia, Vol.~29, Liege International Astrophysical Colloquia, ed.
  B.~{Kaldeich}

\bibitem[{{Hodge} {et~al.}(2013){Hodge}, {Carilli}, {Walter}, {Daddi}, \&
  {Riechers}}]{hodge13}
{Hodge}, J.~A., {Carilli}, C.~L., {Walter}, F., {Daddi}, E., \& {Riechers}, D.
  2013, \apj, 776, 22

\bibitem[{{Hughes} {et~al.}(1998){Hughes}, {Serjeant}, {Dunlop},
  {Rowan-Robinson}, {Blain}, {Mann}, {Ivison}, {Peacock}, {Efstathiou}, {Gear},
  {Oliver}, {Lawrence}, {Longair}, {Goldschmidt}, \& {Jenness}}]{hughes98}
{Hughes}, D.~H., {Serjeant}, S., {Dunlop}, J., {et~al.} 1998, \nat, 394, 241

\bibitem[{{Ivison} {et~al.}(2011){Ivison}, {Papadopoulos}, {Smail}, {Greve},
  {Thomson}, {Xilouris}, \& {Chapman}}]{ivison11}
{Ivison}, R.~J., {Papadopoulos}, P.~P., {Smail}, I., {et~al.} 2011, \mnras,
  412, 1913

\bibitem[{{Ivison} {et~al.}(2010){Ivison}, {Swinbank}, {Swinyard}, {Smail},
  {Pearson}, {Rigopoulou}, {Polehampton}, {Baluteau}, {Barlow}, {Blain},
  {Bock}, {Clements}, {Coppin}, {Cooray}, {Danielson}, {Dwek}, {Edge},
  {Franceschini}, {Fulton}, {Glenn}, {Griffin}, {Isaak}, {Leeks}, {Lim},
  {Naylor}, {Oliver}, {Page}, {P{\'e}rez Fournon}, {Rowan-Robinson}, {Savini},
  {Scott}, {Spencer}, {Valtchanov}, {Vigroux}, \& {Wright}}]{ivison10}
{Ivison}, R.~J., {Swinbank}, A.~M., {Swinyard}, B., {et~al.} 2010, \aap, 518,
  L35+

\bibitem[{{Ivison} {et~al.}(2013){Ivison}, {Swinbank}, {Smail}, {Harris},
  {Bussmann}, {Cooray}, {Cox}, {Fu}, {Kov{\'a}cs}, {Krips}, {Narayanan},
  {Negrello}, {Neri}, {Pe{\~n}arrubia}, {Richard}, {Riechers}, {Rowlands},
  {Staguhn}, {Targett}, {Amber}, {Baker}, {Bourne}, {Bertoldi}, {Bremer},
  {Calanog}, {Clements}, {Dannerbauer}, {Dariush}, {De Zotti}, {Dunne},
  {Eales}, {Farrah}, {Fleuren}, {Franceschini}, {Geach}, {George}, {Helly},
  {Hopwood}, {Ibar}, {Jarvis}, {Kneib}, {Maddox}, {Omont}, {Scott}, {Serjeant},
  {Smith}, {Thompson}, {Valiante}, {Valtchanov}, {Vieira}, \& {van der
  Werf}}]{ivison13}
{Ivison}, R.~J., {Swinbank}, A.~M., {Smail}, I., {et~al.} 2013, \apj, 772, 137

\bibitem[{{Johansson} {et~al.}(2012){Johansson}, {Horellou}, {Lopez-Cruz},
  {Muller}, {Birkinshaw}, {Black}, {Bremer}, {Wall}, {Bertoldi}, {Castillo}, \&
  {Ibarra-Medel}}]{johansson12}
{Johansson}, D., {Horellou}, C., {Lopez-Cruz}, O., {et~al.} 2012, \aap, 543,
  A62

\bibitem[{{Karim} {et~al.}(2011){Karim}, {Schinnerer},
  {Mart{\'{\i}}nez-Sansigre}, {Sargent}, {van der Wel}, {Rix}, {Ilbert},
  {Smol{\v c}i{\'c}}, {Carilli}, {Pannella}, {Koekemoer}, {Bell}, \&
  {Salvato}}]{karim11}
{Karim}, A., {Schinnerer}, E., {Mart{\'{\i}}nez-Sansigre}, A., {et~al.} 2011,
  \apj, 730, 61

\bibitem[{{Karim} {et~al.}(2013){Karim}, {Swinbank}, {Hodge}, {Smail},
  {Walter}, {Biggs}, {Simpson}, {Danielson}, {Alexander}, {Bertoldi}, {de
  Breuck}, {Chapman}, {Coppin}, {Dannerbauer}, {Edge}, {Greve}, {Ivison},
  {Knudsen}, {Menten}, {Schinnerer}, {Wardlow}, {Wei{\ss}}, \& {van der
  Werf}}]{karim13}
{Karim}, A., {Swinbank}, A.~M., {Hodge}, J.~A., {et~al.} 2013, \mnras, 432, 2

\bibitem[{{Knudsen} {et~al.}(2008){Knudsen}, {van der Werf}, \&
  {Kneib}}]{knudsen08}
{Knudsen}, K.~K., {van der Werf}, P.~P., \& {Kneib}, J.-P. 2008, \mnras, 384,
  1611

\bibitem[{{Knudsen} {et~al.}(2005){Knudsen}, {van der Werf}, {Franx},
  {F{\"o}rster Schreiber}, {van Dokkum}, {Illingworth}, {Labb{\'e}},
  {Moorwood}, {Rix}, \& {Rudnick}}]{knudsen05}
{Knudsen}, K.~K., {van der Werf}, P., {Franx}, M., {et~al.} 2005, \apjl, 632,
  L9

\bibitem[{{Koprowski} {et~al.}(2016){Koprowski}, {Dunlop}, {Micha{\l}owski},
  {Roseboom}, {Geach}, {Cirasuolo}, {Aretxaga}, {Bowler}, {Banerji}, {Bourne},
  {Coppin}, {Chapman}, {Hughes}, {Jenness}, {McLure}, {Symeonidis}, \&
  {Werf}}]{koprowski16}
{Koprowski}, M.~P., {Dunlop}, J.~S., {Micha{\l}owski}, M.~J., {et~al.} 2016,
  \mnras, 458, 4321

\bibitem[{{Lee} {et~al.}(2015){Lee}, {Sanders}, {Casey}, {Toft}, {Scoville},
  {Hung}, {Le Floc'h}, {Ilbert}, {Zahid}, {Aussel}, {Capak}, {Kartaltepe},
  {Kewley}, {Li}, {Schawinski}, {Sheth}, \& {Xiao}}]{lee15}
{Lee}, N., {Sanders}, D.~B., {Casey}, C.~M., {et~al.} 2015, \apj, 801, 80

\bibitem[{{Lehmer} {et~al.}(2005){Lehmer}, {Brandt}, {Alexander}, {Bauer},
  {Schneider}, {Tozzi}, {Bergeron}, {Garmire}, {Giacconi}, {Gilli}, {Hasinger},
  {Hornschemeier}, {Koekemoer}, {Mainieri}, {Miyaji}, {Nonino}, {Rosati},
  {Silverman}, {Szokoly}, \& {Vignali}}]{lehmer05}
{Lehmer}, B.~D., {Brandt}, W.~N., {Alexander}, D.~M., {et~al.} 2005, \apjs,
  161, 21

\bibitem[{{Leroy} {et~al.}(2011){Leroy}, {Bolatto}, {Gordon}, {Sandstrom},
  {Gratier}, {Rosolowsky}, {Engelbracht}, {Mizuno}, {Corbelli}, {Fukui}, \&
  {Kawamura}}]{leroy11}
{Leroy}, A.~K., {Bolatto}, A., {Gordon}, K., {et~al.} 2011, \apj, 737, 12

\bibitem[{{Lestrade} {et~al.}(2011){Lestrade}, {Carilli}, {Thanjavur}, {Kneib},
  {Riechers}, {Bertoldi}, {Walter}, \& {Omont}}]{lestrade11}
{Lestrade}, J.-F., {Carilli}, C.~L., {Thanjavur}, K., {et~al.} 2011, \apjl,
  739, L30

\bibitem[{{Lilly} {et~al.}(1996){Lilly}, {Le Fevre}, {Hammer}, \&
  {Crampton}}]{lilly96}
{Lilly}, S.~J., {Le Fevre}, O., {Hammer}, F., \& {Crampton}, D. 1996, \apjl,
  460, L1

\bibitem[{{Madau} {et~al.}(1996){Madau}, {Ferguson}, {Dickinson}, {Giavalisco},
  {Steidel}, \& {Fruchter}}]{madau96}
{Madau}, P., {Ferguson}, H.~C., {Dickinson}, M.~E., {et~al.} 1996, \mnras, 283,
  1388

\bibitem[{{Magdis} {et~al.}(2011){Magdis}, {Daddi}, {Elbaz}, {Sargent},
  {Dickinson}, {Dannerbauer}, {Aussel}, {Walter}, {Hwang}, {Charmandaris},
  {Hodge}, {Riechers}, {Rigopoulou}, {Carilli}, {Pannella}, {Mullaney},
  {Leiton}, \& {Scott}}]{magdis11}
{Magdis}, G.~E., {Daddi}, E., {Elbaz}, D., {et~al.} 2011, \apjl, 740, L15

\bibitem[{{Magdis} {et~al.}(2012){Magdis}, {Daddi}, {B{\'e}thermin}, {Sargent},
  {Elbaz}, {Pannella}, {Dickinson}, {Dannerbauer}, {da Cunha}, {Walter},
  {Rigopoulou}, {Charmandaris}, {Hwang}, \& {Kartaltepe}}]{magdis12}
{Magdis}, G.~E., {Daddi}, E., {B{\'e}thermin}, M., {et~al.} 2012, \apj, 760, 6

\bibitem[{{Magnelli} {et~al.}(2012){Magnelli}, {Saintonge}, {Lutz}, {Tacconi},
  {Berta}, {Bournaud}, {Charmandaris}, {Dannerbauer}, {Elbaz},
  {F{\"o}rster-Schreiber}, {Graci{\'a}-Carpio}, {Ivison}, {Maiolino}, {Nordon},
  {Popesso}, {Rodighiero}, {Santini}, \& {Wuyts}}]{magnelli12}
{Magnelli}, B., {Saintonge}, A., {Lutz}, D., {et~al.} 2012, \aap, 548, A22

\bibitem[{{McLure} {et~al.}(2013){McLure}, {Dunlop}, {Bowler}, {Curtis-Lake},
  {Schenker}, {Ellis}, {Robertson}, {Koekemoer}, {Rogers}, {Ono}, {Ouchi},
  {Charlot}, {Wild}, {Stark}, {Furlanetto}, {Cirasuolo}, \&
  {Targett}}]{mclure13}
{McLure}, R.~J., {Dunlop}, J.~S., {Bowler}, R.~A.~A., {et~al.} 2013, \mnras,
  432, 2696

\bibitem[{{Micha{\l}owski} {et~al.}(2010){Micha{\l}owski}, {Hjorth}, \&
  {Watson}}]{michalowski10}
{Micha{\l}owski}, M., {Hjorth}, J., \& {Watson}, D. 2010, \aap, 514, A67

\bibitem[{{Micha{\l}owski} {et~al.}(2012){Micha{\l}owski}, {Dunlop},
  {Cirasuolo}, {Hjorth}, {Hayward}, \& {Watson}}]{michalowski12}
{Micha{\l}owski}, M.~J., {Dunlop}, J.~S., {Cirasuolo}, M., {et~al.} 2012, \aap,
  541, A85

\bibitem[{{Miettinen} {et~al.}(2015){Miettinen}, {Smol{\v c}i{\'c}}, {Novak},
  {Aravena}, {Karim}, {Masters}, {Riechers}, {Bussmann}, {McCracken}, {Ilbert},
  {Bertoldi}, {Capak}, {Feruglio}, {Halliday}, {Kartaltepe}, {Navarrete},
  {Salvato}, {Sanders}, {Schinnerer}, \& {Sheth}}]{miettinen15}
{Miettinen}, O., {Smol{\v c}i{\'c}}, V., {Novak}, M., {et~al.} 2015, \aap, 577,
  A29

\bibitem[{{Mocanu} {et~al.}(2013){Mocanu}, {Crawford}, {Vieira}, {Aird},
  {Aravena}, {Austermann}, {Benson}, {B{\'e}thermin}, {Bleem}, {Bothwell},
  {Carlstrom}, {Chang}, {Chapman}, {Cho}, {Crites}, {de Haan}, {Dobbs},
  {Everett}, {George}, {Halverson}, {Harrington}, {Hezaveh}, {Holder},
  {Holzapfel}, {Hoover}, {Hrubes}, {Keisler}, {Knox}, {Lee}, {Leitch},
  {Lueker}, {Luong-Van}, {Marrone}, {McMahon}, {Mehl}, {Meyer}, {Mohr},
  {Montroy}, {Natoli}, {Padin}, {Plagge}, {Pryke}, {Rest}, {Reichardt}, {Ruhl},
  {Sayre}, {Schaffer}, {Shirokoff}, {Spieler}, {Spilker}, {Stalder},
  {Staniszewski}, {Stark}, {Story}, {Switzer}, {Vanderlinde}, \&
  {Williamson}}]{mocanu13}
{Mocanu}, L.~M., {Crawford}, T.~M., {Vieira}, J.~D., {et~al.} 2013, \apj, 779,
  61

\bibitem[{{Momcheva} {et~al.}(2015){Momcheva}, {Brammer}, {van Dokkum},
  {Skelton}, {Whitaker}, {Nelson}, {Fumagalli}, {Maseda}, {Leja}, {Franx},
  {Rix}, {Bezanson}, {Da Cunha}, {Dickey}, {F{\"o}rster Schreiber},
  {Illingworth}, {Kriek}, {Labb{\'e}}, {Ulf Lange}, {Lundgren}, {Magee},
  {Marchesini}, {Oesch}, {Pacifici}, {Patel}, {Price}, {Tal}, {Wake}, {van der
  Wel}, \& {Wuyts}}]{momcheva15}
{Momcheva}, I.~G., {Brammer}, G.~B., {van Dokkum}, P.~G., {et~al.} 2015, ArXiv
  e-prints, arXiv:1510.02106

\bibitem[{{Morris} {et~al.}(2015){Morris}, {Kocevski}, {Trump}, {Weiner},
  {Hathi}, {Barro}, {Dahlen}, {Faber}, {Finkelstein}, {Fontana}, {Ferguson},
  {Grogin}, {Gr{\"u}tzbauch}, {Guo}, {Hsu}, {Koekemoer}, {Koo}, {Mobasher},
  {Pforr}, {Salvato}, {Wiklind}, \& {Wuyts}}]{morris15}
{Morris}, A.~M., {Kocevski}, D.~D., {Trump}, J.~R., {et~al.} 2015, \aj, 149,
  178

\bibitem[{{Noble} {et~al.}(2012){Noble}, {Webb}, {Ellingson}, {Faloon}, {Gal},
  {Gladders}, {Hicks}, {Hoekstra}, {Hsieh}, {Ivison}, {Lemaux}, {Lubin},
  {O'Donnell}, \& {Yee}}]{noble12}
{Noble}, A.~G., {Webb}, T.~M.~A., {Ellingson}, E., {et~al.} 2012, \mnras, 419,
  1983

\bibitem[{{Noeske} {et~al.}(2007){Noeske}, {Weiner}, {Faber}, {Papovich},
  {Koo}, {Somerville}, {Bundy}, {Conselice}, {Newman}, {Schiminovich}, {Le
  Floc'h}, {Coil}, {Rieke}, {Lotz}, {Primack}, {Barmby}, {Cooper}, {Davis},
  {Ellis}, {Fazio}, {Guhathakurta}, {Huang}, {Kassin}, {Martin}, {Phillips},
  {Rich}, {Small}, {Willmer}, \& {Wilson}}]{noeske07}
{Noeske}, K.~G., {Weiner}, B.~J., {Faber}, S.~M., {et~al.} 2007, \apjl, 660,
  L43

\bibitem[{{Ono} {et~al.}(2014){Ono}, {Ouchi}, {Kurono}, \& {Momose}}]{ono14}
{Ono}, Y., {Ouchi}, M., {Kurono}, Y., \& {Momose}, R. 2014, \apj, 795, 5

\bibitem[{{Oteo} {et~al.}(2015){Oteo}, {Zwaan}, {Ivison}, {Smail}, \&
  {Biggs}}]{oteo15}
{Oteo}, I., {Zwaan}, M.~A., {Ivison}, R.~J., {Smail}, I., \& {Biggs}, A.~D.
  2015, ArXiv e-prints, arXiv:1508.05099

\bibitem[{{Pannella} {et~al.}(2009){Pannella}, {Carilli}, {Daddi}, {McCracken},
  {Owen}, {Renzini}, {Strazzullo}, {Civano}, {Koekemoer}, {Schinnerer},
  {Scoville}, {Smol{\v c}i{\'c}}, {Taniguchi}, {Aussel}, {Kneib}, {Ilbert},
  {Mellier}, {Salvato}, {Thompson}, \& {Willott}}]{pannella09}
{Pannella}, M., {Carilli}, C.~L., {Daddi}, E., {et~al.} 2009, \apjl, 698, L116

\bibitem[{{Pannella} {et~al.}(2015){Pannella}, {Elbaz}, {Daddi}, {Dickinson},
  {Hwang}, {Schreiber}, {Strazzullo}, {Aussel}, {Bethermin}, {Buat},
  {Charmandaris}, {Cibinel}, {Juneau}, {Ivison}, {Le Borgne}, {Le Floc'h},
  {Leiton}, {Lin}, {Magdis}, {Morrison}, {Mullaney}, {Onodera}, {Renzini},
  {Salim}, {Sargent}, {Scott}, {Shu}, \& {Wang}}]{pannella15}
{Pannella}, M., {Elbaz}, D., {Daddi}, E., {et~al.} 2015, \apj, 807, 141

\bibitem[{{Planck Collaboration} {et~al.}(2011){Planck Collaboration},
  {Abergel}, {Ade}, {Aghanim}, {Arnaud}, {Ashdown}, {Aumont}, {Baccigalupi},
  {Balbi}, {Banday}, {Barreiro}, {Bartlett}, {Battaner}, {Benabed},
  {Beno{\^i}t}, {Bernard}, {Bersanelli}, {Bhatia}, {Bock}, {Bonaldi}, {Bond},
  {Borrill}, {Bouchet}, {Boulanger}, {Bucher}, {Burigana}, {Cabella},
  {Cardoso}, {Catalano}, {Cay{\'o}n}, {Challinor}, {Chamballu}, {Chiang},
  {Chiang}, {Christensen}, {Colombi}, {Couchot}, {Coulais}, {Crill}, {Cuttaia},
  {Dame}, {Danese}, {Davies}, {Davis}, {de Bernardis}, {de Gasperis}, {de
  Rosa}, {de Zotti}, {Delabrouille}, {Delouis}, {D{\'e}sert}, {Dickinson},
  {Donzelli}, {Dor{\'e}}, {D{\"o}rl}, {Douspis}, {Dupac}, {Efstathiou},
  {En{\ss}lin}, {Finelli}, {Forni}, {Frailis}, {Franceschi}, {Galeotta},
  {Ganga}, {Giard}, {Giardino}, {Giraud-H{\'e}raud}, {Gonz{\'a}lez-Nuevo},
  {G{\'o}rski}, {Gratton}, {Gregorio}, {Grenier}, {Gruppuso}, {Hansen},
  {Harrison}, {Henrot-Versill{\'e}}, {Herranz}, {Hildebrandt}, {Hivon},
  {Hobson}, {Holmes}, {Hovest}, {Hoyland}, {Huffenberger}, {Jaffe}, {Jaffe},
  {Jones}, {Juvela}, {Keih{\"a}nen}, {Keskitalo}, {Kisner}, {Kneissl}, {Knox},
  {Kurki-Suonio}, {Lagache}, {L{\"a}hteenm{\"a}ki}, {Lamarre}, {Lasenby},
  {Laureijs}, {Lawrence}, {Leach}, {Leonardi}, {Leroy}, {Lilje},
  {Linden-V{\o}rnle}, {L{\'o}pez-Caniego}, {Lubin}, {Mac{\'{\i}}as-P{\'e}rez},
  {MacTavish}, {Maffei}, {Mandolesi}, {Mann}, {Maris}, {Marshall},
  {Mart{\'{\i}}nez-Gonz{\'a}lez}, {Masi}, {Matarrese}, {Matthai}, {Mazzotta},
  {McGehee}, {Meinhold}, {Melchiorri}, {Mendes}, {Mennella},
  {Miville-Desch{\^e}nes}, {Moneti}, {Montier}, {Morgante}, {Mortlock},
  {Munshi}, {Murphy}, {Naselsky}, {Natoli}, {Netterfield},
  {N{\o}rgaard-Nielsen}, {Noviello}, {Novikov}, {Novikov}, {Osborne}, {Pajot},
  {Paladini}, {Pasian}, {Patanchon}, {Perdereau}, {Perotto}, {Perrotta},
  {Piacentini}, {Piat}, {Plaszczynski}, {Pointecouteau}, {Polenta}, {Ponthieu},
  {Poutanen}, {Pr{\'e}zeau}, {Prunet}, {Puget}, {Rachen}, {Reach}, {Rebolo},
  {Reich}, {Renault}, {Ricciardi}, {Riller}, {Ristorcelli}, {Rocha}, {Rosset},
  {Rubi{\~n}o-Mart{\'{\i}}n}, {Rusholme}, {Sandri}, {Santos}, {Savini},
  {Scott}, {Seiffert}, {Shellard}, {Smoot}, {Starck}, {Stivoli}, {Stolyarov},
  {Stompor}, {Sudiwala}, {Sygnet}, {Tauber}, {Terenzi}, {Toffolatti}, {Tomasi},
  {Torre}, {Tristram}, {Tuovinen}, {Umana}, {Valenziano}, {Varis}, {Vielva},
  {Villa}, {Vittorio}, {Wade}, {Wandelt}, {Wilkinson}, {Ysard}, {Yvon},
  {Zacchei}, \& {Zonca}}]{planck11a}
{Planck Collaboration}, {Abergel}, A., {Ade}, P.~A.~R., {et~al.} 2011, \aap,
  536, A21

\bibitem[{{Planck Collaboration} {et~al.}(2014){Planck Collaboration}, {Ade},
  {Aghanim}, {Armitage-Caplan}, {Arnaud}, {Ashdown}, {Atrio-Barandela},
  {Aumont}, {Baccigalupi}, {Banday}, \& et~al.}]{planck14}
{Planck Collaboration}, {Ade}, P.~A.~R., {Aghanim}, N., {et~al.} 2014, \aap,
  571, A30

\bibitem[{{Puget} {et~al.}(1996){Puget}, {Abergel}, {Bernard}, {Boulanger},
  {Burton}, {Desert}, \& {Hartmann}}]{puget96}
{Puget}, J.-L., {Abergel}, A., {Bernard}, J.-P., {et~al.} 1996, \aap, 308, L5+

\bibitem[{{Rau} \& {Cornwell}(2011)}]{rau11}
{Rau}, U., \& {Cornwell}, T.~J. 2011, \aap, 532, A71

\bibitem[{{R{\'e}my-Ruyer} {et~al.}(2014){R{\'e}my-Ruyer}, {Madden},
  {Galliano}, {Galametz}, {Takeuchi}, {Asano}, {Zhukovska}, {Lebouteiller},
  {Cormier}, {Jones}, {Bocchio}, {Baes}, {Bendo}, {Boquien}, {Boselli},
  {DeLooze}, {Doublier-Pritchard}, {Hughes}, {Karczewski}, \&
  {Spinoglio}}]{remyruyer14}
{R{\'e}my-Ruyer}, A., {Madden}, S.~C., {Galliano}, F., {et~al.} 2014, \aap,
  563, A31

\bibitem[{{Rhoads} {et~al.}(2009){Rhoads}, {Malhotra}, {Pirzkal}, {Dickinson},
  {Cohen}, {Grogin}, {Hathi}, {Xu}, {Ferreras}, {Gronwall}, {Koekemoer},
  {K{\"u}mmel}, {Meurer}, {Panagia}, {Pasquali}, {Ryan}, {Straughn}, {Walsh},
  {Windhorst}, \& {Yan}}]{rhoads09}
{Rhoads}, J.~E., {Malhotra}, S., {Pirzkal}, N., {et~al.} 2009, \apj, 697, 942

\bibitem[{{Riechers} {et~al.}(2011{\natexlab{a}}){Riechers}, {Hodge}, {Walter},
  {Carilli}, \& {Bertoldi}}]{riechers11b}
{Riechers}, D.~A., {Hodge}, J., {Walter}, F., {Carilli}, C.~L., \& {Bertoldi},
  F. 2011{\natexlab{a}}, \apjl, 739, L31

\bibitem[{{Riechers} {et~al.}(2011{\natexlab{b}}){Riechers}, {Carilli},
  {Maddalena}, {Hodge}, {Harris}, {Baker}, {Walter}, {Wagg}, {Vanden Bout},
  {Wei{\ss}}, \& {Sharon}}]{riechers11c}
{Riechers}, D.~A., {Carilli}, C.~L., {Maddalena}, R.~J., {et~al.}
  2011{\natexlab{b}}, \apjl, 739, L32

\bibitem[{{Riechers} {et~al.}(2013){Riechers}, {Bradford}, {Clements},
  {Dowell}, {P{\'e}rez-Fournon}, {Ivison}, {Bridge}, {Conley}, {Fu}, {Vieira},
  {Wardlow}, {Calanog}, {Cooray}, {Hurley}, {Neri}, {Kamenetzky}, {Aguirre},
  {Altieri}, {Arumugam}, {Benford}, {B{\'e}thermin}, {Bock}, {Burgarella},
  {Cabrera-Lavers}, {Chapman}, {Cox}, {Dunlop}, {Earle}, {Farrah}, {Ferrero},
  {Franceschini}, {Gavazzi}, {Glenn}, {Solares}, {Gurwell}, {Halpern},
  {Hatziminaoglou}, {Hyde}, {Ibar}, {Kov{\'a}cs}, {Krips}, {Lupu}, {Maloney},
  {Martinez-Navajas}, {Matsuhara}, {Murphy}, {Naylor}, {Nguyen}, {Oliver},
  {Omont}, {Page}, {Petitpas}, {Rangwala}, {Roseboom}, {Scott}, {Smith},
  {Staguhn}, {Streblyanska}, {Thomson}, {Valtchanov}, {Viero}, {Wang},
  {Zemcov}, \& {Zmuidzinas}}]{riechers13}
{Riechers}, D.~A., {Bradford}, C.~M., {Clements}, D.~L., {et~al.} 2013, \nat,
  496, 329

\bibitem[{{Rodighiero} {et~al.}(2011){Rodighiero}, {Daddi}, {Baronchelli},
  {Cimatti}, {Renzini}, {Aussel}, {Popesso}, {Lutz}, {Andreani}, {Berta},
  {Cava}, {Elbaz}, {Feltre}, {Fontana}, {F{\"o}rster Schreiber},
  {Franceschini}, {Genzel}, {Grazian}, {Gruppioni}, {Ilbert}, {Le Floch},
  {Magdis}, {Magliocchetti}, {Magnelli}, {Maiolino}, {McCracken}, {Nordon},
  {Poglitsch}, {Santini}, {Pozzi}, {Riguccini}, {Tacconi}, {Wuyts}, \&
  {Zamorani}}]{rodighiero11}
{Rodighiero}, G., {Daddi}, E., {Baronchelli}, I., {et~al.} 2011, \apjl, 739,
  L40

\bibitem[{{Rujopakarn} {et~al.}(2016){Rujopakarn}, {Dunlop}, {Rieke}, {Ivison},
  {Cibinel}, {Nyland}, {Jagannathan}, {Silverman}, {Alexander}, {Biggs},
  {Bhatnagar}, {Ballantyne}, {Dickinson}, {Elbaz}, {Geach}, {Hayward},
  {Kirkpatrick}, {McLure}, {Michalowski}, {Miller}, {Narayanan}, {Owen},
  {Pannella}, {Papovich}, {Pope}, {Rau}, {Robertson}, {Scott}, {Swinbank}, {van
  der Werf}, {van Kampen}, \& {Windhorst}}]{rujopakarn16}
{Rujopakarn}, W., {Dunlop}, J.~S., {Rieke}, G.~H., {et~al.} 2016, ArXiv
  e-prints, arXiv:1607.07710

\bibitem[{{Saintonge} {et~al.}(2013){Saintonge}, {Lutz}, {Genzel}, {Magnelli},
  {Nordon}, {Tacconi}, {Baker}, {Bandara}, {Berta}, {F{\"o}rster Schreiber},
  {Poglitsch}, {Sturm}, {Wuyts}, \& {Wuyts}}]{saintonge13}
{Saintonge}, A., {Lutz}, D., {Genzel}, R., {et~al.} 2013, \apj, 778, 2

\bibitem[{{Sandstrom} {et~al.}(2013){Sandstrom}, {Leroy}, {Walter}, {Bolatto},
  {Croxall}, {Draine}, {Wilson}, {Wolfire}, {Calzetti}, {Kennicutt}, {Aniano},
  {Donovan Meyer}, {Usero}, {Bigiel}, {Brinks}, {de Blok}, {Crocker}, {Dale},
  {Engelbracht}, {Galametz}, {Groves}, {Hunt}, {Koda}, {Kreckel}, {Linz},
  {Meidt}, {Pellegrini}, {Rix}, {Roussel}, {Schinnerer}, {Schruba}, {Schuster},
  {Skibba}, {van der Laan}, {Appleton}, {Armus}, {Brandl}, {Gordon}, {Hinz},
  {Krause}, {Montiel}, {Sauvage}, {Schmiedeke}, {Smith}, \&
  {Vigroux}}]{sandstrom13}
{Sandstrom}, K.~M., {Leroy}, A.~K., {Walter}, F., {et~al.} 2013, \apj, 777, 5

\bibitem[{{Schenker} {et~al.}(2013){Schenker}, {Robertson}, {Ellis}, {Ono},
  {McLure}, {Dunlop}, {Koekemoer}, {Bowler}, {Ouchi}, {Curtis-Lake}, {Rogers},
  {Schneider}, {Charlot}, {Stark}, {Furlanetto}, \& {Cirasuolo}}]{schenker13}
{Schenker}, M.~A., {Robertson}, B.~E., {Ellis}, R.~S., {et~al.} 2013, \apj,
  768, 196

\bibitem[{{Scott} {et~al.}(2008){Scott}, {Austermann}, {Perera}, {Wilson},
  {Aretxaga}, {Bock}, {Hughes}, {Kang}, {Kim}, {Mauskopf}, {Sanders},
  {Scoville}, \& {Yun}}]{scott08}
{Scott}, K.~S., {Austermann}, J.~E., {Perera}, T.~A., {et~al.} 2008, \mnras,
  385, 2225

\bibitem[{{Scott} {et~al.}(2012){Scott}, {Wilson}, {Aretxaga}, {Austermann},
  {Chapin}, {Dunlop}, {Ezawa}, {Halpern}, {Hatsukade}, {Hughes}, {Kawabe},
  {Kim}, {Kohno}, {Lowenthal}, {Monta{\~n}a}, {Nakanishi}, {Oshima}, {Sanders},
  {Scott}, {Scoville}, {Tamura}, {Welch}, {Yun}, \& {Zeballos}}]{scott12}
{Scott}, K.~S., {Wilson}, G.~W., {Aretxaga}, I., {et~al.} 2012, \mnras, 423,
  575

\bibitem[{{Scott} {et~al.}(2002){Scott}, {Fox}, {Dunlop}, {Serjeant},
  {Peacock}, {Ivison}, {Oliver}, {Mann}, {Lawrence}, {Efstathiou},
  {Rowan-Robinson}, {Hughes}, {Archibald}, {Blain}, \& {Longair}}]{scott02}
{Scott}, S.~E., {Fox}, M.~J., {Dunlop}, J.~S., {et~al.} 2002, \mnras, 331, 817

\bibitem[{{Scoville} {et~al.}(2013){Scoville}, {Arnouts}, {Aussel}, {Benson},
  {Bongiorno}, {Bundy}, {Calvo}, {Capak}, {Carollo}, {Civano}, {Dunlop},
  {Elvis}, {Faisst}, {Finoguenov}, {Fu}, {Giavalisco}, {Guo}, {Ilbert},
  {Iovino}, {Kajisawa}, {Kartaltepe}, {Leauthaud}, {Le F{\`e}vre}, {LeFloch},
  {Lilly}, {Liu}, {Manohar}, {Massey}, {Masters}, {McCracken}, {Mobasher},
  {Peng}, {Renzini}, {Rhodes}, {Salvato}, {Sanders}, {Sarvestani}, {Scarlata},
  {Schinnerer}, {Sheth}, {Shopbell}, {Smol{\v c}i{\'c}}, {Taniguchi}, {Taylor},
  {White}, \& {Yan}}]{scoville13}
{Scoville}, N., {Arnouts}, S., {Aussel}, H., {et~al.} 2013, \apjs, 206, 3

\bibitem[{{Scoville} {et~al.}(2014){Scoville}, {Aussel}, {Sheth}, {Scott},
  {Sanders}, {Ivison}, {Pope}, {Capak}, {Vanden Bout}, {Manohar}, {Kartaltepe},
  {Robertson}, \& {Lilly}}]{scoville14}
{Scoville}, N., {Aussel}, H., {Sheth}, K., {et~al.} 2014, \apj, 783, 84

\bibitem[{{Sheth} {et~al.}(2004){Sheth}, {Blain}, {Kneib}, {Frayer}, {van der
  Werf}, \& {Knudsen}}]{sheth04}
{Sheth}, K., {Blain}, A.~W., {Kneib}, J.-P., {et~al.} 2004, \apjl, 614, L5

\bibitem[{{Simpson} {et~al.}(2014){Simpson}, {Swinbank}, {Smail}, {Alexander},
  {Brandt}, {Bertoldi}, {de Breuck}, {Chapman}, {Coppin}, {da Cunha},
  {Danielson}, {Dannerbauer}, {Greve}, {Hodge}, {Ivison}, {Karim}, {Knudsen},
  {Poggianti}, {Schinnerer}, {Thomson}, {Walter}, {Wardlow}, {Wei{\ss}}, \&
  {van der Werf}}]{simpson14}
{Simpson}, J.~M., {Swinbank}, A.~M., {Smail}, I., {et~al.} 2014, \apj, 788, 125

\bibitem[{{Simpson} {et~al.}(2015){Simpson}, {Smail}, {Swinbank}, {Chapman},
  {Geach}, {Ivison}, {Thomson}, {Aretxaga}, {Blain}, {Cowley}, {Chen},
  {Coppin}, {Dunlop}, {Edge}, {Farrah}, {Ibar}, {Karim}, {Knudsen},
  {Meijerink}, {Micha{\l}owski}, {Scott}, {Spaans}, \& {van der
  Werf}}]{simpson15}
{Simpson}, J.~M., {Smail}, I., {Swinbank}, A.~M., {et~al.} 2015, \apj, 807, 128

\bibitem[{{Skelton} {et~al.}(2014){Skelton}, {Whitaker}, {Momcheva}, {Brammer},
  {van Dokkum}, {Labb{\'e}}, {Franx}, {van der Wel}, {Bezanson}, {Da Cunha},
  {Fumagalli}, {F{\"o}rster Schreiber}, {Kriek}, {Leja}, {Lundgren}, {Magee},
  {Marchesini}, {Maseda}, {Nelson}, {Oesch}, {Pacifici}, {Patel}, {Price},
  {Rix}, {Tal}, {Wake}, \& {Wuyts}}]{skelton14}
{Skelton}, R.~E., {Whitaker}, K.~E., {Momcheva}, I.~G., {et~al.} 2014, \apjs,
  214, 24

\bibitem[{{Smail} {et~al.}(1997){Smail}, {Ivison}, \& {Blain}}]{smail97}
{Smail}, I., {Ivison}, R.~J., \& {Blain}, A.~W. 1997, \apjl, 490, L5+

\bibitem[{{Smail} {et~al.}(2002){Smail}, {Ivison}, {Blain}, \&
  {Kneib}}]{smail02}
{Smail}, I., {Ivison}, R.~J., {Blain}, A.~W., \& {Kneib}, J.-P. 2002, \mnras,
  331, 495

\bibitem[{{Smolcic} {et~al.}(2012){Smolcic}, {Aravena}, {Navarrete},
  {Schinnerer}, {Riechers}, {Bertoldi}, {Feruglio}, {Finoguenov}, {Salvato},
  {Sargent}, {McCracken}, {Albrecht}, {Karim}, {Capak}, {Carilli},
  {Cappelluti}, {Elvis}, {Ilbert}, {Kartaltepe}, {Lilly}, {Sanders}, {Sheth},
  {Scoville}, \& {Taniguchi}}]{smolcic12}
{Smolcic}, V., {Aravena}, M., {Navarrete}, F., {et~al.} 2012, ArXiv e-prints,
  arXiv:1205.6470

\bibitem[{{Strandet} {et~al.}(2016){Strandet}, {Weiss}, {Vieira}, {de Breuck},
  {Aguirre}, {Aravena}, {Ashby}, {B{\'e}thermin}, {Bradford}, {Carlstrom},
  {Chapman}, {Crawford}, {Everett}, {Fassnacht}, {Furstenau}, {Gonzalez},
  {Greve}, {Gullberg}, {Hezaveh}, {Kamenetzky}, {Litke}, {Ma}, {Malkan},
  {Marrone}, {Menten}, {Murphy}, {Nadolski}, {Rotermund}, {Spilker}, {Stark},
  \& {Welikala}}]{strandet16}
{Strandet}, M.~L., {Weiss}, A., {Vieira}, J.~D., {et~al.} 2016, \apj, 822, 80

\bibitem[{{Swinbank} {et~al.}(2010){Swinbank}, {Smail}, {Longmore}, {Harris},
  {Baker}, {De Breuck}, {Richard}, {Edge}, {Ivison}, {Blundell}, {Coppin},
  {Cox}, {Gurwell}, {Hainline}, {Krips}, {Lundgren}, {Neri}, {Siana},
  {Siringo}, {Stark}, {Wilner}, \& {Younger}}]{swinbank10}
{Swinbank}, A.~M., {Smail}, I., {Longmore}, S., {et~al.} 2010, \nat, 464, 733

\bibitem[{{Tacconi} {et~al.}(2010){Tacconi}, {Genzel}, {Neri}, {Cox}, {Cooper},
  {Shapiro}, {Bolatto}, {Bouch{\'e}}, {Bournaud}, {Burkert}, {Combes},
  {Comerford}, {Davis}, {Schreiber}, {Garcia-Burillo}, {Gracia-Carpio}, {Lutz},
  {Naab}, {Omont}, {Shapley}, {Sternberg}, \& {Weiner}}]{tacconi10}
{Tacconi}, L.~J., {Genzel}, R., {Neri}, R., {et~al.} 2010, \nat, 463, 781

\bibitem[{{Tacconi} {et~al.}(2013){Tacconi}, {Neri}, {Genzel}, {Combes},
  {Bolatto}, {Cooper}, {Wuyts}, {Bournaud}, {Burkert}, {Comerford}, {Cox},
  {Davis}, {F{\"o}rster Schreiber}, {Garc{\'{\i}}a-Burillo}, {Gracia-Carpio},
  {Lutz}, {Naab}, {Newman}, {Omont}, {Saintonge}, {Shapiro Griffin}, {Shapley},
  {Sternberg}, \& {Weiner}}]{tacconi13}
{Tacconi}, L.~J., {Neri}, R., {Genzel}, R., {et~al.} 2013, \apj, 768, 74

\bibitem[{{Thomson} {et~al.}(2012){Thomson}, {Ivison}, {Smail}, {Swinbank},
  {Weiss}, {Kneib}, {Papadopoulos}, {Baker}, {Sharon}, \& {van
  Moorsel}}]{thomson12}
{Thomson}, A.~P., {Ivison}, R.~J., {Smail}, I., {et~al.} 2012, \mnras, 425,
  2203

\bibitem[{{Tremonti} {et~al.}(2004){Tremonti}, {Heckman}, {Kauffmann},
  {Brinchmann}, {Charlot}, {White}, {Seibert}, {Peng}, {Schlegel}, {Uomoto},
  {Fukugita}, \& {Brinkmann}}]{tremonti04}
{Tremonti}, C.~A., {Heckman}, T.~M., {Kauffmann}, G., {et~al.} 2004, \apj, 613,
  898

\bibitem[{{Vieira} {et~al.}(2010){Vieira}, {Crawford}, {Switzer}, {Ade},
  {Aird}, {Ashby}, {Benson}, {Bleem}, {Brodwin}, {Carlstrom}, {Chang}, {Cho},
  {Crites}, {de Haan}, {Dobbs}, {Everett}, {George}, {Gladders}, {Hall},
  {Halverson}, {High}, {Holder}, {Holzapfel}, {Hrubes}, {Joy}, {Keisler},
  {Knox}, {Lee}, {Leitch}, {Lueker}, {Marrone}, {McIntyre}, {McMahon}, {Mehl},
  {Meyer}, {Mohr}, {Montroy}, {Padin}, {Plagge}, {Pryke}, {Reichardt}, {Ruhl},
  {Schaffer}, {Shaw}, {Shirokoff}, {Spieler}, {Stalder}, {Staniszewski},
  {Stark}, {Vanderlinde}, {Walsh}, {Williamson}, {Yang}, {Zahn}, \&
  {Zenteno}}]{vieira10}
{Vieira}, J.~D., {Crawford}, T.~M., {Switzer}, E.~R., {et~al.} 2010, \apj, 719,
  763

\bibitem[{{Voss} {et~al.}(2006){Voss}, {Bertoldi}, {Carilli}, {Owen}, {Lutz},
  {Holdaway}, {Ledlow}, \& {Menten}}]{voss06}
{Voss}, H., {Bertoldi}, F., {Carilli}, C., {et~al.} 2006, \aap, 448, 823

\bibitem[{{Walter} {et~al.}(2012){Walter}, {Decarli}, {Carilli}, {Bertoldi},
  {Cox}, {da Cunha}, {Daddi}, {Dickinson}, {Downes}, {Elbaz}, {Ellis}, {Hodge},
  {Neri}, {Riechers}, {Weiss}, {Bell}, {Dannerbauer}, {Krips}, {Krumholz},
  {Lentati}, {Maiolino}, {Menten}, {Rix}, {Robertson}, {Spinrad}, {Stark}, \&
  {Stern}}]{walter12}
{Walter}, F., {Decarli}, R., {Carilli}, C., {et~al.} 2012, \nat, 486, 233

\bibitem[{{Walter} {et~al.}(2016){Walter}, {Decarli}, {Aravena}, {Carilli},
  {Bouwens}, {da Cunha}, {Daddi}, {Ivison}, {Riechers}, {Smail}, {Swinbank},
  {Weiss}, {Anguita}, {Assef}, {Bacon}, {Bauer}, {Bell}, {Bertoldi}, {Chapman},
  {Colina}, {Cortes}, {Cox}, {Dickinson}, {Elbaz}, {G{\'o}nzalez-L{\'o}pez},
  {Ibar}, {Inami}, {Infante}, {Hodge}, {Karim}, {Le Fevre}, {Magnelli}, {Neri},
  {Oesch}, {Ota}, {Popping}, {Rix}, {Sargent}, {Sheth}, {van der Wel}, {van der
  Werf}, \& {Wagg}}]{walter16}
{Walter}, F., {Decarli}, R., {Aravena}, M., {et~al.} 2016, ArXiv e-prints,
  arXiv:1607.06768

\bibitem[{{Wang} {et~al.}(2011){Wang}, {Cowie}, {Barger}, \&
  {Williams}}]{wang11}
{Wang}, W.-H., {Cowie}, L.~L., {Barger}, A.~J., \& {Williams}, J.~P. 2011,
  \apjl, 726, L18

\bibitem[{{Webb} {et~al.}(2004){Webb}, {Brodwin}, {Eales}, \& {Lilly}}]{webb04}
{Webb}, T.~M.~A., {Brodwin}, M., {Eales}, S., \& {Lilly}, S.~J. 2004, \apj,
  605, 645

\bibitem[{{Wei{\ss}} {et~al.}(2009){Wei{\ss}}, {Kov{\'a}cs}, {Coppin}, {Greve},
  {Walter}, {Smail}, {Dunlop}, {Knudsen}, {Alexander}, {Bertoldi}, {Brandt},
  {Chapman}, {Cox}, {Dannerbauer}, {De Breuck}, {Gawiser}, {Ivison}, {Lutz},
  {Menten}, {Koekemoer}, {Kreysa}, {Kurczynski}, {Rix}, {Schinnerer}, \& {van
  der Werf}}]{weiss09}
{Wei{\ss}}, A., {Kov{\'a}cs}, A., {Coppin}, K., {et~al.} 2009, \apj, 707, 1201

\bibitem[{{Wei{\ss}} {et~al.}(2013){Wei{\ss}}, {De Breuck}, {Marrone},
  {Vieira}, {Aguirre}, {Aird}, {Aravena}, {Ashby}, {Bayliss}, {Benson},
  {B{\'e}thermin}, {Biggs}, {Bleem}, {Bock}, {Bothwell}, {Bradford}, {Brodwin},
  {Carlstrom}, {Chang}, {Chapman}, {Crawford}, {Crites}, {de Haan}, {Dobbs},
  {Downes}, {Fassnacht}, {George}, {Gladders}, {Gonzalez}, {Greve},
  {Halverson}, {Hezaveh}, {High}, {Holder}, {Holzapfel}, {Hoover}, {Hrubes},
  {Husband}, {Keisler}, {Lee}, {Leitch}, {Lueker}, {Luong-Van}, {Malkan},
  {McIntyre}, {McMahon}, {Mehl}, {Menten}, {Meyer}, {Murphy}, {Padin},
  {Plagge}, {Reichardt}, {Rest}, {Rosenman}, {Ruel}, {Ruhl}, {Schaffer},
  {Shirokoff}, {Spilker}, {Stalder}, {Staniszewski}, {Stark}, {Story},
  {Vanderlinde}, {Welikala}, \& {Williamson}}]{weiss13}
{Wei{\ss}}, A., {De Breuck}, C., {Marrone}, D.~P., {et~al.} 2013, \apj, 767, 88

\bibitem[{{Whitaker} {et~al.}(2012){Whitaker}, {van Dokkum}, {Brammer}, \&
  {Franx}}]{whitaker12}
{Whitaker}, K.~E., {van Dokkum}, P.~G., {Brammer}, G., \& {Franx}, M. 2012,
  \apjl, 754, L29

\bibitem[{{Whitaker} {et~al.}(2014){Whitaker}, {Franx}, {Leja}, {van Dokkum},
  {Henry}, {Skelton}, {Fumagalli}, {Momcheva}, {Brammer}, {Labb{\'e}},
  {Nelson}, \& {Rigby}}]{whitaker14}
{Whitaker}, K.~E., {Franx}, M., {Leja}, J., {et~al.} 2014, \apj, 795, 104

\bibitem[{{Xu} {et~al.}(2007){Xu}, {Pirzkal}, {Malhotra}, {Rhoads}, {Mobasher},
  {Daddi}, {Gronwall}, {Hathi}, {Panagia}, {Ferguson}, {Koekemoer},
  {K{\"u}mmel}, {Moustakas}, {Pasquali}, {di Serego Alighieri}, {Vernet},
  {Walsh}, {Windhorst}, \& {Yan}}]{xu07}
{Xu}, C., {Pirzkal}, N., {Malhotra}, S., {et~al.} 2007, \aj, 134, 169

\bibitem[{{Yabe} {et~al.}(2014){Yabe}, {Ohta}, {Iwamuro}, {Akiyama}, {Tamura},
  {Yuma}, {Kimura}, {Takato}, {Moritani}, {Sumiyoshi}, {Maihara}, {Silverman},
  {Dalton}, {Lewis}, {Bonfield}, {Lee}, {Curtis-Lake}, {Macaulay}, \&
  {Clarke}}]{yabe14}
{Yabe}, K., {Ohta}, K., {Iwamuro}, F., {et~al.} 2014, \mnras, 437, 3647

\bibitem[{{Yamaguchi} {et~al.}(2016){Yamaguchi}, {Tamura}, {Kohno}, {Aretxaga},
  {Dunlop}, {Hatsukade}, {Hughes}, {Ikarashi}, {Ishii}, {Ivison}, {Izumi},
  {Kawabe}, {Kodama}, {Lee}, {Makiya}, {Matsuda}, {Nakanishi}, {Ohta},
  {Rujopakarn}, {Tadaki}, {Umehata}, {Wang}, {Wilson}, {Yabe}, \&
  {Yun}}]{yamaguchi16}
{Yamaguchi}, Y., {Tamura}, Y., {Kohno}, K., {et~al.} 2016, ArXiv e-prints,
  arXiv:1607.02331

\bibitem[{{Younger} {et~al.}(2007){Younger}, {Fazio}, {Huang}, {Yun}, {Wilson},
  {Ashby}, {Gurwell}, {Lai}, {Peck}, {Petitpas}, {Wilner}, {Iono}, {Kohno},
  {Kawabe}, {Hughes}, {Aretxaga}, {Webb}, {Mart{\'{\i}}nez-Sansigre}, {Kim},
  {Scott}, {Austermann}, {Perera}, {Lowenthal}, {Schinnerer}, \& {Smol{\v
  c}i{\'c}}}]{younger07}
{Younger}, J.~D., {Fazio}, G.~G., {Huang}, J.-S., {et~al.} 2007, \apj, 671,
  1531

\bibitem[{{Yun} {et~al.}(2012){Yun}, {Scott}, {Guo}, {Aretxaga}, {Giavalisco},
  {Austermann}, {Capak}, {Chen}, {Ezawa}, {Hatsukade}, {Hughes}, {Iono},
  {Johnson}, {Kawabe}, {Kohno}, {Lowenthal}, {Miller}, {Morrison}, {Oshima},
  {Perera}, {Salvato}, {Silverman}, {Tamura}, {Williams}, \& {Wilson}}]{yun12}
{Yun}, M.~S., {Scott}, K.~S., {Guo}, Y., {et~al.} 2012, \mnras, 420, 957

\bibitem[{{Zahid} {et~al.}(2014){Zahid}, {Dima}, {Kudritzki}, {Kewley},
  {Geller}, {Hwang}, {Silverman}, \& {Kashino}}]{zahid14}
{Zahid}, H.~J., {Dima}, G.~I., {Kudritzki}, R.-P., {et~al.} 2014, \apj, 791,
  130

\end{thebibliography}

\label{lastpage}

\end{document}